\documentclass{article}

\usepackage[margin=1in]{geometry}
\usepackage{adjustbox}
\usepackage[OT1]{fontenc} %
\usepackage{graphicx} %
\usepackage[table]{xcolor}%
\usepackage{caption} %
\usepackage{subcaption}
\usepackage{ctable} %
\usepackage{titlesec} %
\usepackage{multirow}
\usepackage{longtable}[=v4.13] %
\usepackage{tabu}
\usepackage{placeins}
\usepackage{lineno}
\usepackage{ulem}
\usepackage{tabu}
\usepackage{array}
\newcolumntype{?}{!{\vrule width 1pt}}
\usepackage{authblk}
\usepackage[symbol]{footmisc}
\usepackage{natbib}
\usepackage{bibunits}
\usepackage{placeins}
\usepackage{etoolbox}
\usepackage{amsfonts}
\usepackage{algorithm,algpseudocode,algorithmicx}
\usepackage{listings}
\usepackage{pifont}
\newcommand{\cmark}{\ding{51}}%
\newcommand{\xmark}{\ding{55}}%
\usepackage{makecell}
\usepackage{stmaryrd}
\usepackage{bm} %
\usepackage{dsfont} %
\usepackage{xcolor,colortbl}
\usepackage{siunitx}
\usepackage{amsmath,amssymb,amsfonts}%
\usepackage{amsthm}%
\usepackage{mathrsfs}%
\usepackage{textcomp}%
\usepackage{manyfoot}%
\usepackage{booktabs}\let\cline\cmidrule%

\usepackage[draft]{pgf}

\definecolor{LightGreen}{rgb}{0.8274509803921568, 0.9725490196078431, 0.8274509803921568}
\definecolor{white}{rgb}{1, 1, 1}

\newcolumntype{g}{>{\columncolor{LightGreen}}c}

\PassOptionsToPackage{hyphens}{url}\usepackage{hyperref}

\makeatletter
\newcommand*{\newbibstartnumber}[1]{%
  \apptocmd{\thebibliography}{%
    \global\c@NAT@ctr #1\relax
    \addtocounter{NAT@ctr}{-1}%
  }{}{}%
}
\makeatother

\definecolor{LightYellow}{rgb}{1,1,0.88}

\graphicspath{{figures/}}
\setcitestyle{square,comma,numbers,sort}

\newcommand{\Var}{\mathrm{Var}}

\usepackage{cleveref}
\definecolor{deepblue}{rgb}{0,0,0.5}
\definecolor{deepred}{rgb}{0.6,0,0}
\definecolor{deepgreen}{rgb}{0,0.5,0}

\DeclareFixedFont{\ttb}{T1}{txtt}{bx}{n}{12} %

\lstdefinestyle{mystyle}{
  language=Python,
  commentstyle=\color{deepgreen},%
  keywordstyle=\color{deepblue},
  basicstyle=\ttfamily\footnotesize,
  breakatwhitespace=false,
  breaklines=true,
  keepspaces=true,
  numbers=left,
  numbersep=5pt,
  showspaces=false,
  showstringspaces=false,
  showtabs=false,
  tabsize=4
}

\newcommand\totalwordcount{
    }
\newcommand\sectionwordcount{
}

\newcommand{\RNum}[1]{\uppercase\expandafter{\romannumeral #1\relax}}
\titleformat{\paragraph}
{\normalfont\normalsize\bfseries}{\theparagraph}{1em}{}
\titlespacing*{\paragraph}
{0pt}{3.25ex plus 1ex minus .2ex}{1.5ex plus .2ex}

\title{FedECA: Federated External Control Arms for Causal Inference with Time-To-Event Data in Distributed Settings}
\author[1$\dagger$]{Jean Ogier du Terrail\thanks{Corresponding author, e-mail: jean.duterrail.scientific.contact@gmail.com}}
\author[1$\dagger$]{Quentin Klopfenstein}
\author[1$\dagger$]{Honghao Li}
\author[1]{Imke Mayer}
\author[1]{Nicolas Loiseau}
\author[1]{Mohammad Hallal}
\author[1]{Michael Debouver}
\author[1]{Thibault Camalon}
\author[1]{Thibault Fouqueray}
\author[1]{Jorge Arellano Castro}
\author[1]{Zahia Yanes}
\author[2]{Laëtitia Dahan}
\author[3]{Julien Taïeb}
\author[4, 5]{Pierre Laurent-Puig}
\author[6]{Jean-Baptiste Bachet}
\author[4]{Shulin Zhao}
\author[7]{Remy Nicolle}
\author[8]{Jérôme Cros}
\author[9]{Daniel Gonzalez}
\author[10]{Robert Carreras-Torres}
\author[11, 10]{Adelaida Garcia Velasco}
\author[12]{Kawther Abdilleh}
\author[12]{Sudheer Doss}
\author[1]{Félix Balazard}
\author[1]{Mathieu Andreux}

\affil[1]{Owkin, Inc., New York, NY, USA}
\affil[2]{Department of Digestive Oncology, Hôpital la Timone, Marseille, France}
\affil[3]{GI oncology department Georges Pompidou European Hospital, Université Paris Cité, CARPEM CCC, 20 rue leblanc 75015 Paris, APHP, Paris, France}
\affil[4]{Centre de Recherche des Cordeliers, Sorbonne Université, Inserm, Université Paris Cité, Paris, France}
\affil[5]{Institut du Cancer Paris CARPEM, AP-HP Centre, Hôpital Européen Georges Pompidou, Paris, France}
\affil[6]{Sorbonne University, Hepatogastroenterology and digestive oncology department, Pitié Salpêtrière hospital, APHP, Paris, France}
\affil[7]{Université Paris Cité, Centre de Recherche sur l'Inflammation (CRI), INSERM, U1149, CNRS, ERL 8252, F-75018 Paris, France}
\affil[8]{Department of Pathology, Université Paris Cité - FHU MOSAIC, Beaujon Hospital, Clichy, France}
\affil[9]{Fédération Francophone de Cancérologie Digestive, Dijon, France}
\affil[10]{Institut d'Investigació Biomèdica de Girona (IDIBGI), Girona, Catalonia, Spain}
\affil[11]{Department of Medical Oncology, Catalan Institute of Oncology, Doctor Josep Trueta University Hospital, Girona, Catalonia, Spain}
\affil[12]{Pancreatic Cancer Action Network, El Segundo, CA, USA}

\begin{document}

\captionsetup[figure]{labelfont={bf},labelsep=period}

\maketitle

\totalwordcount
\footnotetext[2]{These authors contributed equally.}

\sectionwordcount
\begin{abstract}
External control arms can inform early clinical development of experimental drugs
and provide efficacy evidence for regulatory approval.
However, accessing sufficient real-world or historical clinical trials data is challenging.
Indeed, regulations protecting patients' rights by strictly controlling data processing
make pooling data from multiple sources in a central server often difficult.
To address these limitations, we develop a method that leverages federated learning to
enable inverse probability of treatment weighting for time-to-event outcomes on separate
cohorts without needing to pool data.
To showcase its potential, we apply it in different settings
of increasing complexity, culminating with a real-world use-case in which our method is
used to compare the treatment effect of two approved chemotherapy regimens using data from
three separate cohorts of patients with metastatic pancreatic cancer.
By sharing our code, we hope it will foster the creation of federated research
networks and thus accelerate drug development.
\end{abstract}

\sectionwordcount
\begin{bibunit}[unsrt]
\section{Introduction \label{sec:intro}}
Development of innovative drugs is a long and challenging task with increasing costs~\cite{dimasi2016innovation}.
The probability of success of a new drug is low, with about 10\% of drugs that enter clinical trials reaching FDA approval~\cite{hay2014clinical}.
Phase III randomized trials, that aim at
establishing clinical efficacy, fail approximately in one case out of two~\cite{dowden2019trends}.
Improving this rate is crucial to overcome those obstacles and accelerate the access to new drugs while reducing the development costs.

Statistical methods were developed to compare the efficacy of a treatment to a control group
that is built with data from external sources to the current trial (External Control Arm - ECA).
ECA methods take into account the potential bias introduced by the non-randomized nature of the control group,
and enable early assessment of treatment efficacy,
which can inform the transition from a single-arm phase II to a phase III clinical trial~\cite{ventz2019design,yin2023historic}.
ECA are increasingly used in clinical applications~\cite{wang2023current}, and are receiving more and more attention from
regulatory agencies (FDA, EMA)~\cite{fda2023ECA,EMA2023ECA}.
Externally controlled trials may also substitute randomized controlled trials (RCT) in specific situations where an RCT would be deemed unfeasible or untimely.
This is the case for rare diseases where patient recruitment is difficult and time-consuming~\cite{khachatryan2023external},
as well as in some oncology cases involving specific patient subgroups~\cite{mishrakalyani2022,lambert2022enriching,wang2023current,przepiorka2015fda}.

Statistically, the lack of randomization between the treatment group and the external control group
makes a naive comparison unreliable due to confounding bias.
Therefore, assuming that the treatment effect is identifiable~\cite{robins2001data},
statistical or machine learning methods~\cite{austin2011introduction,lunceford2004stratification,austin2016variance,robins1986new,chatton2020g,chernozhukov2018double,loiseau2022external}
are needed in this context to provide valid estimates and inferences of the treatment effect.
Despite advances in statistical methods, data sharing is a major obstacle to the feasibility of ECA.
Due to their sensitivity, health data are strictly regulated, e.g., by the General Data Protection Regulation (GDPR)
in the EU and the Health Insurance Portability and Accountability Act (HIPAA) in the US.
Even after careful pseudonymization~\cite{ohmann2017sharing}, sharing health
data remains a complex endeavor, notably because of data ownership and liability issues.
Thereby, in practice, it is difficult to set up an ECA involving phase II data from a pharmaceutical company and
potentially real-world data from different hospitals or medical institutions.

To address this data sharing challenge, various machine learning techniques have been proposed in recent years.
Among them, federated learning (FL)~\cite{mcmahan2017communication}, a privacy-enhancing technology (PET),
makes it possible to extract knowledge and train models from multiple institutions without pooling data.
It has already been used with success in similar settings to connect pharmaceutical companies~\cite{Oldenhof2023} and hospitals~\cite{pati2022federated, du2023federated}
in federated research networks.
Herein, we investigate the use of FL to build ECA, focusing on time-to-event outcomes such as
progression-free survival (PFS) or overall survival (OS), which are predominant
in oncology RCTs~\cite{le2021time}.

In this work, we present FedECA, a federated external control arm method which is
a federated version of the well-known inverse probability of treatment weighting (IPTW) statistical method~\cite{rosenbaum1983central} for time-to-event outcomes.
FedECA facilitates ECAs for pharmaceutical companies
by giving them access to real-world control patients from multiple institutions, while
limiting patient data exposure thanks to FL.
We first demonstrate the efficacy of our approach on synthetic data
created using realistic data generation processes, both with in-RAM simulations
and on a federated network deployed with 10 distant synthetic centers
located in the cloud.
We show that FedECA achieves identical conclusions to IPTW on pooled data as well as better statistical
power compared to a competing method based on federated analytics (FA), while also controlling for moment
differences between the two arms.
Furthermore, we showcase two examples of FedECA applied to real patient
data, starting with a FL simulation using real data from trials before moving on to
the end-to-end deployment of a real federated research network between three research institutions
located in different countries.

\sectionwordcount
\section{Results \label{sec:res}}

\subsection{FedECA, a federated ECA method}

Here we describe our federated extension of the IPTW method for time-to-events outcomes: FedECA.
FedECA estimates the treatment effect by comparing the experimental drug arm
stored in one center, with a control arm defined by external data held within different centers,
as illustrated in Figure~\ref{fig:graphical_abstract}.
FedECA consists of three main steps, all performed via FL.
It first trains a propensity score model using logistic regression to obtain weights,
then fits a weighted time-to-event Cox model to correct for potential confounding bias,
and finally computes an aggregated statistic which allows to test the treatment effect. See Methods~\ref{sec:method_overview} for more details.
Simultaneously we develop alongside FedECA FA methods in order to both visualize and validate the results of FedECA (see Methods~\ref{sec:fed_analytics}) as one would in an equivalent pooled ECA analysis.

Figure \ref{fig:graphical_abstract} illustrates the advantages of our proposed method,
where data can remain on the premises of the participating centers and only aggregated information is shared across a number of communication rounds.
Herein, an aggregator node is responsible for orchestrating the training process, aggregating and redistributing the results to all centers,
without directly seeing the raw data itself.
This is in contrast to the classical ECA analysis, where data is pooled into a single center and data privacy is not an issue.
In this privacy-enhanced context that we tackle, and that we describe in details in Methods~\ref{sec:dp}, all kinds of data analyses from simple statistics
such as computing global variance or histograms to more complex methods such as IPTW are markedly more difficult to apply as one can only share aggregated data.

\subsection{FedECA is equivalent to a standard IPTW model trained on pooled data \label{sec:fed_vs_pooled}}

In spite of such difficulties, we demonstrate both mathematically and numerically FedECA's equivalence to the pooled IPTW model.
Mathematical proof of the equivalence under proper assumptions can be found in Methods~\ref{sec:iptw_webdisco}.
In particular, an important assumption underlying this derivation is the use of the Breslow's approximation for tied-times
when constructing the partial likelihood of the Cox model.
In addition to mathematical equivalence, we also study numerical equivalence,
which could be affected by the propagation of machine precision errors during repeated computations.
We demonstrate this equivalence between pooled and federated IPTW analyses on realistic simulated data
with right-censored events, 10 normally distributed and correlated covariates,
and a treatment allocation depending on the covariates.

We refer to Methods~\ref{subsec:data_sim} for details on the data generation process.
In the following numerical experiment, we compare the results obtained from a classical IPTW analysis
where data are pooled into the same place with those obtained from FedECA operating in distributed
settings.
Here we monitor four key metrics:
the hazard ratio of the treatment allocation covariate estimated from a Cox model,
the partial likelihood of the Cox model,
the p-value associated to the hazard ratio (HR), derived from a two-sided Wald test that
assumes under the null hypothesis an asymptotic chi-squared distribution with one degree of freedom,
and the propensity scores estimated from the logistic regression.
We repeat this simulation 100 times and report the relative errors
between pooled IPTW and FedECA in Figure~\ref{fig:pooled_equivalent}.
It should be noted that the relative error we report also takes into account
potential differences depending on the optimizer used to train the propensity model.
For instance, in our implementation for the pooled reference, we rely on \verb+sklearn+'s default optimizer,
which is the Limited memory Broyden-Fletcher-Goldfarb-Shanno (LBFGS) optimizer,
and thus differs slightly from the exact Newton-Raphson steps used in FedECA.

The results in Figure~\ref{fig:pooled_equivalent} show that the relative errors between FedECA and the pooled IPTW
are negligeable, not exceeding 0.2\%, illustrating the effectiveness of the proposed
federated optimization process.
Moreover, \ref{fig:pooled_equivalent_nb_clients}
shows that the number of centers among which data is split has no impact on FedECA's performance.
Hence, we illustrate that up to a negligible error, due to finite precision
numerical errors in the optimization process, FedECA provides results that are
equivalent to the classical IPTW despite not having access to all data in the same
location.

\subsection{FedECA and MAIC both control SMD}
To assess the performance of FedECA, we compare it with another more naïve federated method
in terms of the ability to correctly detect the presence of a treatment effect.
We start by focusing on the reweighting step of FedECA, an important step to correct for the confounding bias,
as measured by the standardized mean difference (SMD) of covariates between the two patient groups after reweighting.
The SMD is a coarse univariate measure which summarizes for each covariate
the balance between the two groups by only looking at the first and second order
moments, see Methods Section~\ref{sec:smd}. However SMD is expected by regulators~\cite{us2023considerations},
which ask ECA methods to control SMD below a threshold (usually $10\%$).
In this context, the main competitor of FedECA is the matching-adjusted indirect comparison (MAIC)~\cite{signorovitch2012matching},
a method that allows to reweight the individual patient data (IPD) from a treatment arm
to match the mean and the standard deviation of data from an external control arm
for which the IPD are not accessible.
The two reweighting approaches differ mainly in two ways.
First, since MAIC's reweighting procedure involves communicating statistics only once between the centers,
it is essentially a federated analytic method (FA), whereas FedECA's reweighting uses FL.
Second, while MAIC explicitly enforces perfect matching of low order moments irrespective of the covariate shift,
leading to zero SMD by design (see Methods Section~\ref{sec:maic}), FedECA does multivariate balancing
through the propensity scores.

To compare the two methods, we consider scenarios with different levels of covariate shift,
which is a parameter that controls the intensity of the
confounding factors on the treatment allocation variable biasing the two groups.
A covariate shift of zero is equivalent to a random allocation in the treatment arms
(see Methods~\ref{subsec:data_sim} for more details).
In this experiment, in one end of the spectrum, treatment allocation does not
depend on the covariates: there is no covariate shift. Therefore the SMDs of all covariates
are small, even before reweighting. On the other end of the spectrum, treatment allocation depends more and more
on the covariates values (details of the data generation are given in Section~\ref{subsec:data_sim}). Therefore
we expect weighting to be necessary to control SMD.
The results are illustrated in Figure~\ref{fig:smd_a} where we show the mean absolute SMD
as a function of the covariate shift for FedECA, MAIC and the unweighted baseline.
For low covariate shift ($\leq0.5$) all three methods control the SMDs of the covariates,
while for medium to high covariate shifts ($>0.5$), the unweighted method fails to control the SMDs while both MAIC and FedECA
control SMD. More details per-covariate are available Figure~\ref{fig:smd_b} for the two extremes.

\subsection{FedECA outperforms MAIC in power to detect a treatment effect \label{sec:power_analysis}}
Following the comparison using SMD, we now compare the effect of the different reweighting offered by FedECA and MAIC
in terms of treatment effect estimation, as measured by type I error and statistical power.
After reweighting, FedECA trains a Cox model using weighted time-to-event data of both arms.
The presence of a treatment effect is determined by the hazard ratio estimated from the Cox model,
as well as the associated p-value. A p-value less than $0.05$ is considered significant.
For the type I error experiment, we generate synthetic data with no treatment effect between the two arms,
while for the statistical power experiment, we generate synthetic data with a true treatment effect.
Note that for MAIC, unlike for FedECA, the training of the Cox model assumes the pooling of
all IPD on treatment allocation, time to event, as well as censoring.
While this assumption seemingly contradicts the ECA's setup, it can be considered as an idealization
of a real-world use case of ECA analysis using MAIC, as detailed in Section~\ref{sec:maic}.

Figure \ref{fig:power_typeI} shows the estimated type I error and statistical power
for FedECA, MAIC and the unweighted baseline, under varying covariate shift and number of samples.
One of the key factors influencing the type I error and statistical power estimation is the
variance estimation method applied for each treatment effect point estimation.
Here, we compared three variance estimation methods: the bootstrap estimator,
the robust sandwich estimator, and the naive estimator based on the
inversion of the observed Fisher information~\cite{austin2016variance}.
For FedECA, only the bootstrap variance estimator successfully controls the type I
error at around $5\%$. In comparison, the robust variance estimator systematically
overestimates the variance, leading to an inflated p-value and thus a conservative empirical type I error rate.
Finally, the naive variance estimator fails to control the type I error.
These results are consistent with previous work~\cite{austin2016variance}.
For MAIC with bootstrap variance estimation, resampling with replacement is performed
only on the treatment group, since the pseudo IPD of the control arm are fixed by design.
Compared with FedECA, it only controls the type I error for small covariate shifts,
and loses control when the covariate shift increases.
The robust variance estimator shows similar variance overestimation to FedECA.
For the unweighted baseline, as it cannot account for the confounding bias
introduced by covariate shift, it quickly loses control of
the type I error as soon as the covariate shift is no longer zero.
Next, for those methods that successfully control the type I error, we compare their statistical power.
FedECA with bootstrap variance estimator shows the best performance,
followed by FedECA with the robust variance estimator. Both FedECA variants outperform
MAIC with the robust variance estimator, as the covariate shift and number of samples change.
Based on the above results, we choose the bootstrap variance estimator
for all experiments on treatment effect estimation.

\subsection{FedECA can be used in real-world conditions on synthetic data \label{sec:real_world_analysis}}

We host up to 11 ``servers" in the cloud (10 "centers" and 1 aggregation server) and deploy the Substra~\cite{galtier2019substra}
software on all centers. Details of the cloud setup are available in Section~\ref{subsec:real_world}.
For each experiment, we use the first of the servers as the trusted third party performing the aggregation (the ``server") and the other
servers as data owners holding a different part of the data (the ``centers").
Each ``center" has a different set of credentials, giving it different permissions
over the assets created in the federated network.
Each center registers a predefined subset of the synthetic data as if it were its own through
the Substra system.
FedECA can be run on the deployed network simply by changing the type of backend
used and specifying the identifiers of the datasets (hashes) registered into the Substra platform as inputs
to the fit method following \verb_scikit-learn_'s fit API~\cite{pedregosa2011scikit}.
An example of FedECA python API is given in~\ref{list:code_example}.
We monitor the runtime of the full pipeline when running in-RAM and on the cloud
as a function of the number of centers. We give conservative estimates by setting
a large target number of rounds ($20$) for the training of the propensity model and the Cox
model, and by computing the federated robust sandwich estimator,
which adds overhead and is not necessary if using the bootstrap variance estimator.

In-RAM experiments take a few seconds with $10$ centers, which is to be compared with IPTW on pooled data, which has a
below-second runtime. This slowdown is mainly due to (1) the sequential processing of each client
(2) the static nature of the Substra framework, which forces the execution of a higher target number of rounds
than needed for convergence (see Section~\ref{sec:static} for further explanation).
While (2) is a fundamental limitation of Substra, (1) could be improved by using Python multiprocessing.
The real-world runtime is almost constant with respect to the number of clients
and is under $2$ hours ($1$h$18$min on average, with a standard deviation
of around $3$ minutes). A complete breakdown of the different runtimes across settings is shown in ~\ref{tab:real_world_synth}.
Insofar as $10$ centers is already a large number in the considered cross-silo FL setup,
this result suggests good scalability in terms of speed, provided that an appropriate infrastructure
can be deployed across the different centers. Our result is consistent with previous Substra deployments~\cite{Oldenhof2023}.

\subsection{FedECA can be used on real world use-cases: Application on real prostate cancer data with simulated federated learning \label{subsec:prostate}}

We access data from two phase III trials in metastatic castration-resistant prostate cancer
from the Yale University Open Data Access (YODA)
project~\cite{krumholz2016yale, ross2018overview}.
In all cases, we simulate an FL setup in which data are held by two synthetic centers, one
holding the treated arm and the other the remaining patients. We note that, according to our experiments
~\ref{fig:pooled_equivalent_nb_clients}, the number of centers has no impact on FedECA's performance
(nor does the way the data are distributed across the different centers),
and we therefore expect roughly the same results if we had chosen different data splits.
Full details of cohort construction can be found in Methods~\ref{sec:yoda_cohort_construction}.

In the following, we present results focused on estimating the average treatment effect
on the radiographic progression-free survival (rPFS) endpoint
by comparing regimens in different trials.
We first show the results reported for each trial found in the associated publications.
Then, we conduct simulated ECA studies by replacing the abiraterone acetate + prednisone (AA-P) arm of
each trial with the same arm from the other trial. In addition, we compare the two AA-P arms to test
the exchangeability of the two study populations and to validate previous ECA analyses.
Finally, we compare the apalutamide + abiraterone acetate + prednisone (Apa-AA-P) arm with the prednisone (P) arm,
which is not seen in the literature and demonstrates the potential usefulness of our
method to provide additional evidence that is otherwise unavailable or costly to produce in terms of time and resources.
In Table~\ref{tab:real_world_stats_yoda}, we show the consistency of the
treatment effect estimations with the published results, as well as the non-significance of the treatment
effect when comparing two AA-P arms. In addition, we show a significant treatment
effect of Apa-AA-P versus P, which is reasonable considering the
superiority of Apa-AA-P over AA-P, and of AA-P over P.

\subsection{FedECA can be used on real world use-cases: Application on real metastatic pancreatic cancer data in a real deployed federated research network \label{subsec:pancreas}}

We access metastatic pancreatic cancer data in three cancer centers:
the Fédération Francophone de Cancérologie Digestive (FFCD), which holds data from two completed RCTs,
the Institut d'Investigació Biomèdica de Girona (IDIBGI)
and the Pancreatic Cancer Action Network (PanCAN), which hold data from clinical practice.
Full details of cohort construction can be found in Methods~\ref{sec:cohort_construction}.
Regarding FL setup, we deploy a Substra-based federated network across the three centers.
Contrary to the experiment in Section~\ref{subsec:prostate} with a simulated FL setup,
in order to perform any statistical analysis involving data from multiple centers now one has to go through the Substra federated network infrastructure and thus
requires executing Substra's ``compute plans'', which comes with overhead and imposes constraints on what can be computed safely (see Section~\ref{sec:dp}).
More details of the FL setup can also be found in Section~\ref{subsec:real_world_pancreatic}.

We aim to estimate the treatment effect of FOLFIRINOX over gemcitabine and nab-paclitaxel. Given that all three centers have patients in both treatment groups,
we begin by testing a key assumption required to combine cohorts from different centers receiving the same treatment: the exchangeability between pairs of centers.
For each treatment group, we compare patients from two centers and test if there is a center effect after correcting for measured confounders with IPTW
(e.g. we test if patients under FOLFIRINOX have different outcomes between FFCD and IDIBGI). Such comparisons are done for all three pairs of centers.
Unfortunately we observe a strong center effect at the PanCAN center whose population has significantly better outcomes than the other centers in both treatment groups
(Supplementary \ref{tab:exchang}, \ref{fig:exchang_km}).
We suspect that there is a residual immortal time bias not addressed by the correction we performed in this cohort (see Methods~\ref{sec:cohort_construction}).
Consequently, we exclude the entire PanCAN cohort from the treatment effect analysis that we present in the remainder of this section,
and leave the results of a federated analysis including data from all three centers in Supplementary
(\ref{tab:real_world_all_centers}, \ref{fig:real_world_all_and_pancan}) for illustrative purposes.

Using the combined FFCD and IDIBGi cohort in FedECA, we compare the efficacy of FOLFIRINOX versus gemcitabine and nab-paclitaxel on overall survival (OS).
Figure~\ref{fig:real_world}(a) demonstrates that the propensity model in FedECA, trained with FL, effectively balances the
two patient groups over the group of selected covariates, reducing the SMD between the two arms to below 10\% for all covariates,
which was not achieved before reweighting.
Table~\ref{tab:real_world_stats} shows the estimated treatment effect on overall survival of FOLFIRINOX over gemcitabine and nab-paclitaxel
with an HR of $0.84\,(0.68, 1.04)$ and an associated p-value of $0.118$. While this result does not reach statistical significance,
it is consistent with the literature using IPTW on pooled data (e.g. HR $=0.77\,(0.70, 0.85)$\cite{chan2020real}),
and goes in the direction of superiority of FOLFIRINOX over gemcitabine and nab-paclitaxel.
For comparison, local analyses at each center (Table~\ref{tab:real_world_stats}) result in broader confidence intervals of the estimated HR.
Figure~\ref{fig:real_world}(c)-(d) highlight the above results by displaying propensity-weighted Kaplan-Meier curves.
Note that in Figure~\ref{fig:real_world} and \ref{fig:real_world_all_and_pancan}, results on the combined cohorts are obtained
through the application of federated analytics (FA) without pooling data, see Sec~\ref{sec:fed_analytics} and \ref{fig:fed_kaplan_graph}
and \ref{fig:fed_smd_graph}.

\sectionwordcount
\section{Discussion\label{sec:disc}}

In this work we have introduced FedECA, a federated extension of the IPTW method for estimating treatment effects in the context of external control arms.
Our results demonstrate that FedECA replicates its pooled-equivalent counterpart IPTW up to machine precision (see Figure~\ref{fig:pooled_equivalent}), ensuring the same statistical properties as IPTW.
Compared to the simpler federated analytic baseline MAIC, FedECA shows superior statistical power and controls the type I error while effectively adjusting for confounding factors as shown by the standardized mean difference below 10\% (see Figure~\ref{fig:power_typeI}).
Unlike many stratified competitors~\cite{buchanan2014worth, shu2020inverse, luo2022odach,li2022distributed, park2022wicox}, FedECA is well-suited for drug development settings, where treated and control patients are in separate locations with pharmaceutical companies holding only treated patient data and control patients spread across multiple institutions.
It also remains applicable in scenarios where both treatment arms are distributed across different institutions, as demonstrated in the metastatic pancreatic adenocarcinoma experiment.
While MAIC can only estimate the average treatment effect on the control (ATC), FedECA supports the estimation of the ATE, the average treatment effect on the treated (ATT), as well as the ATC by changing the weights used in the estimation.
Furthermore, FedECA also enables covariate adjustment through adjusted IPTW, providing researchers with greater flexibility in choosing 
between a marginal effect or a conditional one ~\cite{daniel2021making}.
This distinction is important in the context of time-to-event outcomes due to the non-collapsibility of the hazard ratio.

FedECA is not only an algorithm but also a software solution that has been deployed and tested in real-world settings.
A first experiment demonstrates that FedECA can reproduce results of individual clinical trial~\cite{saad2021nct02257736, ryan2013nct00887198,ryan2015nct00887198}, while also enabling analyses that would not be feasible if each trial was analyzed separately.
Although the data used in this experiment were not physically distributed in different locations, the results of those simulations are representative of what could be achieved on similar data in a real federated setting.
In a second experiment, we deployed a Substra federated research network across three cancer centers in three different countries: the Fédération Française de Cancérologie Digestive (FFCD, France), the institut d'investigació biomèdica de Girona (IDIBGI, Spain) and the Pancreatic Cancer Action Network (PanCAN, USA).
This study aimed to estimate the ATE of FOLFIRINOX (leucovorin and fluorouracil plus irinotecan and oxaliplatin) over gemcitabine and nab-paclitaxel~\cite{pusceddu2019comparative, chiorean2019real, williet2019folfirinox, klein2022comparison}.
After excluding data from one center affected by immortal time bias, our results, while not reaching statistical significance, are consistent with the findings of meta-analyses and pooled analyses from the literature. 
While the results of our analysis rely on a smaller sample size ($n=378$) than the largest previous efforts~\cite{hegewisch2019tpk, chan2020real, riedl2021gemcitabine, chun2021comparison, klein2022comparison}, it brings additional evidence on this topic of rising medical interest~\cite{hepatology2023cause}. 

One of the key challenges of real-world data is handling missing values or missing features, as encountered in our metastatic pancreatic adenocarcinoma experiment.
While extensive research exists on missing data imputation in machine learning and its impact on causal inference analyses~\cite{pantanowitz2009missing, stekhoven2012missforest, cerda2018similarity, le2020neumiss, mayer2020doubly, le2021sa}, it remains underexplored in the context of federated learning.
In this study, we used a naive solution by applying MissForest~\cite{stekhoven2012missforest} independently at each site.
Imputation per-site is suboptimal and could possibly further increase heterogeneity and biases.
Addressing these limitations requires future work on federated missing data imputation methods.
Beyond missing data, another important aspect is the choice of confounding factors when building an ECA based on propensity scores~\cite{yao2021survey}.
FedECA remains sensitive to misspecification in the propensity score method as its pooled version, IPTW.
When building an ECA, one should carefully select the confounders to include in the propensity score method and should consider the possibility of unmeasured confounders as well as ways to perform sensitivity analysis to assess the robustness of the results~\cite{imbens2003sensitivity}.
Additionally, when performing an ECA analysis, it is crucial to ensure consistent data collection across centers, particularly regarding the definition of endpoints.
For example, progression free survival (PFS) is not defined in a standardized way in clinical practice, which can lead to bias and errors in the estimation of the treatment effect.

In this work, we focus on the (log) hazard ratio as the treatment effect estimand as it is commonly used with time-to-event endpoints~\cite{chan2020real,saad2021nct02257736,ryan2013nct00887198,ryan2015nct00887198}.
Other effect measures have been proposed for time-to-event outcomes such as contrasts of restricted mean survival time (RMST)~\cite{zhao2016restricted}.
The latter has the advantage of being collapsible and offers better clinical relevance and interpretability in certain applications~\cite{pak2017interpretability}.
An IPTW-based estimator for the difference of RMST has been introduced~\cite{conner2019adjusted} and could guide an extension of FedECA to RMST-based effect estimation in a federated setting.

Another important open research direction that we leave to future work is to study more deeply the security profile of standalone FedECA.
Indeed, the federation of both the propensity score model and the Cox proportional hazards (PH) model currently requires transmitting aggregated information to the central server.
In the absence of local differential privacy (DP) mechanisms, this transmitted information could, in theory, be exploited by adversarial attacks (we sketch how such attacks could be performed in Methods Sec.~\ref{sec:dp}).
Because of this reason, in addition to providing the pseudo-code of our FedECA algorithms, we traced in our implementation all quantities exchanged across centers when
performing the different federated learning and analytics algorithms involved in FedECA allowing a full privacy audit of the implementation by experts using informations reported in Methods~\ref{sec:dp}.
We tested the application of DP to the first part of the FedECA training, which is the training of the propensity model. However, it was shown to be already
detrimental to the statistical analysis (see \ref{fig:dp_results} and \ref{fig:fedeca_dp_graph} with ~\ref{tab:dp_prop_vars}).
This DP experiment raises an interesting question which is whether the use of DP in the clinical trials of tomorrow is warranted as it trades off accuracy of treatment effect estimation for data privacy.
We do not pretend to answer this question in this work.

Another privacy enhancing layer that could be added to FedECA would be to use secure aggregation (SA)~\cite{bonawitz2017practical} to hide individual contributions through cryptographic operations.
This would demonstrably hide potentially sensitive information such as per-client risk sets and would also allow private set unions (PSU)~\cite{bloom1970space} to compute the global event times or
could also be used to secure the aggregator node~\cite{bonawitz2017practical, NEURIPS2022_ed3c686f}.

To conclude, FedECA is a federated method for real-world distributed ECA analysis.
FedECA is particularly suited for the scenario where treated patient data are in a distinct center and the external control arm is split across different centers that cannot share their data.
FedECA is a federated extension of IPTW that reproduces the result of a pooled analysis, yielding similar treatment effect estimation and similar statistical guarantees.
FedECA enables causal inference in distributed ECA settings while limiting IPD exposure.
We demonstrated that FedECA is not only a simulation tool but a valid method for real-world applications showcasing its ability in two clinically different contexts.

Implementing federated methods in real-world healthcare environments, as we did for the metastatic pancreatic cancer use case, still present major technical and operational challenges that are absent from simulated FL.
Our implementation of FedECA is based on an open source FL software hosted by the LFAI, which had already been successfully used in the targeted high-security healthcare setting with both pharmaceutical companies and cancer centers~\cite{du2023federated,Oldenhof2023}.
Having a trusted implementation like this one, whose security was audited and that is compatible with heterogeneous environments is a prerequisite for building real FL networks as the implementation needs to be vetted by the different IT (Information Technology) teams from all partner institutions.
While we focus the scope of this paper on the technical and methodological challenges, we want to emphasize that the non-algorithmic challenges associated with setting up any FL networks between real institutions are still to this day potentially the main bottlenecks in such analyses as we touch upon in Methods~\ref{sec:related_works} . 

We hope FedECA, alongside other real-world applications of federated learning, can shift perspectives and drive collaborations between hospitals, medical centers, and pharmaceutical companies, demonstrating that medical discoveries are achievable while minimizing patient data exposure.

\FloatBarrier

\FloatBarrier
\newpage
\clearpage
\begin{figure*}[!h]
  \centering
  \begin{subfigure}[b]{\textwidth}
    \centering
    \includegraphics[width=.49\linewidth,trim={6cm 4cm 6cm 4cm},clip]{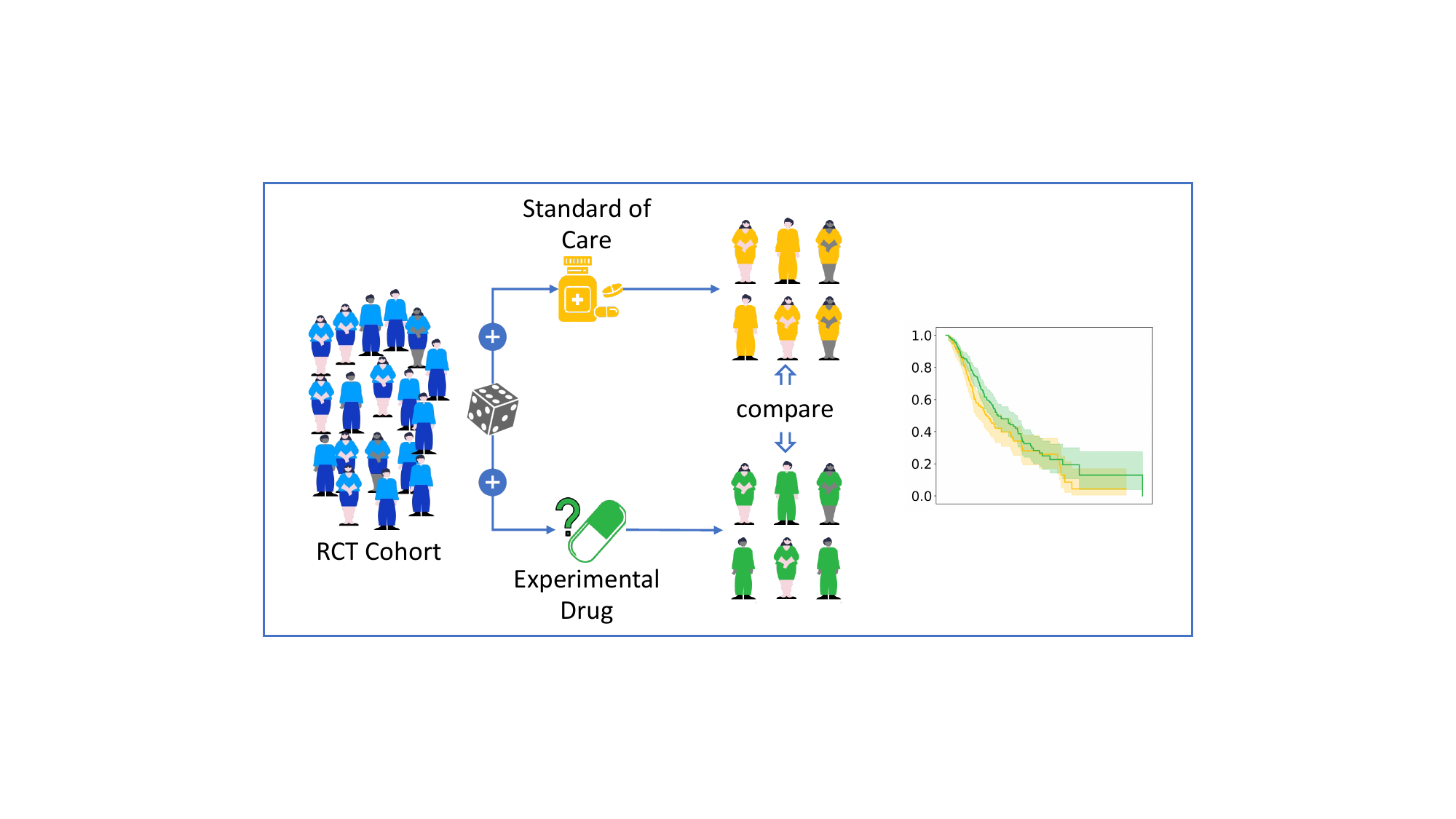}
    \includegraphics[width=.49\linewidth,trim={6cm 4cm 6cm 4cm},clip]{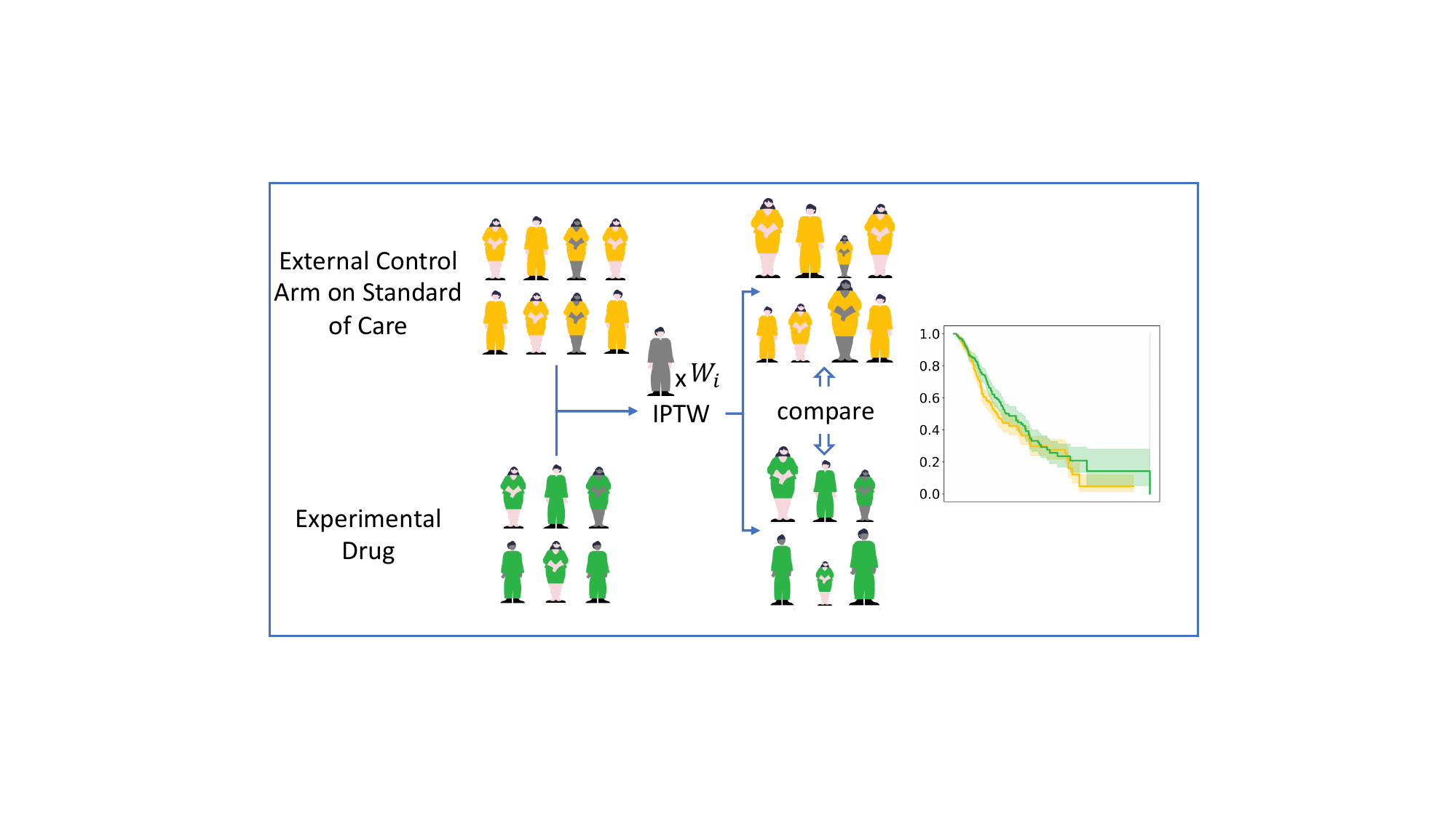}
    \includegraphics[width=.49\linewidth,trim={6cm 4cm 6cm 4cm},clip]{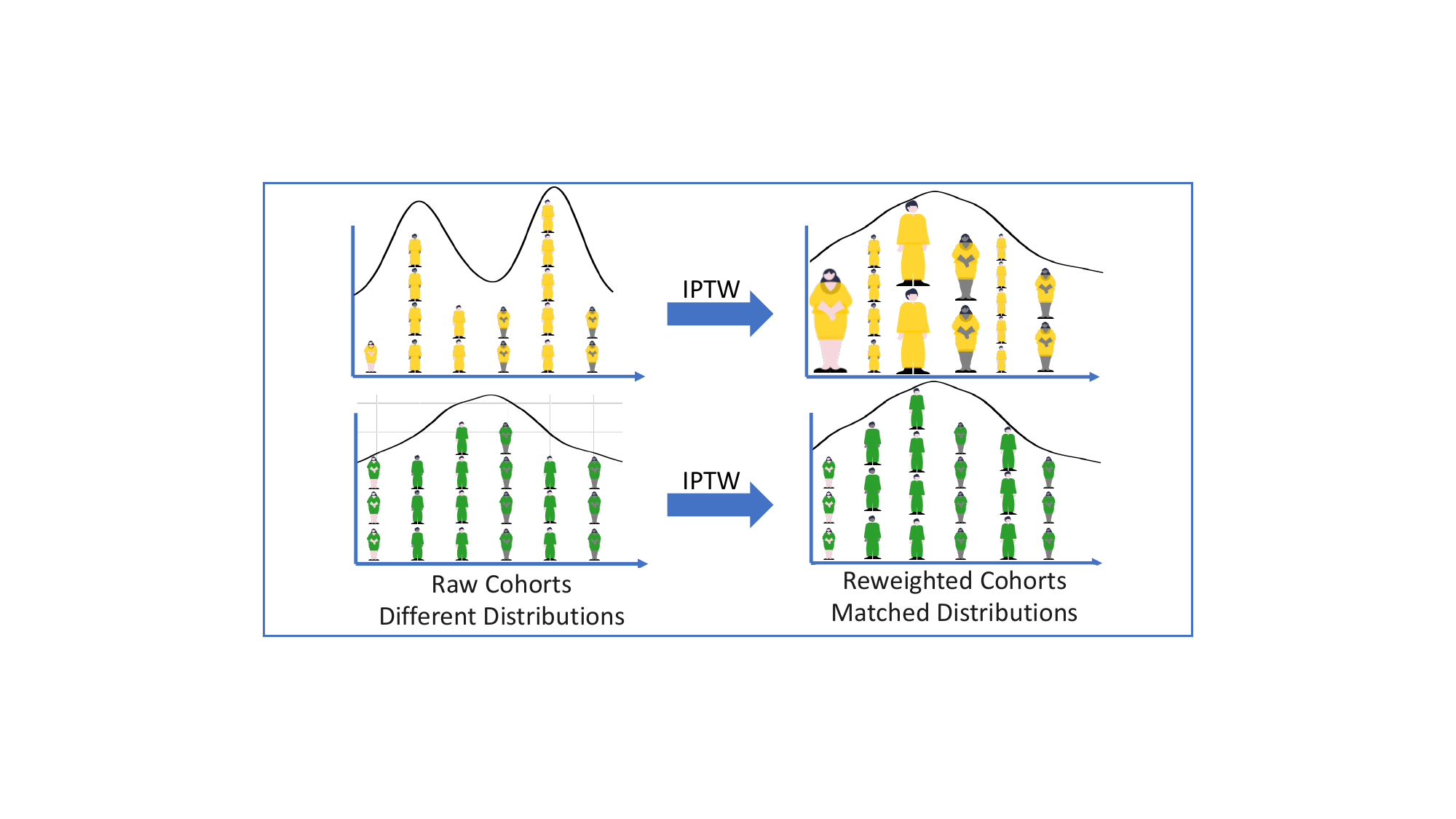}
  \end{subfigure}
  \begin{subfigure}[b]{0.8\textwidth}
    \centering
    \includegraphics[width=1.\linewidth]{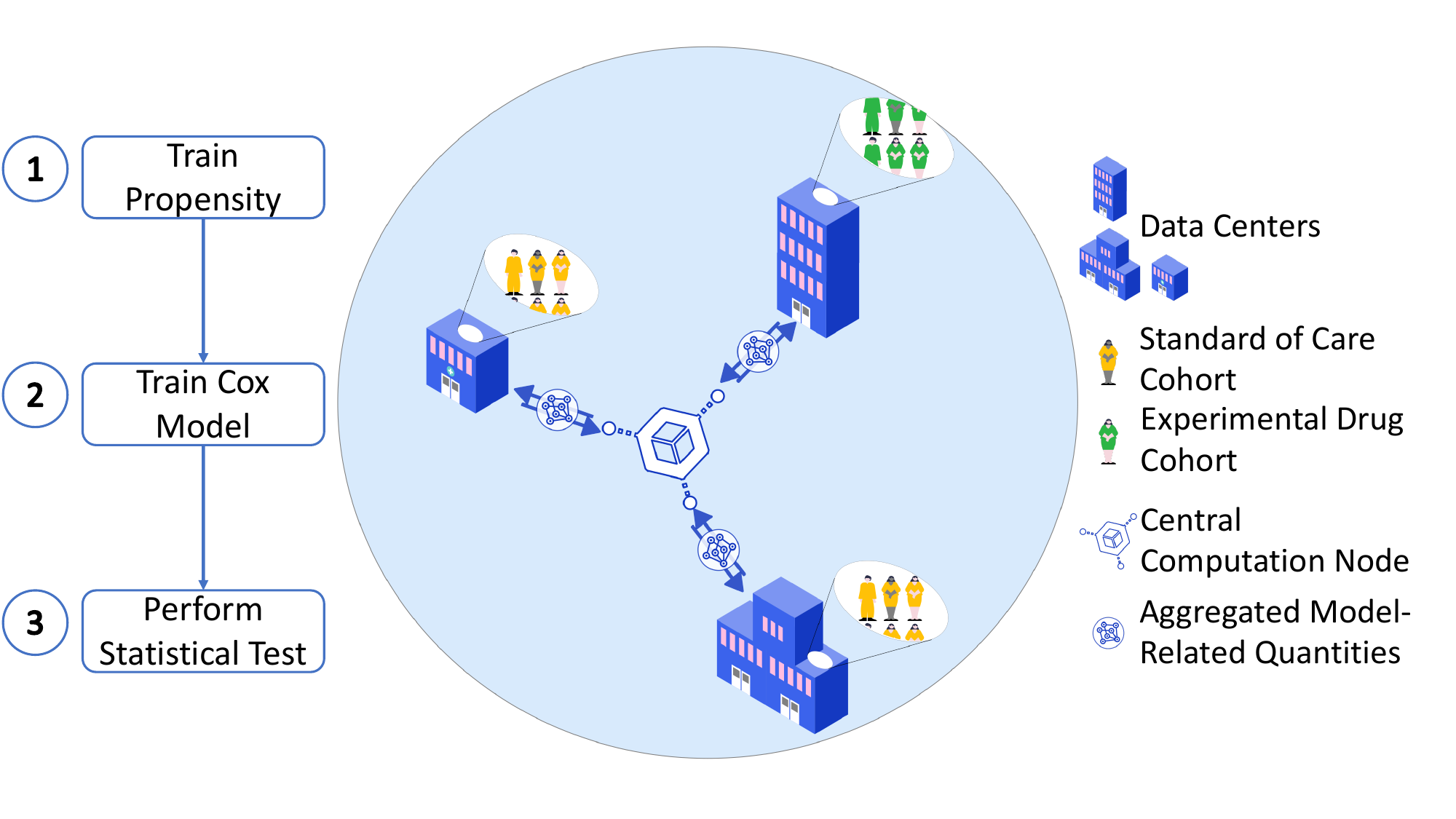}
  \end{subfigure}
\end{figure*}
\clearpage
\begin{figure*}[tb]
    \centering
    \includegraphics[width=1.\linewidth]{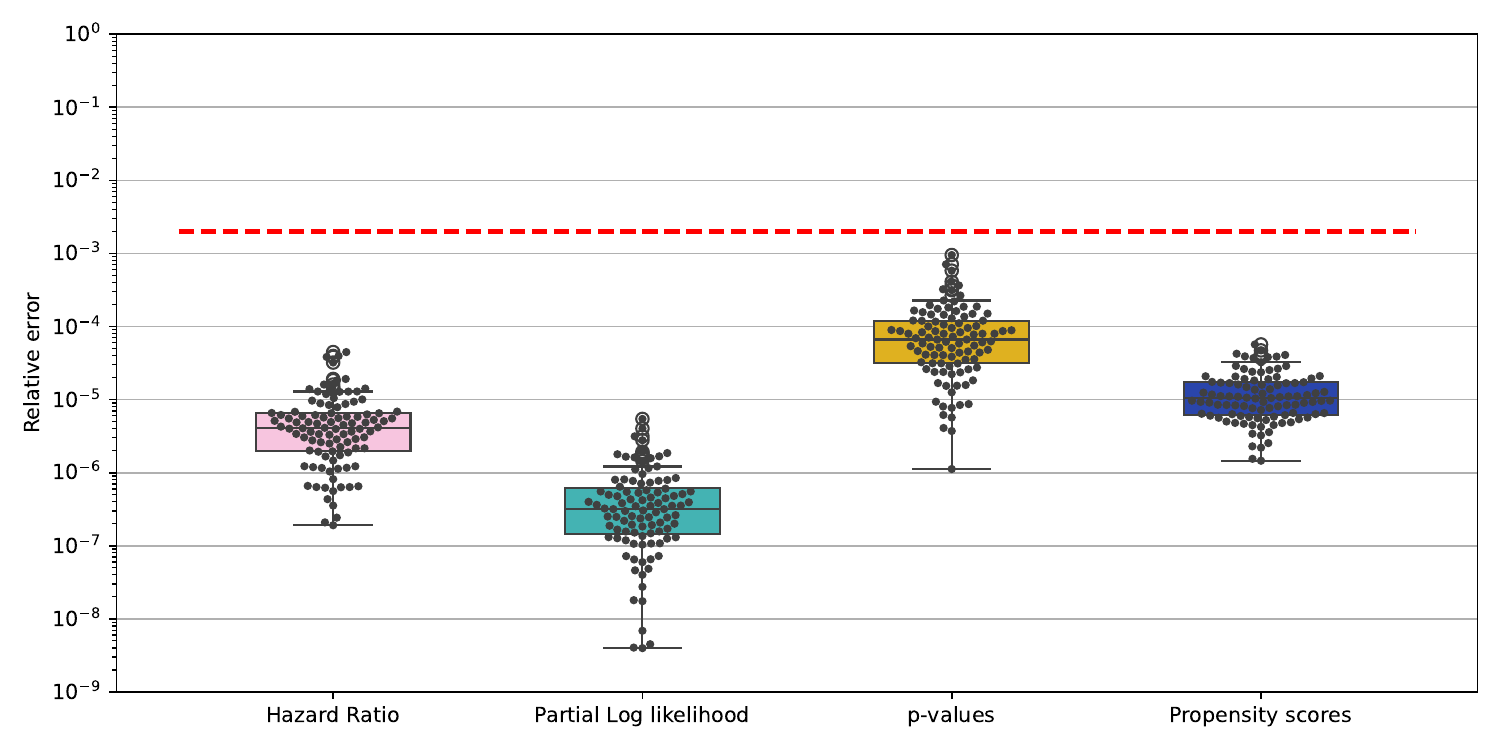}
\end{figure*}

\clearpage
\begin{figure*}
  \centering
  \captionsetup{justification=centering}
  \medskip

    \centering
    \begin{subfigure}{.39\linewidth}
    \includegraphics[width=\linewidth]{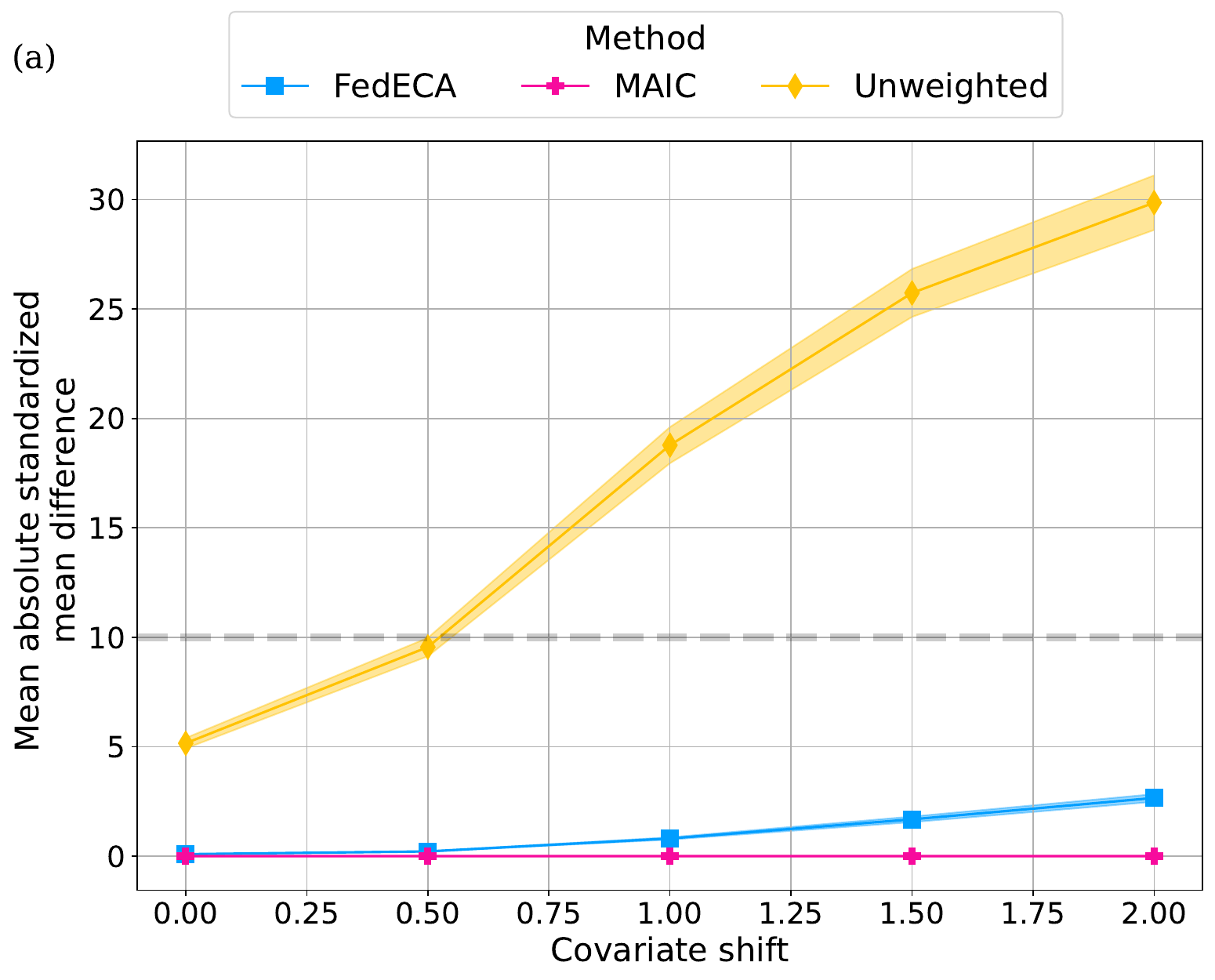}
    \end{subfigure}
    \begin{subfigure}{.6\linewidth}
    \includegraphics[width=\linewidth]{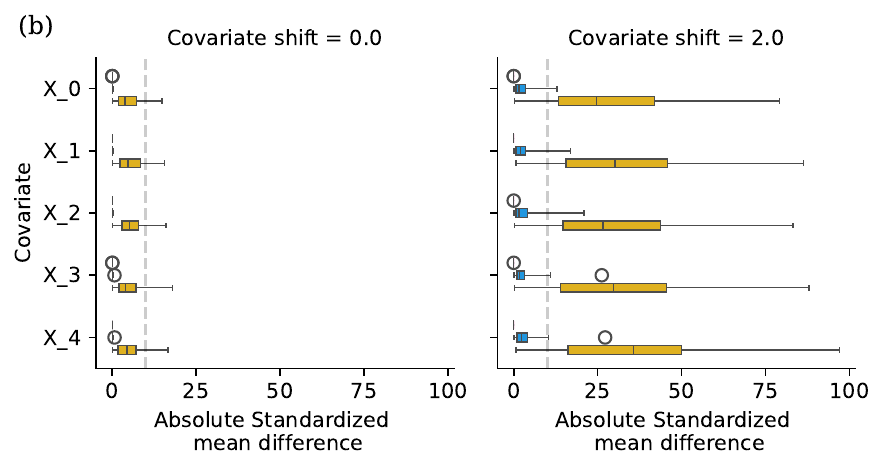}
    \end{subfigure}
    \begin{subfigure}{.8\linewidth}
    \includegraphics[width=\linewidth]{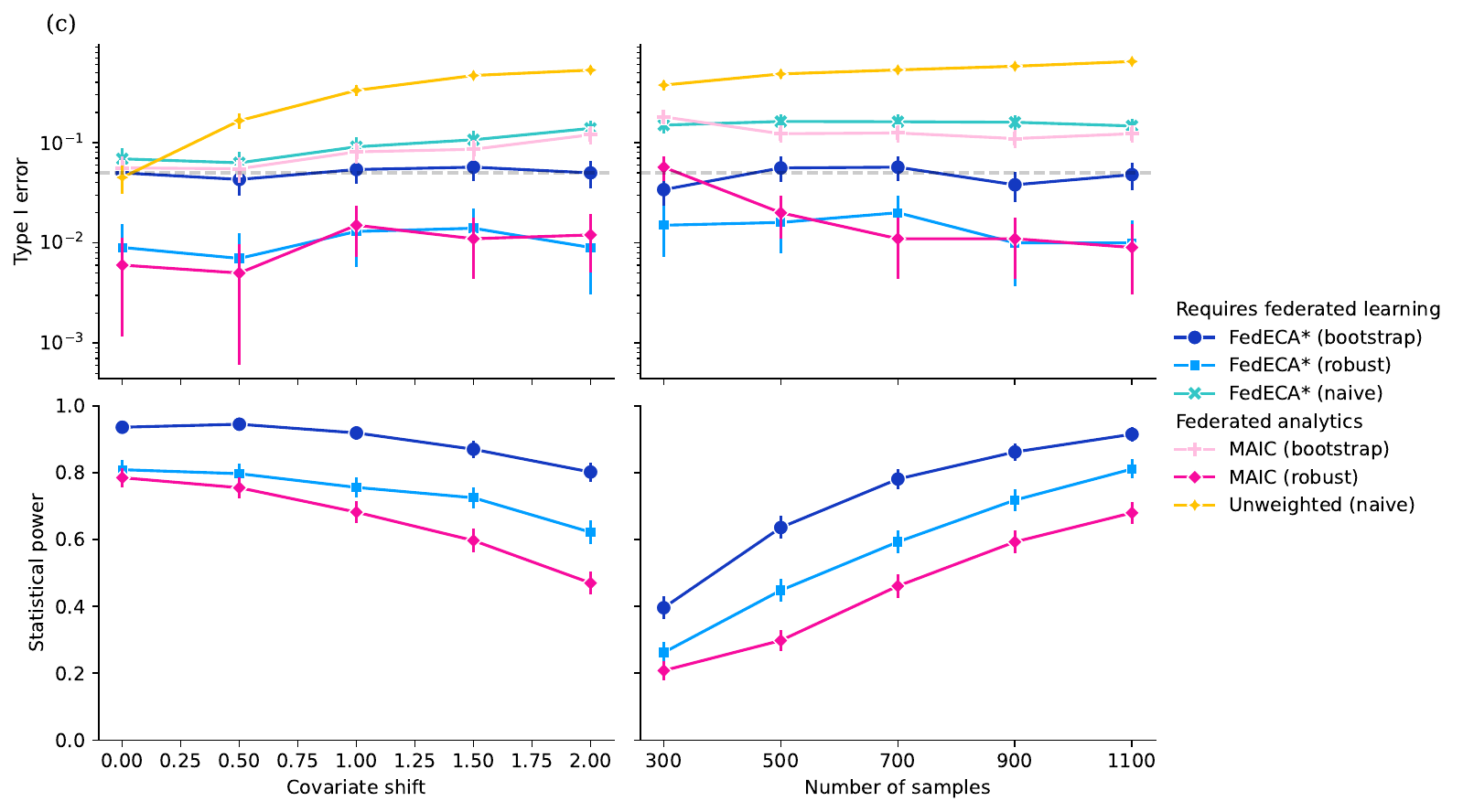}
    \end{subfigure}

\end{figure*}
\clearpage
\begin{figure*}
  \centering
  \includegraphics[width=\linewidth]{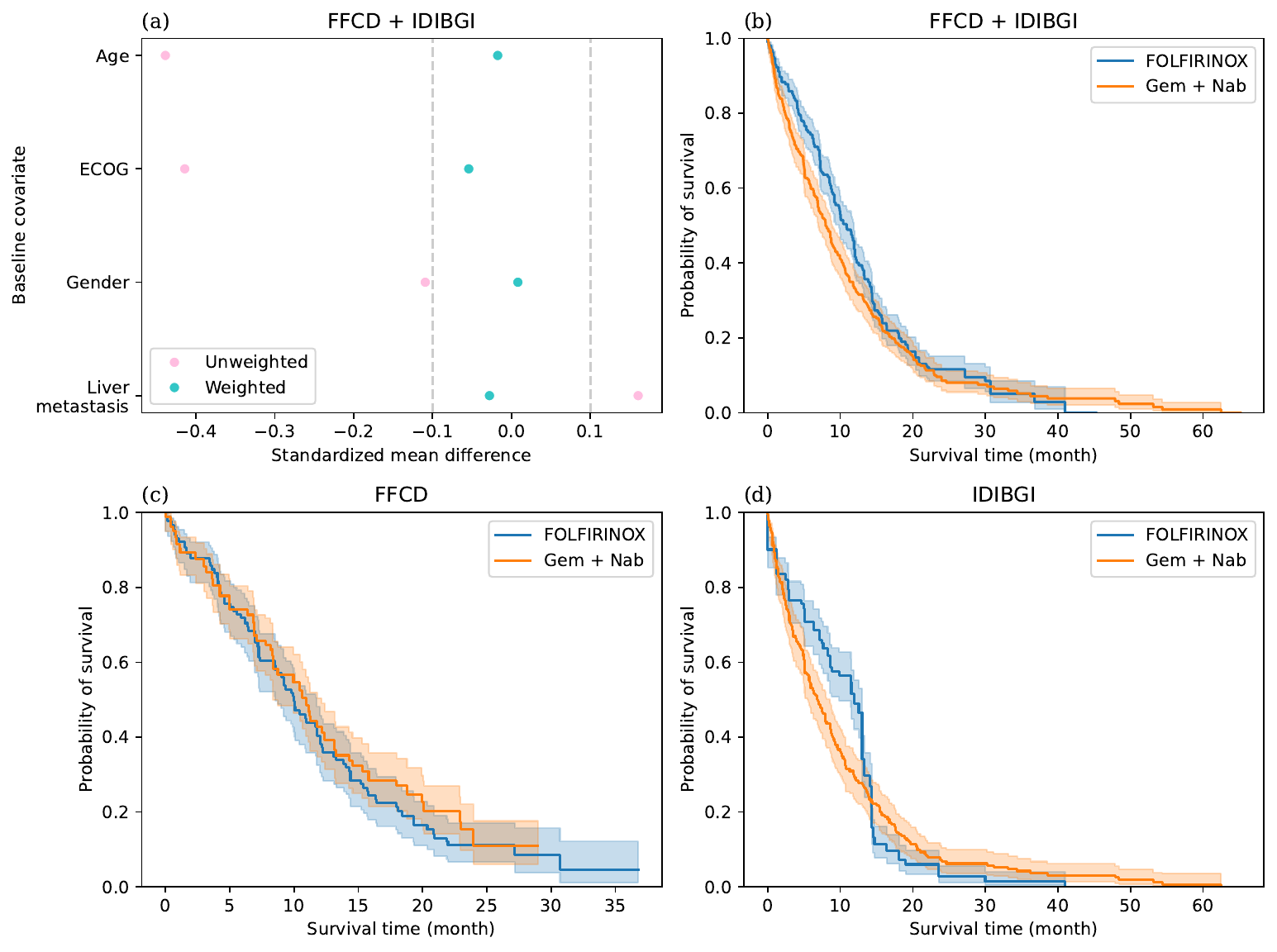}
\end{figure*}

\clearpage
\FloatBarrier

\sectionwordcount

\section{Methods \label{sec:methods}}
\sectionwordcount
\subsection{Inclusion and ethics statement}
We provide below the inclusion and ethics statements related to the patient data we access.
Note that we only access data from past studies in a retrospective fashion.

The ethics of this retrospective study on clinical data collected during care and from past clinical trials were validated
by each institution according to corresponding local regulations. We list below the
corresponding statements from each of the participating cancer centers.

Regarding FFCD data, the study was conducted in accordance with the ethical principles
outlined in the Declaration of Helsinki, International Council for Harmonisation of
Technical Requirements for Pharmaceuticals for Human Use (ICH) requirements and Good Clinical
Practice guidelines; it received authorization from the French national medicines agency
(ASNM), and independent ethics committee (number 214-R18 and 14-12-79 respectively for
PRODIGE 35 and PRODIGE 37). The study was both registered in clinicaltrials.gov
(NCT02352337 for PRODIGE 37 and NCT02827201 for PRODIGE 35) and EudraCT 2014-004449-28.

For IDIBGI data the study was approved by the Comitè d’Ètica d’Investigació amb Medicaments
CEIM GIRONA the 8th of August 2023 (Acta 11/2023) under reference CEIM code 2023.165
with principal investigators ADELAIDA GARCIA VELASCO and ROBERT CARRERAS TORRES
and SANOFI-AVENTIS SA as promoter.

Finally for PanCAN data the sponsor of the IRB was the Pancreatic Cancer Action Network (\# KYT001)
the IRB reference is 20192301, study 1265508 with main investigator Matrisian, Lynn.

For FFCD and IDIBGI, an informed consent form with nonopposition
principle for the reuse of their health data for research purposes was communicated to the patient at the time of admission
in each of the study centers in accordance with european regulations.
For PanCAN's data, authorization was obtained to use patient data through PanCAN’s Know Your Tumor program.
All data used are fully anonymized and as a result, did not require consent.

Participants were not compensated.

\subsection{Problem statement}

We consider a setting where one center, e.g., a pharmaceutical company, hosts data of all
treated patients and approaches several other centers to use their data as control to define
a distributed ECA. Although FedECA also works in the more general case where there
is no constraint on patient mixing within the participating centers.

We suppose that FDA guidelines for ECA~\cite{us2023considerations}
have been applied to direct data harmonization so that variables, assigned or
received treatments, data formats, variable ranges, outcome definitions and
inclusion criteria match across centers. However we touch on the practical challenges
associated with such a requirement when studying the metastatic pancreatic adenocarcinoma use-case
in Section~\ref{sec:cohort_construction}.
We assume that variables are not missing and relegate discussing data
imputation questions associated with real-world data to Section~\ref{sec:cohort_construction}
as well.

We consider as well that all centers arrived at a consensus on a common list of
confounding factors that influence both the exposure and the outcome of interest,
we give examples of such lists in Section~\ref{sec:yoda_cohort_construction} and Section~\ref{sec:cohort_construction}.

Moreover, the studied treatment effect is the average treatment effect (ATE)
evaluated using the hazard ratio (HR) with time-to-event outcomes. See the discussion
regarding this design choice.

Finally we assume the deployment of a federated solution such as Substra~\cite{galtier2019substra}
between the centers as well as a trusted third party or aggregator. This setup is detailed for the real-world deployment use-cases in Section~\ref{subsec:real_world}.

We note that because of the scope of this article, we do not necessarily dwell
on such technicalities; in practice however, they are a crucial aspect of
FL projects and should not be underestimated~\cite{kairouz2021advances, pati2022federated, bujotzek2024real}.
\subsection{Federated External Control Arms (FedECA)}
\subsubsection{Method overview}\label{sec:method_overview}
The ECA methodology we use relies on 3 main steps: training a propensity score model,
fitting a weighted Cox model, and testing the parameter related to the treatment.
We first introduce them here
in a pooled-level fashion, before explaining in detail how we adapted them
to the federated setting in the next sections.

\paragraph*{Setup and notations}
Each patient is represented by covariates $\bm{X} \in \mathbb{R}^p$.
It undergoes treatment $A \in \lbrace 0, 1\rbrace$, corresponding either to the treated ($A=1$)
or control ($A=0$) arm.
We denote $\bm{x}_i$ the covariates of the $i$-th patient, and $a_i$ its treatment allocation.
Following treatment, the patient has an event of interest (e.g., death or disease relapse)
at a random time $T^*$.
The patient may leave the arm before the event of interest is actually observed,
a phenomenon called censoring: we denote the observed time $T$,
whose realizations are denoted $t_i$.
We note~$\delta_i = 1$ if this corresponds to a true event, resp. $\delta_i=0$ if censorship took place.
Additionally, we define the observed outcome $Y_i=(T_i, \delta_i)$.
Let $n$ denote the total number of patients, indexed by $i$.

Let~$\mathcal{S}$ denote the finite set of all observed times,
i.e.~$\mathcal{S} = \lbrace t_i \rbrace_{i=1}^n$.
At a given time~$s$, let~$\mathcal{D}_s$ denote the set of patients with an event at this time,
i.e.
\begin{equation}
    \forall s \in \mathcal{S}, \mathcal{D}_s = \lbrace i | t_i = s, \delta_i = 1 \rbrace ,
\end{equation}
and let~$\mathcal{R}_s$ denote the set of patients at risk at this time,
i.e.
\begin{equation}
    \forall s \in \mathcal{S}, \mathcal{R}_s = \lbrace i | t_i \geq s \rbrace .
\end{equation}
Further, let $\mathring{\mathcal{S}}$ denote the set of times where at least
one true event occurs, i.e.
\begin{equation}
    \mathring{\mathcal{S}} = \lbrace s \in \mathcal{S} | \mathcal{D}_s \neq \emptyset \rbrace .
\end{equation}

Data is distributed among~$K$ different centers, with~$n_k$ samples per center.
We denote~$\bm{x}_{i, k}$ the~$i$-th covariate vector from the~$k$-th center;
accordingly,~$a_{i, k}$ denotes the treatment allocation,~$y_{i, k}=(t_{i, k}, \delta_{i, k})$ the
observed outcome, where~$t_{i, k}$ is the observed time event,
and~$\delta_{i, k}$ whether a true event took place.
Similarly, for each time $s$ and center $k$, we define the subset~$\mathcal{D}_{s, k}$
and~$\mathcal{R}_{s, k}$ as the respective restrictions of~$\mathcal{D}_s$
and~$\mathcal{R}_s$ to center $k$.

\paragraph*{Propensity score model training}
Due to the lack of randomization, for each sample, the probability of being assigned
the treatment $A$ might depend on the covariates $\bm{X}$.
We train a propensity score model $p_{\bm{\theta}}$ with parameters $\bm{\theta}$ such that
\begin{equation}\label{eq:propensity_score}
    p_{\bm{\theta}}(\bm{x}) \approx \mathbb{P}[A | \bm{X} = \bm{x}].
\end{equation}
We use a logistic model for $p_{\bm{\theta}}$, i.e.,
\begin{equation}\label{eq:logistic_propensity_score}
    p_{\bm{\theta}}(\bm{x}) = \frac{1}{1+\exp(-\bm{\theta}^T \bm{x})}.
\end{equation}
Its negative log-likelihood is given by
\begin{equation}
    \mathcal{J}(\bm{\theta}) = \sum_{i = 1}^{n}\left\lbrace a_{i} \log p_{\bm{\theta}}(\bm{x}_{i})
    +(1-a_{i}) \log ( 1 - p_{\bm{\theta}}(\bm{x}_{i}))
    \right\rbrace.
\end{equation}

In Section~\ref{sec:fed_propensity}, we explain how this model is trained in a federated setting.

\paragraph*{Inverse Probability Weighted Treatment (IPTW)}
For each sample $i$, we define an IPTW weight~$w_i \in (0, +\infty)$ based on the propensity score model trained
in the previous step as
\begin{equation}
    w_{i}=
\begin{cases}
    \frac{1}{\max(p_{\theta}(\bm{x}_{i}), \varepsilon)} & \text{if } a_{i} = 1,\\
    \frac{1}{\max(1-p_{\theta}(\bm{x}_{i}), \varepsilon)}              & \text{otherwise.}
\end{cases}
\end{equation}
In order to avoid overflow errors, $\varepsilon>0$ was set to $10^{-16}$ in our experiments.

We note that it might be that in case of insufficient overlap, weights would take extreme values. 
In our cases it is not what we observe from looking at per-center histograms of propensity scores in ~\ref{fig:hist_weights}.
In the general case future work might be needed to deal appropriately with extreme values.

We then train a weighted Cox proportional hazards (CoxPH) model with parameters $\bm{\beta} \in \mathbb{R}^q$,
related to patient-specific variables $\bm{z}_i \in \mathbb{R}^q$.
We stress that the variables~$\bm{z}_i$ are not the same as the covariates~$\bm{x}_i$.
More precisely, for the vanilla IPTW method, the sole covariate used is the treatment allocation,
i.e.,~$\bm{z}_i = a_i$.
In the general case of the adjusted IPTW (adjIPTW) method, one may use additional covariates, especially if they are
known confounders.
We note that our federated framework can support both classical IPTW and adjIPTW unlike in ~\cite{shu2020inverse},
although we choose to illustrate our results with IPTW for the sake of simplicity.

The CoxPH model is fitted by maximizing a data-fidelity term consisting in
the partial likelihood $L(\bm{\beta})$ with Breslow's approximation for tied times~\cite{breslow1975}:

\begin{equation}\label{eq:partial-likelihood-breslow}
L(\bm{\beta}) = \prod_{i: \delta_i = 1}
\left(
    \frac{
        e^{\bm{\beta}^T \bm{z}_j}
    }{
        \underset{j: t_j \geq t_i}{\sum} w_j e^{\bm{\beta}^T \bm{z}_j}
    }
\right)^{w_i}
=
\prod_{s \in \mathring{\mathcal{S}}}
    \prod_{i \in \mathcal{D}_s}
    \left(
    \frac{
        e^{\bm{\beta}^T \bm{z}_i}
    }{
        \underset{j \in \mathcal{R}_s}{\sum} w_j e^{\bm{\beta}^T \bm{z}_j}
    }
\right)^{w_i},
\end{equation}
where the second equation has been rewritten using the sets $\mathcal{D}_s$ and $\mathcal{R}_s$.
For numerical stability, we use the negative log-likelihood $\ell(\beta) = \log L(\beta)$, which reads
\begin{equation}\label{eq:log_likelihood_ipt_weighted_coxph}
    \ell(\bm{\beta}) = -\sum_{s \in \mathring{\mathcal{S}}} \sum_{i \in \mathcal{D}_s} \left \lbrace w_i
    \bm{\beta}^T \bm{z}_i - w_i \log\left(\underset{j \in \mathcal{R}_s}{\sum} w_j e^{\bm{\beta}^T \bm{z}_j}  \right) \right\rbrace.
\end{equation}
While $\ell(\bm{\beta})$ represents a data-fidelity term, we also add
a regularization $\psi(\bm{\beta})$ with strength $\gamma > 0$, leading to the full loss
\begin{equation}\label{eq:full_loss_iptw_coxph}
    \mathcal{L}(\bm{\beta}) = \ell(\bm{\beta}) + \gamma \psi(\bm{\beta}).
\end{equation}
In Section~\ref{sec:iptw_webdisco}, we describe how we minimize the loss $\mathcal{L}$ in a federated setting,
which is one of the main technical innovations of this paper.

\paragraph*{Variance estimation and statistical testing}
Once the weights are fitted, we estimate the variance matrix of $\bm{\hat{\beta}}$ using
a robust variance estimator~\cite{binder1992fitting}.
Let us denote
\begin{eqnarray}
    \zeta^0_s(\bm{\beta}) &=& \sum_{j \in \mathcal{R}_s} w_j e^{\bm{\beta}^T \bm{z}_j},\\
    \bm{\zeta}^1_s(\bm{\beta}) &=& \sum_{j \in \mathcal{R}_s} w_j e^{\bm{\beta}^T \bm{z}_j}\bm{z}_j,\\
    \bm{\zeta}^2_s(\bm{\beta}) &=& \sum_{j \in \mathcal{R}_s} w_j e^{\bm{\beta}^T \bm{z}_j}\bm{z}_j \bm{z}_j^T,
\end{eqnarray}
and $\hat{\zeta}^0_s(\bm{\beta})$, $\bm{\hat{\zeta}}^1_s(\bm{\beta})$, $\bm{\hat{\zeta}}^2_s(\bm{\beta})$ the
analogous quantities using the estimated weights $\{\hat{w}_i\}_{i=1}^n$.

Following~\cite{binder1992fitting,shu2020inverse}, the robust variance estimator of
the variance of $\bm{\hat{\beta}}$ takes the following form:

\begin{equation}
    \widehat{\bm{Var}}(\bm{\hat{\beta}}) = \bm{H}^{-1}\bm{Q} (\bm{H}^{-1})^T,
\end{equation}
where
\begin{align}\label{eq:H_pooled}
    \bm{H} = & \,  \sum_{s \in \mathring{\mathcal{S}}} \sum_{i \in \mathcal{D}_{s}} \hat{w}_i \left(\frac{\bm{\hat{\zeta}}^2_{s}(\bm{\hat{\beta}})}{\hat{\zeta}^0_{s}(\bm{\hat{\beta}})} - \frac{\bm{\hat{\zeta}}^1_{s}(\bm{\hat{\beta}})\bm{\hat{\zeta}}^1_{s}(\bm{\hat{\beta}})^T}{\hat{\zeta}^0_{s}(\bm{\hat{\beta}})^2} \right), \\
    \label{eq:Q_pooled}
    \bm{Q} = &\,   \sum_{i=1}^{n}\bm{\hat{\varphi}}_i(\bm{\hat{\beta}})\bm{\hat{\varphi}}_i(\bm{\hat{\beta}})^T, \\
    \begin{split}
        \label{eq:varphi}
        \bm{\hat{\varphi}}_i(\bm{\hat{\beta}}) = & \, \delta_{i} \hat{w}_i \left(\bm{z}_i - \frac{\bm{\hat{\zeta}}^1_{s}(\bm{\hat{\beta}})}{\hat{\zeta}^0_{s}(\bm{\hat{\beta}})}\right)
        - \hat{w}_i \exp(\bm{\hat{\beta}}^T\bm{z}_i)\bm{z}_i \sum_{s' \in \mathring{\mathcal{S}}} \sum_{j \in \mathcal{D}_{s'}} \frac{\hat{w}_j\mathds{1}_{\{s'\leq s\}}}{\hat{\zeta}^0_{s'}(\bm{\hat{\beta}})}\\
        & \, + \hat{w}_i \exp(\bm{\hat{\beta}}^T\bm{z}_i) \sum_{s' \in \mathring{\mathcal{S}}} \sum_{j \in \mathcal{D}_{s'}} \frac{\hat{w}_j\mathds{1}_{\{s'\leq s\}}\bm{\hat{\zeta}}^1_{s'}(\bm{\hat{\beta}})}{\hat{\zeta}^0_{s'}(\bm{\hat{\beta}})^2}, \text{ for all } i \in \mathcal{D}_{s}, s\in\mathring{\mathcal{S}}, 
        \end{split}
\end{align}
with $\mathds{1}_{\{s'\leq s\}}$ the indicator function that has the value $1$
on all times $s'$ (with events) and is $0$ otherwise.

Eventually, a Wald test is performed on the entry of $\bm{\hat{\beta}}$ corresponding
to the treatment allocation, assuming a $\chi^2$ distribution with 1 degree of freedom~\cite{klein2003survival}.

\subsubsection{Related works}\label{sec:related_works}  

Before diving into the details of the federation of the propensity score model and the weighted Cox model,
we provide some context explaining the position of FedECA in the literature. 

\paragraph{Methodology}

In the case of binary or continuous outcomes, inverse probability of treatment weighting (IPTW)
can be directly federated, and has been explored extensively~\cite{toh2018combining, xiong2021federated, han2021federated, han2023multiply, tarumi2023personalized, almodovar2024propensity}.
In contrast, to the best of our knowledge, few works have explored the federation of ML model training compatible with ECA
for time-to-event outcomes.

The difficulty of this federation is that the straightforward application of FL algorithms such as Federated~Averaging~\cite{mcmahan2017communication}
to time-to-event ML models is impossible due to the non-separability of the Cox proportional hazards (PH) loss~\cite{lu2015webdisco, andreux2020federated}.
Careful federation of the training of ML models capable of handling time-to-event outcomes is possible~\cite{lu2015webdisco, andreux2020federated}
but often requires either to use tree-based models~\cite{wang2022survmaximin, archetti2023federated},
approximations~\cite{andreux2020federated, ogier2022flamby} or can only be performed
in stratified settings~\cite{buchanan2014worth, shu2020inverse, luo2022odach,li2022distributed, park2022wicox},
which limits the applicability of such federated analyses for ECA analyses.  

Indeed, existing stratified federated IPTW methods such as~\cite{shu2020inverse} cannot be applied to ECA as,
in the realistic setting we consider, the treatment variable is constant within each center and
thus comparison between the treated and untreated groups cannot be done locally from within a single center.
A recent work proposed a propensity score method to estimate hazard ratios in a federated weighted Cox PH model~\cite{huang2023covariate}.
The main difference with our work lies in the fact that they have considered propensity scores based on the combination
of local propensity scores (computed in each center) and global ones demonstrating superior performance than the global
scores alone. However, as previously stated, in this paper, we consider a setting where local propensity
score models cannot be trained locally to predict treatment allocation since the
variable to predict is constant in each center.

In particular, we extend WebDISCO~\cite{lu2015webdisco}, which, alongside~\cite{huang2023covariate}, is to the best of our knowledge, 
one of the few exact methods enabling federated learning of time-to-event models in ECA contexts.
Our method can also be seen as an extension of stratified IPTW for time-to-event outcomes
from~\cite{shu2020inverse, huang2023covariate} to the non-stratified case that
allows application to ECA.

Our methodological contributions do not stop there as FedECA also 1. supports adjusted IPTW using any sets of covariates for the training of the Cox model,
2. proposes a federated algorithm for robust distributed estimation 3. develops an efficient bootstrap implementation in FL as well as 4.
provides two federated analytics estimators (Federated SMD and Federated Kaplan-Meier) allowing to perform an end-to-end ECA study in a federated setting.

Other lines of work tackle the federated analytics setting where no learning is involved
and propose to use aggregated data (AD), such as matching-adjusted indirect comparison (MAIC)~\cite{signorovitch2012matching} to perform
direct comparisons in combination with the available individual patients data.  

Finally, another popular research direction is to propose private representations of
patient covariates~\cite{rassen2010multivariate, lee2018privacy, kawamata2022collaborative, imakura2023dc}
that can be pooled into a central server. These methods have the drawback of not yielding pooled-equivalent results.
Furthermore, centralizing these representations increases the potential leakage risks associated
with a successful, even if unlikely, attack, compared to a federated storage system.  

We summarize in ~\ref{tab:relatedworks} the differences between FedECA and methods from the literature.

\paragraph{Real-world Federated Learning}
 
Besides the technical methodology differences, FedECA stands out from related works thanks to its actual implementation on a real federated learning network connecting
three clinical centers Section~\ref{subsec:pancreas}. This practical application demonstrates how our method can effectively address questions about treatment efficacy
in clinical practice in the case of metastatic pancreatic cancer while enhancing the privacy of individual patients' data. 
We believe this is FedECA's most important contribution as most of the literature on FL only studies simulated scenarios. The main reason why most FL research is theoretical is
that real-world federated networks are still, to this day, highly complex to set up and operate.

This is due to several factors notably:
\begin{itemize}
\item The need for trust, which is a critical factor to build collaborations between competing institutions and FL providers. In practice, trust is often established through successful
prior partnerships between pairs of actors or facilitated when the principal investigator has strong credentials to show, which help them engage with new stakeholders. Open-source code is also a key enabler for trust in such collaborations.
\item The need for a contractual and legal framework within which one can deploy a federated network between different legal entities respecting local jurisdictions
\item The federated learning (FL) solution must be compatible with potentially heterogeneous IT systems as some institutions may
refuse to store their data in normalized cloud environments hosted in specific countries due to concerns over data ownership.
\item The federated network implementation has to be vetted by each of the IT team of the partner institution to make sure there is no data leakage.
\item FL collaborations also require harmonizing the data beforehand, which in practice is often mostly manual and requires the help of data engineers and doctors
onsite as well as AI specialists across all participating centers and coordinated usually through e-mails by the principal investigator.
\item The usual data-science workflow is rendered much more complex by constraints on data access and sharing, which limits the kind of analyses that can be run.
\end{itemize}  

The software we are using, Substra, is dockerized, has been audited for its security and is easily integrated into existing infrastructure although
it usually requires DevOps (Development Operations) teams in each hospital or institution to be successfully deployed.

With this in mind, we go on to precisely describe the federation scheme we employ by
specifying all quantities that are communicated between the centers and the aggregator.

\subsubsection{Federated propensity model training}\label{sec:fed_propensity}
Our goal is to fit a model for the propensity score~\eqref{eq:logistic_propensity_score} based on distributed data
$\lbrace (\bm{x}_{i, k}, a_{i, k})_i \rbrace_{k=1}^K$.
Let $\mathcal{J}$ denote the full negative log-likelihood of the model, and
$\mathcal{J}_k$ the negative log-likelihood for each center, i.e.,
\begin{equation}
    \mathcal{J}_k(\bm{\theta}) = \sum_{i = 1}^{n_k}\left\lbrace a_{i, k} \log p_{\bm{\theta}}(\bm{x}_{i, k})
    +(1-a_{i, k}) \log ( 1 - p_{\bm{\theta}}(\bm{x}_{i, k}))
    \right\rbrace.
\end{equation}
Due to the separability of each loss term in per-sample terms~\cite{yang2018applied}, we have
\begin{equation}\label{eq:separability_propensity}
    \mathcal{J}(\bm{\theta}) = \sum_{k=1}^K \mathcal{J}_k(\bm{\theta}).
\end{equation}
Using the separability~\eqref{eq:separability_propensity}, it is straightforward to optimize~$\mathcal{J}$
using a second-order method, since its gradient and Hessian can be computed from the sum of local quantities,
see Section 2.1 of~\cite{islamov2021distributed}.
We call this naïve strategy~\textsc{FedNewtonRaphson}: its pseudocode is provided in Algorithm~\ref{alg:fednewton}.
This algorithm has a hyperparameter corresponding to the number of steps: in our numerical experiments, we noted
that $E = 10$ is sufficient to obtain proper convergence.

The strategy~\textsc{FedNewtonRaphson} requires to compute full batch gradients and Hessians, in time $O(n_k)$ on each center,
and each communication with the aggregator requires the exchange of $O(p^2)$ floating numbers.
In the setting of ECAs, we usually have both $n_k \leq 10^3$ and $p\leq 10^3$, making such a second-order approach tractable.
We note that for larger data settings, several improvements could be considered following~\cite{li2019feddane, islamov2021distributed},
which would reduce the quantities of transmitted parameters. We leave such improvements to future work.

\begin{minipage}{0.6\textwidth}
    \centering
    \begin{algorithm}[H]
       \caption{\textsc{FedNewtonRaphson}}
       \label{alg:fednewton}
    \begin{algorithmic}[1]
    \State Initialize $\bm{\theta}_{0} = 0$
        \For{$e=1$ {\bfseries to} $E$}
            \State Aggregator sends $\bm{\theta}_{e-1}$ to each center
             \For{$k=1$ {\bfseries to} $K$ {\bfseries in parallel} }\Comment{On each center}
             \State $\bm{g}_{e, k} = \bm{\nabla}_{\bm{\theta}} J_k(\theta_{e-1})$
             \State $\bm{H}_{e,k} = \bm{\nabla}_{\bm{\theta}}^2 J_k(\theta_{e-1})$
             \State Send $\bm{g}_{e, k}$ and $\bm{H}_{e, k}$ to the aggregator
             \EndFor
        \State $\bm{g}_{e} = \frac{1}{K} \sum_{k=1}^K \bm{g}_{e, k}$\Comment{Aggregator-side}
        \State $\bm{H}_{e} = \frac{1}{K}\sum_{k=1}^K \bm{H}_{e, k}$
       \State $\bm{\theta}_{e} = \bm{\theta}_{e-1} - (\bm{H}_{e})^{-1} \bm{g}_e$
        \EndFor\\
        \Return $\bm{\theta}_{E}$

    \end{algorithmic}
\end{algorithm}
\end{minipage}
\subsubsection{Inverse probability weighted WebDISCO}\label{sec:iptw_webdisco}
Here we propose a method to minimize the regularized weighted CoxPH model~\eqref{eq:full_loss_iptw_coxph}
in a federated fashion.
Since the non-separability of the weighted CoxPH log-likelihood~$\ell(\bm{\beta})$ prevents
the use of vanilla FL algorithms, we inspire ourselves from
WebDISCO~\cite{lu2015webdisco} to build a pooled-equivalent second-order method.
It should be noted, however, that the method can only be applied to partial likelihood
under the Breslow's approximation for tied times~\eqref{eq:partial-likelihood-breslow},
as opposed to the Efron's approximation.

\paragraph*{Non-separability}
Compared to the logistic propensity score model, the main difficulty of
federating Equation~\eqref{eq:log_likelihood_ipt_weighted_coxph}
stems from the non-separability of the log-likelihood, i.e., the cross-center terms.
Indeed, for any time $s$, the risk set $\mathcal{R}_s$ is a union of per-center terms, i.e.
\begin{equation}\label{eq:union_risk_sets}
    \mathcal{R}_s = \cup_{k = 1}^K \mathcal{R}_{s, k}.
\end{equation}
Thus, the aggregated Equation~\eqref{eq:log_likelihood_ipt_weighted_coxph} can be rewritten as
\begin{equation}
    \ell(\bm{\beta}) = - \sum_{k=1}^K \sum_{s \in \mathring{\mathcal{S}}} \sum_{i \in \mathcal{D}_{s, k}} \left \lbrace w_i
    \bm{\beta}^T \bm{z}_{i, k} - w_i \log\left(\underset{j \in \mathcal{R}_{s, k}}{\sum} w_j e^{\bm{\beta}^T \bm{z}_{j, k}} +
    \sum_{k' \neq k} \underset{j \in \mathcal{R}_{s, k'}}{\sum} w_j e^{\bm{\beta}^T \bm{z}_{j, k'}}  \right) \right\rbrace\ ,
\end{equation}
where the loss for each sample~$i$ of each center~$k$ involves terms from other samples~$j$ in other centers~$k' \neq k$.
The non-separability of the CoxPH loss is a well-known issue in a federated setting and previous works
have investigated reformulations to make it amenable to vanilla federated learning solvers~\cite{andreux2020federated}.
Here we instead adapt the WebDISCO method~\cite{lu2015webdisco} to the weighted case in order
to keep pooled-equivalent results and benefit from second-order acceleration.

\paragraph*{Federated computation of $\bm{\nabla}_{\bm{\beta}} \ell(\bm{\beta})$ and $\bm{\nabla}^2_{\bm{\beta}} \ell(\bm{\beta})$}
Our method consists in performing an iterative server-level Newton-Raphson descent on $\mathcal{L}$.
The gradient~$\bm{\nabla}_{\bm{\beta}} \ell(\bm{\beta})$ and Hessian~$\bm{\nabla}^2_{\bm{\beta}} \ell(\bm{\beta})$ thus need
to be computed in a federated fashion.
These quantities can be computed in closed-form as
\begin{equation}\label{eq:gradient_log_likelihood_pooled}
    \bm{\nabla}_{\bm{\beta}} \ell(\bm{\beta}) =
    - \sum_{s \in \mathring{\mathcal{S}}} \sum_{i \in \mathcal{D}_s} \left ( w_i \bm{z}_i
        -w_i \frac{
            \sum_{j \in \mathcal{R}_s}w_j e^{\bm{\beta}^T \bm{z}_j}\bm{z}_j
        }{
            \sum_{j \in \mathcal{R}_s} w_j e^{\bm{\beta}^T \bm{z}_j}
        }\right ),
\end{equation}
and
\begin{equation}\label{eq:hessian_log_likelihood_pooled}
    \bm{\nabla}^2_{\bm{\beta}} \ell(\bm{\beta}) =
    \sum_{s \in \mathring{\mathcal{S}}} \sum_{i \in \mathcal{D}_s} w_i \left\lbrace \frac{
        \sum_{j \in \mathcal{R}_s} w_j e^{\bm{\beta}^T \bm{z}_j} \bm{z}_j \bm{z}_j^T
    }{
        \sum_{j \in \mathcal{R}_s} w_j e^{\bm{\beta}^T \bm{z}_j}
    }
    - \frac{
        \left(\sum_{j \in \mathcal{R}_s} w_j e^{\bm{\beta}^T \bm{z}_j} \bm{z}_j\right)
        \left(\sum_{j' \in \mathcal{R}_s} w_{j'} e^{\bm{\beta}^T \bm{z}_{j'}} \bm{z}_{j'}^T\right)
    }{
        (\sum_{j \in \mathcal{R}_s} w_j e^{\bm{\beta}^T \bm{z}_j})^2
    }
    \right\rbrace.
\end{equation}
Note that the Hessian evaluated at $\bm{\beta}=\hat{\bm{\beta}}$, $\bm{\nabla}^2_{\bm{\beta}}\ell(\hat{\bm{\beta}})$,
corresponds, up to a sign, to the quantity $\bm{H}$ defined in~\eqref{eq:H_pooled} for the robust variance estimator.
We now define the local counterparts $\zeta^h_{s, k}(\bm{\beta})$ of the previously introduced quantities where the sum is restricted to the risk set $\mathcal{R}_{s, k}$,
\begin{eqnarray}
    \label{eq:zeta_0_k}
    \zeta^0_{s, k}(\bm{\beta}) &=& \sum_{j \in \mathcal{R}_{s, k}} w_j e^{\bm{\beta}^T \bm{z}_j},\\
    \label{eq:zeta_1_k}
    \bm{\zeta}^1_{s, k}(\bm{\beta}) &=& \sum_{j \in \mathcal{R}_{s, k}} w_j e^{\bm{\beta}^T \bm{z}_j}\bm{z}_j,\\
    \label{eq:zeta_2_k}
    \bm{\zeta}^2_{s, k}(\bm{\beta}) &=& \sum_{j \in \mathcal{R}_{s, k}} w_j e^{\bm{\beta}^T \bm{z}_j}\bm{z}_j \bm{z}_j^T.
\end{eqnarray}
Further, let us denote
\begin{eqnarray}
    W_{s} &=& \sum_{i \in \mathcal{D}_{s}} w_i,\\
    \bm{Z}_{s} &=& \sum_{i \in \mathcal{D}_{s}} w_i \bm{z}_i,
\end{eqnarray}
and
\begin{eqnarray}
    \label{eq:w_s_k}
    W_{s, k} &=& \sum_{i \in \mathcal{D}_{s, k}} w_i,\\
    \label{eq:z_s_k}
    \bm{Z}_{s, k} &=& \sum_{i \in \mathcal{D}_{s, k}} w_i \bm{z}_i,
\end{eqnarray}
where by convention, in all cases, the sum is set to $0$ in case of an empty set.
Equations~\eqref{eq:gradient_log_likelihood_pooled} and~\eqref{eq:hessian_log_likelihood_pooled} can be respectively rewritten
as
\begin{equation}
    \bm{\nabla}_{\bm{\beta}} \ell(\bm{\beta}) =
    - \sum_{s \in \mathring{\mathcal{S}}} \bm{Z}_s
        -W_s \frac{
            \bm{\zeta}^1_s(\bm{\beta})
        }{
            \zeta^0_s(\bm{\beta})
        },
\end{equation}
\begin{equation}
    \bm{\nabla}^2_{\bm{\beta}} \ell(\bm{\beta}) =
    \sum_{s \in \mathring{\mathcal{S}}} W_s \left\lbrace \frac{
        \bm{\zeta}^2_s(\bm{\beta})
    }{
        \zeta^0_s(\bm{\beta})
    }
    - \frac{
        \bm{\zeta}^1_s(\bm{\beta})\bm{\zeta}^1_s(\bm{\beta})^T
    }{
        \zeta^0_s(\bm{\beta})^2
    }
    \right\rbrace.
\end{equation}

Using these equations, we can rewrite
\begin{equation}\label{eq:gradient_computation_iptw_fed}
    \bm{\nabla}_{\bm{\beta}} \ell(\bm{\beta}) =
    - \sum_{k=1}^K \left\lbrace
    \sum_{s \in \mathring{\mathcal{S}}} \bm{Z}_{s, k}
        - W_{s, k} \frac{
            \sum_{k'} \bm{\zeta}^1_{s, k'}(\bm{\beta})
        }{
            \sum_{k'} \zeta^0_{s, k'}(\bm{\beta})
        }
    \right\rbrace
    ,
\end{equation}
\begin{equation}\label{eq:hessian_computation_iptw_fed}
    \bm{\nabla}^2_{\bm{\beta}} \ell(\bm{\beta}) =
    \sum_{k=1}^K \sum_{s \in \mathring{\mathcal{S}}} W_{s, k} \left\lbrace \frac{
        \sum_{k'}\bm{\zeta}^2_{s, k'}(\bm{\beta})
    }{
        \sum_{k'}\zeta^0_{s, k'}(\bm{\beta})
    }
    - \frac{
        \left(\sum_{k'} \bm{\zeta}^1_{s, k'}(\bm{\beta}) \right) \left(\sum_{k'}\bm{\zeta}^1_{s, k'}(\bm{\beta})\right)^T
    }{
        \left(\sum_{k'}\zeta^0_{s, k'}(\bm{\beta}) \right)^2
    }
    \right\rbrace.
\end{equation}

Assuming the set of all true event times~$\mathring{\mathcal{S}}$ is known to all centers,
we see that it is possible to reconstruct the full gradient~$\bm{\nabla}_{\bm{\beta}} \ell(\bm{\beta})$
and Hessian~$\bm{\nabla}^2_{\bm{\beta}} \ell(\bm{\beta})$ based
on the 5-uplet~$\lbrace (W_{s, k}, \bm{Z}_{s, k}, \zeta^0_{s, k}(\bm{\beta}), \bm{\zeta}^1_{s, k}(\bm{\beta}), \bm{\zeta}^2_{s, k}(\bm{\beta})) \rbrace_{s, k}$.
Algorithm~\ref{alg:iptWebdisco} sums up this algorithm.

\begin{minipage}{0.6\textwidth}
    \centering
    \begin{algorithm}[H]
       \caption{\textsc{FedCoxComp}} %
       \label{alg:iptWebdisco}
    \begin{algorithmic}[1]
    \Require Weights $\bm{\beta}$, set~$\mathring{\mathcal{S}}$
    \State Aggregator sends $\bm{\beta}$ to each center
    \For{$k=1$ {\bfseries to} $K$ {\bfseries in parallel} }\Comment{On each center}
        \For{$s \in \mathring{\mathcal{S}}$}
            \State Compute $W_{k, s}$ with~\eqref{eq:w_s_k}\Comment{$0$ if $\mathcal{D}_{s, k} = \emptyset$}
            \State Compute $\bm{Z}_{k, s}$ with~\eqref{eq:z_s_k}.
        \EndFor
        \For{$s \in \mathring{\mathcal{S}}$ s.t. $W_{k, s} > 0$} \Comment{$0$ otherwise}
            \State Compute $\zeta^0_{s, k}(\bm{\beta})$ with~\eqref{eq:zeta_0_k}
            \State Compute $\bm{\zeta}^1_{s, k}(\bm{\beta})$ with~\eqref{eq:zeta_1_k}
            \State Compute $\bm{\zeta}^2_{s, k}(\bm{\beta})$ with~\eqref{eq:zeta_2_k}
        \EndFor
        \State Send back $\lbrace (W_k, \bm{Z}_k, \zeta^0_{s, k}(\bm{\beta}), \bm{\zeta}^1_{s, k}(\bm{\beta}),
        \bm{\zeta}^2_{s, k}(\bm{\beta})) \rbrace_{s \in \mathring{\mathcal{S}}}$
    \EndFor
    \State Compute $\bm{\nabla}_{\bm{\beta}} \ell(\bm{\beta})$ with~\eqref{eq:gradient_computation_iptw_fed} \Comment{On the server}
    \State Compute $\bm{\nabla}_{\bm{\beta}}^2 \ell(\bm{\beta})$ with~\eqref{eq:hessian_computation_iptw_fed}\\
    \Return $\bm{\nabla}_{\bm{\beta}} \ell(\bm{\beta}), \bm{\nabla}_{\bm{\beta}}^2 \ell(\bm{\beta})$
    \end{algorithmic}
\end{algorithm}
\end{minipage}

\begin{minipage}{1\textwidth}
    \centering
    \begin{algorithm}[H]
       \caption{\textsc{non-robust FedECA}} %
       \label{alg:fediptw}
    \begin{algorithmic}[1]
    \Require Maximal number of steps $E$, LR schedule $(\alpha_e)_e$, regularization $\gamma$
    \State Initialization $\bm{\beta}_0 = 0$
    \For{$e=1$ to $E$}
        \State $\bm{\nabla}_{\bm{\beta}} \ell(\bm{\beta}_{e-1}), \bm{\nabla}^2_{\bm{\beta}} \ell(\bm{\beta}_{e-1}) = \mathrm{FedCoxComp}(\bm{\beta}_{e-1})$ \Comment{Communication between server and centers}
        \State $\bm{\nabla}_{\bm{\beta}} \mathcal{L}(\bm{\beta}_{e-1}) = \bm{\nabla}_{\bm{\beta}} \ell(\bm{\beta}_{e-1}) + \gamma \bm{\nabla}_{\bm{\beta}} \psi(\bm{\beta}) $
        \State $\bm{\nabla}_{\bm{\beta}}^2 \mathcal{L}(\bm{\beta}_{e-1}) = \bm{\nabla}_{\bm{\beta}}^2 \ell(\bm{\beta}_{e-1}) + \gamma \bm{\nabla}_{\bm{\beta}}^2 \psi(\bm{\beta})$
        \State $\bm{\beta}_e = \bm{\beta}_{e - 1} - \alpha_e \left(\bm{\nabla}_{\bm{\beta}}^2 \mathcal{L}(\bm{\beta}_{e-1}) \right)^{-1}\bm{\nabla}_{\bm{\beta}} \mathcal{L}(\bm{\beta}_{e-1})$
        \If{Stopping criterion}
        $e=E$
        \EndIf
    \EndFor \\
    \Return $\bm{\beta}_E$
    \end{algorithmic}
\end{algorithm}
\end{minipage}

\paragraph*{Non-robust FedECA}
To optimize the full loss~\eqref{eq:full_loss_iptw_coxph}, we can now leverage the computation
of the gradient and Hessian of the weighted CoxPH loss~$\ell$ to perform a second-order Newton-Raphson
descent. We follow the hyperparameters of \verb_lifelines_~\cite{davidson2019lifelines} for this optimization.
In particular, we use the same learning rate strategy, the same regularizer and the same stopping criterion.
Indeed, as \verb_lifeline_'s regularizer does not depend on data and is smooth,
its gradient and Hessian can be computed on the server's side deriving twice the following equation:
\begin{align}
    \mathcal{L}(\bm{\beta}) &= \ell(\bm{\beta}) + \gamma \psi(\bm{\beta}),  \\
\end{align}
with $\gamma$ the strength of the regularization.

In more details for the regularizer $\psi(\bm{\beta})$, we use a soft elastic-net regularization~\cite{zou2005regularization}
with hyperparameters $\lambda >0 $ and $\alpha>0$:
\begin{equation}
    \psi(\bm{\beta}) = \lambda \left( \sum_{r} \phi_{\alpha}(\bm{\beta}_r) \right) + \frac{1-\lambda}{2}\| \bm{\beta} \|_2^2 ,
\end{equation}
where $\phi_{\alpha}$ is a smooth approximation of the absolute value that is progressively sharpened with the round $e$.
\begin{align}
    \alpha &= 1.3^{e},  \\
    \phi_{\alpha}(x) &= \frac{1}{\alpha}\left( \log(1 + \exp(\alpha x)) + \log(1 + \exp(- \alpha x))\right).
\end{align}
We also allow for constant learning rates as in \verb_scikit-survival_~\cite{polsterl2020scikit}. We note that
implementing different learning rate strategies or regularizers should be straightforward with our implementation.
Algorithm~\ref{alg:fediptw} summarizes the full algorithm used.

We note that in practice, due to the linearity of both models and as covariates are often low dimensional in clinical trials, tuning hyper-parameters
such as learning rates and regularizations parameters is superfluous. In fact we use the default settings without regularization in all of our
experiments unless explicitly stated. We also note that regularization in linear models for treatment effect estimation must be imposed 
carefully to avoid the over-shrinking effect described in ~\cite{hahn2018regularization}.
We still give practitioners ways to tune such hyper-parameters, as it is already the case in non-distributed softwares, in order not to loose flexibility.
This allows to accomodate potential future workflows such as using deep-learning based covariates, which might be high-dimensional~\cite{courtiol2019deep}
and thus optimizing the Cox loss might, in this case, require the use of ridge regularization to keep the hessian from being ill-conditioned.
Regarding federated hyper-parameter tuning in general see Section~\ref{sec:fedhp}.

\subsubsection{Statistical test}
\paragraph{Federated robust variance estimation}

The robust variance estimator can be obtained by aggregating local quantities as we demonstrate in the following.
We assume that each client has access to $\zeta^0_s(\bm{\hat{\beta}})$ and $\bm{\zeta^1_s}(\bm{\hat{\beta}})$
for all $s \in \mathring{\mathcal{S}}$.
This can be achieved by simply allowing the server to transmit the quantities $\zeta^0_{s, k}(\bm{\hat{\beta}})$ and $\bm{\zeta^1_{s, k}}(\bm{\hat{\beta}})$
to the centers in addition to $\bm{H}$.

The global goal is to compute the robust estimator of the variance given by
\begin{equation}
    \widehat{\bm{Var}}(\bm{\hat{\beta}}) = \bm{H}^{-1}\bm{Q} (\bm{H}^{-1})^T,
\end{equation}
where $\bm{H}$ \eqref{eq:H_pooled} corresponds to the Hessian $\bm{\nabla}^2_{\bm{\beta}} \ell(\hat{\bm{\beta}})$ and $\bm{Q}$ is defined in \eqref{eq:Q_pooled}.
We note that through FedECA (\ref{alg:fediptw}) each client already has access to $\bm{H}$.

Let us define $\bm{M}_{k}$ as
\begin{equation}
    \bm{M}_{k} = \sum_{i=1}^{n_{k}} (\bm{H}^{-1} \bm{\hat{\varphi}}_i(\bm{\hat{\beta}}))\bm{\hat{\varphi}}_i(\bm{\hat{\beta}})^T (\bm{H}^{-1})^T \enspace ,
\end{equation}
where the sum is on all indices belonging to client $k$.

Then we have, 
\begin{equation}
    \widehat{\bm{Var}}(\bm{\hat{\beta}}) = \sum_{k=1}^{K} \bm{M}_{k}.
\end{equation}
Indeed, if we let $\bm{\hat{\Phi}}(\hat{\beta}) \in \mathbb{R}^{n, p}$ be the matrix
whose rows are the $\bm{\hat{\varphi}}_i(\bm{\hat{\beta}})$ for all $i \in \llbracket 1, n \rrbracket$.
then we can write the variance as
\begin{align}
    \widehat{\bm{Var}}(\bm{\hat{\beta}}) &= \bm{H}^{-1}\bm{\hat{\Phi}}(\hat{\beta})^{T}\bm{\hat{\Phi}}(\hat{\beta})(\bm{H}^{-1})^T,  \\
    \bm{\hat{\Phi}}(\hat{\beta})^{T}\bm{\hat{\Phi}}(\hat{\beta})_{i,j} &= \sum_{k=1}^{n}\left(\bm{\hat{\varphi}}_k(\bm{\hat{\beta}})\right)_{i} \cdot \left(\bm{\hat{\varphi}}_k(\bm{\hat{\beta}})\right)_{j} = \sum_{k=1}^{K}\sum_{m=1}^{n_{k}}\left(\bm{\hat{\varphi}}_m(\bm{\hat{\beta}})\right)_{i} \cdot \left(\bm{\hat{\varphi}}_m(\bm{\hat{\beta}})\right)_{j}, \\
    \widehat{\bm{Var}}(\bm{\hat{\beta}}) &= \bm{H}^{-1}\bm{\hat{\Phi}}(\bm{\hat{\beta}})^{T}\bm{\hat{\Phi}}(\bm{\hat{\beta}})(H^{-1})^T = \sum_{k=1}^{K} \bm{M}_{k}.
\end{align}

Each client can compute $\bm{\hat{\varphi}}_i(\hat{\bm{\beta}})$ with Eq.~\eqref{eq:varphi} for all its samples $i$ ($\forall s, i \in \mathcal{D}_{s,k}$)
as long as it has access to $\zeta^0_{s, k}(\bm{\hat{\beta}})$ and $\bm{\zeta^1_{s, k}}(\bm{\hat{\beta}})$ for all $s \in \mathring{\mathcal{S}}$.
Therefore each client can compute the corresponding $\bm{M}_{k}$.

This leads us to the full robust algorithm of FedECA in \ref{alg:fedeca}.
Once the variance is estimated using the above expression, we can perform inference using, e.g., a Z-test.
Note that as in \verb_lifelines_~\cite{davidson2019lifelines} we use the Hessian of the regularized function.
Therefore to accommodate the computation of the variance we modify non-robust FedECA as depicted in Alg.~\ref{alg:fedeca}.
Privacy-wise this modification (a) gives each client the same knowledge as the server on the last round and (b) communicates
an additional $\bm{M}_{k}$ matrix by center, which is reasonable. In addition, in the IPTW case the matrix only the treatment allocation is used as a covariate and hence $\bm{M}_{k}$ is a scalar $M_{k}$.
\FloatBarrier
\begin{minipage}{1.\textwidth}
    \centering
    \begin{algorithm}[H]
       \caption{\textsc{RobustFedCoxComp}} %
       \label{alg:robustiptWebdisco}
    \begin{algorithmic}[1]
    \Require Weights $\bm{\beta}$, set~$\mathring{\mathcal{S}}$
    \State Aggregator sends $\bm{\beta}$ to each center
    \For{$k=1$ {\bfseries to} $K$ {\bfseries in parallel} }\Comment{On each center}
        \For{$s \in \mathring{\mathcal{S}}$}
            \State Compute $W_{k, s}$ with~\eqref{eq:w_s_k}\Comment{$0$ if $\mathcal{D}_{s, k} = \emptyset$}
            \State Compute $\bm{Z}_{k, s}$ with~\eqref{eq:z_s_k}.
        \EndFor
        \For{$s \in \mathring{\mathcal{S}}$ s.t. $W_{k, s} > 0$} \Comment{$0$ otherwise}
            \State Compute $\zeta^0_{s, k}(\bm{\beta})$ with~\eqref{eq:zeta_0_k}
            \State Compute $\bm{\zeta}^1_{s, k}(\bm{\beta})$ with~\eqref{eq:zeta_1_k}
            \State Compute $\bm{\zeta}^2_{s, k}(\bm{\beta})$ with~\eqref{eq:zeta_2_k}
        \EndFor
        \State Send back $\lbrace (W_k, \bm{Z}_k, \zeta^0_{s, k}(\bm{\beta}), \bm{\zeta}^1_{s, k}(\bm{\beta}),
        \bm{\zeta}^2_{s, k}(\bm{\beta})) \rbrace_{s \in \mathring{\mathcal{S}}}$
    \EndFor
    \State Compute $\bm{\nabla}_{\bm{\beta}} \ell(\bm{\beta})$ with~\eqref{eq:gradient_computation_iptw_fed} \Comment{On the server}
    \State Compute $\bm{\nabla}_{\bm{\beta}}^2 \ell(\bm{\beta})$ with~\eqref{eq:hessian_computation_iptw_fed}\\
    \Return $\bm{\nabla}_{\bm{\beta}} \ell(\bm{\beta}), \bm{\nabla}_{\bm{\beta}}^2 \ell(\bm{\beta})$ \Comment{And if it's the last round} \Return $\forall s \in \mathring{\mathcal{S}} , \zeta^0_{s, k}(\bm{\beta}), \bm{\zeta}^1_{s, k}(\bm{\beta}), W_{s}$
    \end{algorithmic}
\end{algorithm}
\end{minipage}

\begin{minipage}{1\textwidth}
    \centering
    \begin{algorithm}[H]
       \caption{\textsc{FedECA}} %
       \label{alg:fedeca}
    \begin{algorithmic}[1]
    \Require Maximal number of steps $E$, LR schedule $(\alpha_e)_e$, regularization $\gamma$
    \State Initialization $\bm{\beta}_0 = 0$
    \For{$e=1$ to $E$}
        \State $\bm{\nabla}_{\bm{\beta}} \ell(\bm{\beta}_{e-1}), \bm{\nabla}^2_{\bm{\beta}} \ell(\bm{\beta}_{e-1}) = \mathrm{RobustFedCoxComp}(\beta_{e-1})$ \Comment{Communication between server and centers}
        \State $\bm{\nabla}_{\bm{\beta}} \mathcal{L}(\bm{\beta}_{e-1}) = \bm{\nabla}_{\bm{\beta}} \ell(\bm{\beta}_{e-1}) + \gamma \bm{\nabla}_{\bm{\beta}} \psi(\bm{\beta}) $
        \State $\bm{\nabla}_{\bm{\beta}}^2 \mathcal{L}(\bm{\beta}_{e-1}) = \bm{\nabla}_{\bm{\beta}}^2 \ell(\bm{\beta}_{e-1}) + \gamma \bm{\nabla}_{\bm{\beta}}^2 \psi(\bm{\beta})$
        \State $\bm{\beta}_e = \bm{\beta}_{e - 1} - \alpha_e \left(\bm{\nabla}_{\bm{\beta}}^2 \mathcal{L}(\bm{\beta}_{e-1}) \right)^{-1}\bm{\nabla}_{\bm{\beta}} \mathcal{L}(\bm{\beta}_{e-1})$
        \If{Stopping criterion}
        $e=E$
        \EndIf
    \EndFor \\
    \Return $\bm{\beta}_E$
    \State Define $\bm{\hat{\beta}} = \bm{\beta}_E$
    \For{$k=1$ {\bfseries to} $K$ {\bfseries in parallel} }\Comment{On each center}
        \State Send back $\bm{M}_k \bm{M}_k^T$ where $\bm{M}_k = \sum_{s \in \mathring{\mathcal{S}}} \sum_{i \in \mathcal{D}_{s, k}} \bm{H}^{-1} \bm{\hat{\varphi}}_i(\bm{\hat{\beta}})$.
    \EndFor
    \State Compute $\widehat{\bm{Var}}(\bm{\hat{\beta}}) = \sum_{k=1}^{K}\bm{M}_k \bm{M}_k^T$ \Comment{On the server}\\
    \Return $\widehat{\bm{Var}}(\bm{\hat{\beta}})$
    \end{algorithmic}
\end{algorithm}
\end{minipage}

\paragraph{Federated bootstrap estimation}
When implementing bootstrap in federated setups, the most straightforward implementation is to bootstrap samples
per-center. However this creates some edge cases if centers have small
sample sizes (e.g. $\exists k | n_k=1$) and risks underestimating the variance compared to
the pooled case.
In order to circumvent this issue we label all
samples from $1$ to $n$ and label clients’ samples irrespective of their order. This requires only
sharing the number of samples by clients and nothing else. In practice we label the centers
from $1$ to $K$ randomly and then assign the number from $1$ to $n_1$ to the first client’s samples and
so forth. As each client knows its indices we can then ask the server to sample indices from $1$ to
$n$ as we would in the pooled case and then send the current set of indices chosen for each
bootstrap to all clients that would then use it to bootstrap its own cohort. This way there is no
bias in sampling even for arbitrarily small clients. We refer to this alternative as global bootstrap.

We therefore implement both but use in our experiments this global bootstrap where
we sample with replacement the global distributed cohort as if it were pooled.

Distributed computation with Substra introduces an overhead per atomic task executed
locally on each partner's machine mainly due to docker image building. This overhead is not negligible and can become a bottleneck
when performing bootstrapping, which naïvely necessitate to execute $O(n_{rounds} * n_{bootstraps})$
tasks per-client. To alleviate this issue we implement a more efficient bootstrapping strategy
where we only have to run $O(n_{rounds})$ tasks. This is achieved by employing hooks so that each task,
instead of executing its normal code, bootstraps itself and then executes all its bootstraped versions
producing a list of bootstraped results per task.
This requires to also modify the aggregation steps to be able to aggregate each bootstrap run independently
and then redispatch all bootstraped aggregations to each client.
Details of this non-trivial implementation trick can be found in the
\href{https://github.com/owkin/fedeca/blob/main/fedeca/strategies/bootstraper.py}{bootstraper.py} script.

Note that we could also add another layer of parallelization inside each task by
using Python multi-processing as each bootstrap run is independent of each other.
However, the impact of this optimization would be negligible with respect to the overhead
introduced by the distributed constraints.
With this optimization, a Substra experiment with $200$ bootstraps lasts less than an hour instead of $\approx 200$ hours naïvely without this parallelization layer
which would have made the bootstrap variance estimation impractical irrespective of the size of the federated network. 
Note that other kinds of parallelization schemes could also be undertaken such as running multiple training jobs (so-called ``Compute Plans" in Substra) 
in parallel as was done in MELLODDY~\cite{Oldenhof2023}. However this option necessitates scaling servers' 
computational ressources (CPUs, RAM) linearly with the number of training jobs in parallel which is impractical.

\subsubsection{Privacy of FedECA \label{sec:dp}}
We consider that time-to-event and censorship are safe to share, this is a strong assumption
but is often used in clinical trials as KM curves are released~\cite{liu2021ipdfromkm}.
More generally, every federated computational graph involved in FedECA and created by our implementation of the above algorithms 
as well the ones underpinning the FA methods of Section~\ref{sec:fed_analytics} can be audited easily thanks to \ref{fig:fedeca_graph}, \ref{fig:robust_cox_var}, \ref{fig:fed_kaplan_graph} and \ref{fig:fed_smd_graph}.
Each individual variable name in those graphs can be understood thanks to the associated tables~\ref{tab:webdisco_vars}, \ref{tab:cox_robust_var}, \ref{tab:prop_vars}, \ref{tab:fed_kaplan_vars} and \ref{tab:fed_smd_vars}.
Details on how this tracing step is performed are available in \ref{sec:software}.

Regarding the security of the covariates, we place ourselves in the ``honest-but-curious" threat model,
described in more detail in Substra's documentation~\cite{SubstraPrivacyStrategy}.

The only covariate used when doing IPTW is the treatment allocation, which is known throughout centers.
Therefore the only quantities tied to the covariates that are communicated
are (1) the gradients of the propensity model, and (2) the scalar product of covariates
and propensity model weights that are exposed through the propensity scores, averaged on risk sets
and on distinct event times.
Regarding the first point we propose an implementation of a differentially private version
of the propensity model training that we describe in the next paragraph.
Regarding the second point we assume that the dimension $p$ of the covariate vector is such that $p>>1$
and therefore that leaking scalar products is an acceptable risk in this context;
This is a strong assumption.
In the general case it could theoretically allow for attacks such as membership attacks~\cite{shokri2017membership}.
Making the pipeline end-to-end differential private (DP) is an open problem.
One could in principle rely again on DP to either add noise to the scalar
products themselves or to the propensity scores when training the Cox PH model. However,
this would affect the result even more than when applying DP only to the propensity
model training, which already has a strong effect see~\ref{fig:dp_results}.
Another research avenue would be to increase the average/minimum size of
the per-client risk sets by discretizing the times and applying random quantization
mechanisms (RQM)~\cite{youn2023randomized}.
We note that in this second case another downside would be that, in addition to
destabilizing the training of the Cox model, it would artificially create more ties in the data,
which would in return affect the quality of the Breslow estimator.

Because it exposes sums of additional covariates, studying the privacy of the federation
of adjusted IPTW requires a specific treatment that we leave to future work.

\paragraph*{Differential privacy of the propensity model}

We list here some properties of DP that are relevant to our implementation and refer the reader
to the work of ~\cite{dwork2014algorithmic} or ~\cite{desfontainesblog20211001}
for a more complete exposition of the topic:

DP provides slack parameters ($\epsilon,\delta$), which
allow to strike a trade-off between model accuracy and privacy of individual contributions.

A process $\mathcal{M}$ is ($\epsilon, \delta$)-DP if and only if $\forall
D, D'$ adjacent (differing by one element), we have:
\begin{equation}
    p(\mathcal{M}(D) \in S) \leq  p(\mathcal{M}(D') \in S) \cdot \exp(\epsilon) + \delta
\end{equation}
Perfect privacy guarantees are only obtained by taking $(\epsilon,\delta)=(0,0)$ which
makes the process $\mathcal{M}$ provably indistinguishable from the addition or removal of one
individual. In practice in real-world deployments it seems $\epsilon$ between $0.1$ and $50$ are used
depending on the application~\cite{desfontainesblog20211001} with different values of $\delta$.
DP benefits from nice composability properties~\cite{dwork2014algorithmic} and can thus be applied easily to
ML training methods that are iterative by nature and can therefore be applied to FL as well~\cite{abadi2016deep}.

We use the Opacus library~\cite{opacusyousefpour2021}, which implements the privacy
accountant method of~\cite{abadi2016deep} to train the propensity model within FedECA with differential
privacy (DP) with various ($\epsilon, \delta$) couples.

Our implementation is available in the script \href{https://github.com/owkin/fedeca/blob/main/fedeca/algorithms/torch_dp_fed_avg_algo.py}{torch\_dp\_fed\_avg\_algo.py}
and uses R{\'e}nyi differential privacy (RDP)~\cite{mironov2017renyi} which gives tighter
bounds alongside with Poisson sampling.

\subsection{Federated Analytics for end-to-end federated ECA analysis\label{sec:fed_analytics}}

Regulators ask for SMD and Kaplan-Meier survival curves~\cite{kaplan1958nonparametric}
in addition to the hazard ratio (HR), the associated confidence intervals (CI) and p-value in order to
validate the reweighting of the propensity model and to be able to describe the patient
population's time-to-events distribution within the two groups.

Therefore we also implement two federated analytics methods that we call Fed-Kaplan
and Fed-SMD in order to compute such quantities globally on distributed data without
compromising data.

It is to be noted that Fed-Kaplan can be directly derived from FedECA as it requires
the same quantities, namely the risk sets and number of occurrence of events. However,
Fed-SMD requires the communication of additional second order terms which increases
the attack surface of FedECA.

\subsubsection{Federated Kaplan-Meier estimator}

We follow FedECA implementation to compute per-center and communicate the unique
times of events $\mathcal{S}_{k}$, the (weighted) risk set $\mathcal{R}_{s, k}$
and the (weighted) number of deaths occurring at these times $\mathcal{D}_{s, k}$
for each arm.

This enables to compute in the server $\mathcal{R}_{s}$ and $\mathcal{D}_{s}$
which then allows to compute the Kaplan-Meier estimator at each time $t$ of a predefined
grid for each arm as well as the Greenwood and exponential Greenwood confidence
intervals~\cite{sawyer2003greenwood}.

For completeness, we remind the reader of these well-known formulas that we rewrite
using our notations:
\begin{align*}
    \hat{S}(t) &= \prod_{s \in \mathring{\mathcal{S}} | s \leq t} \left(1 - \frac{\underset{j \in \mathcal{D}_s}{\sum} w_j}{\underset{k \in \mathcal{R}_s}{\sum} w_k}\right), \\
    \Var(\hat{S}) &= \hat{S}(t)^{2} \prod_{s \in \mathring{\mathcal{S}} | s \leq t} \frac{\underset{j \in \mathcal{D}_s}{\sum} w_j}{(\underset{k \in \mathcal{R}_s}{\sum} w_k) \times \left(\underset{k \in \mathcal{R}_s}{\sum} w_k - \underset{j \in \mathcal{D}_s}{\sum} w_j \right)}, \\
    Z(t) &= \log ( \log \hat{S}(t)), \\
    \hat{\Var[Z(t)]} &= \frac{1}{(\log \hat{S}(t))^{2}} \prod_{s \in \mathring{\mathcal{S}} | s \leq t} \frac{\underset{j \in \mathcal{D}_s}{\sum} w_j}{(\underset{k \in \mathcal{R}_s}{\sum} w_k) \times \left(\underset{k \in \mathcal{R}_s}{\sum} w_k - \underset{j \in \mathcal{D}_s}{\sum} w_j \right)}.
\end{align*}
With $\hat{S}(t)$ the Kaplan-Meier estimator of the survival function and $\Var(\hat{S})$
and $\hat{\Var[Z(t)]}$ respectively the Greenwood and exponential Greenwood estimators
of the variance of the Kaplan-Meier estimator at time $t$.
In practice exponential Greenwood shall be used~\cite{sawyer2003greenwood} and this
is what we display in the results.

\subsubsection{SMD estimator \label{sec:smd}}
\paragraph{Computing SMD in a federated setting}
We compute the standardized mean difference (SMD) for each covariate before and
after weighting as defined by:
\begin{equation}
    SMD = \frac{\bar{x}_1 - \bar{x}_2}{\sqrt{\frac{s_1^2 + s_2^2}{2}}}.
\end{equation}
Where $\bar{x}_1$ and $\bar{x}_2$ are the means of the covariate in the two arms
and $s_1^2$ and $s_2^2$ are the variances of the covariate in the two arms.
As explained in \cite{greifer2023cobalt, austin2008critical} we use the
variance of the groups before weighting as a normalizer.
We compute this quantity in a federated fashion efficiently in two aggregation rounds
by developing the variance following~\cite{pebay2016numerically}:
\begin{equation}
    \frac{1}{n - 1} \sum_{i=1}^{n} (x_i - \bar{x})^2 = \frac{1}{n - 1} \left( \sum_{i=1}^{n} x_i^2 - n \bar{x}^2 \right).
\end{equation}
Effectively each center transmits uncentered moments of order 1 and 2 and the
server uses them to derive the centered moments of order 1 and 2.

\subsection{Datasets and cohorts construction}

\subsubsection{Synthetic data generating model of time-to-event outcome \label{subsec:data_sim}}
To illustrate the performance of our proposed FL implementation, we rely on simulations
with synthetic data. We simulate covariates and related time-to-event outcomes respecting
the proportional hazards (PH) assumption, with the baseline hazard
function derived from a Weibull distribution. For simplicity we assume a constant treatment
effect across the population. The data generation process consists of
several consecutive steps that we describe below assuming our target is a dataset with
$p$ covariates and $n$ samples.

First, a design matrix $\bm{X} = [\bm{X}^{(1)},\ldots, \bm{X}^{(p)}]\in \mathbb{R}^{n\times p} \sim \mathcal{N}(0, \bm{\Sigma})$ is
drawn from a multivariate normal distribution to obtain (baseline) observations for $n$
individuals described by $p$ covariates. The covariance matrix
$\bm{\Sigma}$ is taken to be a Toeplitz matrix such that the covariances between pairs $(\bm{X}^{(i)}, \bm{X}^{(j)})$ of
covariates decay geometrically. In other words, for a fixed $\rho>0$, we have
$\textnormal{cov}(\bm{X}^{(i)}, \bm{X}^{(j)}) = \rho^{|i-j|}$.
Such a covariance matrix implies a locally and hierarchically grouped structure
underlying the covariates, which we choose to mimic the potentially complex structure
of real-world data.
To reflect the varying correlations of the covariates with the outcome
of interest, the coefficients $\bm{\beta}_i$ of the linear combination used to build the
hazard ratio are drawn from a standard normal distribution.
\begin{equation}
  \label{eq-data-covariate}
  \begin{split}
    &\bm{\Sigma} = \textnormal{Toeplitz}(1, \rho, \rho^2, \cdots, \rho^{p-1}), \\
    &\bm{X}\in \mathbb{R}^{n\times p} \sim \mathcal{N}(0, \bm{\Sigma}), \\
    &\bm{\beta}\in \mathbb{R}^p \sim \mathcal{N}(0, 1).
  \end{split}
\end{equation}

In the context of clinical trials with external control arms, which implies
non-randomized treatment allocation, we simulate the treatment allocation in such a way
that it depends on the covariates. More precisely, we introduce the treatment allocation
variable $A$ that follows a Bernoulli distribution, where the probability of being
treated (the propensity score) $q$ depends on
a linear combination of the covariates, connected by a logit link function $g$. The
coefficients $\bm{\alpha}_i$ of the linear combination are drawn from a uniform distribution,
where the range $k \ge 0$ is symmetric around $0$ and is normalized by the number of
covariates. The degree of influence of the covariates on $A$ can be regulated by
adjusting the value of $k$. The greater the value of $k$, the stronger
the influence, and therefore the lower the degree of overlap between the distributions
of propensity scores of the treated and (external) control groups. Conversely, $k=0$
removes the dependence, leading to a randomized treatment allocation.
\begin{equation}
  \label{eq-data-propensity}
  \begin{split}
    &\bm{\alpha}\in \mathbb{R}^p \sim p^{-1/2}U(-k, k), \\
    &q_i = g^{-1}(\bm{\alpha}^T \bm{X}_i) = (1 + e^{-\bm{\alpha}^T \bm{X}_i})^{-1}, \\
    &a_i|\bm{X}_i \sim \textnormal{Bern}(q_i).
  \end{split}
\end{equation}

Once drawn, the treatment allocation variable $A_i$ is composed with the constant treatment
effect, defined here as the hazard ratio $\mu$, to obtain the final hazard ratio $h_i$
for each individual. The time-to-event $T^*_i$ of each sample is then drawn from a
Weibull distribution with shape $\nu$ and the scale depending on $h_i$ and $\nu$.
Meanwhile, for all samples we assume a constant dropout (or censoring) rate $d$ across
time, resulting in a censoring time that follows an exponential distribution.

\begin{equation}
  \label{eq-data-gen}
  \begin{split}
    &h_i(a_i) = \mu^{a_i}\exp(\bm{\beta}^T\bm{X}_{i}), \\
    &T^*_i \sim \mathcal{W}(h_i(a_i)^{-\frac{1}{\nu}}, \nu), \\
    &C_i \sim \mathcal{E}(d)
  \end{split}
\end{equation}
Finally, the event indication variable $\delta_i$ can be derived from $T^*_i$ and $C_i$:
$\delta_i=\mathds{1}_{T^*_i\leq C_i}$. And the observed outcome $Y_i$ for the $i$th
individual is defined as the couple $Y_i=(T_i=\min(T^*_i, C_i), \delta_i)$, i.e., it
corresponds to the observed time and the information on whether an event is observed.

\subsubsection{Prostate cancer cohort construction \label{sec:yoda_cohort_construction}}

We access data of two phase III randomized clinical trials from the Yale University Open Data Access (YODA) project~\cite{krumholz2016yale, ross2018overview}
of patients with metastatic castration-resistant prostate cancer. The first
trial, NCT02257736, has apalutamide, abiraterone acetate and prednisone
(Apa-AA-P) for the treatment arm, and placebo, abiraterone acetate and
prednisone (AA-P) for the control arm. The primary outcome is radiographic
progression-free survival (rPFS). The second trial, NCT00887198, has AA-P for
the treatment arm, and placebo and prednisone (P) for the control arm.
The primary outcomes are overall survival (OS) and rPFS. We thus artificially
distribute the data of the first trial to one ``client" (simulated server) and
the data of the second arm to the second client replicating natural splits such
as in~\cite{ogier2022flamby} to simulate a federated learning setup.

While all patients are randomized in each trial, it is still necessary to
correct for potential confounding when comparing arms from different trials.
First, the inclusion/exclusion criteria of both trials were aligned, patients
in NCT02257736 with present visceral metastases at randomization were excluded
to match the exclusion criterion of NCT00887198.
Then a group of variables of patient's baseline characteristics were chosen for
propensity-weighting based on literature review as well as on their
availability in both trials. The chosen covariates are age, body-mass index (BMI),
eastern cooperative oncology group (ECOG), brief pain inventory (BPI) score and
bone-metastasis-only. We present baseline characteristics for each trial in
~\ref{tab:nct02257736} and \ref{tab:nct00887198}.

We then filter these patients to remove non-informative patients and patients with
missing survival information. The final full cohort consists of $n=1927$
patients ($n=839$ for NCT02257736 and $n=1088$ for NCT00887198) in three
treatment arms (Apa-AA-P, AA-P and P). We infer the missing covariates on a
per-center basis using MissForest~\cite{stekhoven2012missforest}. The flow
diagram of the cohort construction is present in ~\ref{fig:yoda_flow},
including different ECA experiments conducted in this study.

We note that, since we submitted our research plan proposal to YODA (provided in \ref{fig:yoda_propal})
in order to access the data, we departed from the original plan in the following ways:
\begin{itemize}
    \item IPTW is studied instead of G-computation
    \item we do not study conformal prediction
    \item time-to-event endpoints are studied instead of change in SLD or change in PSA
\end{itemize}

\subsubsection{Pancreatic adenocarcinoma cohort construction \label{sec:cohort_construction}}
\paragraph*{Cohort construction}
We access data from three different sources: the Fédération Francophone de Cancérologie Digestive (FFCD),
the Institut d'Investigació Biomèdica de Girona (IDIBGI), and the Pancreatic Cancer Action Network (PanCAN).
The FFCD data consists of a subset of two clinical trials: PRODIGE 35~\cite{dahan2021randomized}
and PRODIGE 37~\cite{rinaldi2020gemcitabine} that respectively compare the first line efficacy
of, for PRODIGE 35, 6 months of FOLFIRINOX (arm A), 4 months of FOLFIRINOX followed
by leucovorin plus fluorouracil maintenance treatment for controlled patients (arm B),
and a sequential treatment alternating gemcitabine and fluorouracil, leucovorin,
and irinotecan every 2 months (arm C) and for PRODIGE 37: alternately receive
gemcitabine + nab-paclitaxel for 2 months then FOLFIRI.3 for 2 months (arm A),
or gemcitabine + nab-paclitaxel alone until progression (arm B).
We use both the FOLFIRINOX arm A from PRODIGE 35 ($n=92$) and the gemcitabine + nab paclitaxel
arm B from PRODIGE 37 with ($n=61$).
The inclusion criteria of this new subset is thus metastatic pancreatic adenocarcinoma
patients with a performance status eastern cooperative oncology group (ECOG)
of either 0, 1 or 2.
We select patients with the same inclusion criteria treated with FOLFIRINOX or
gemcitabine + nab-paclitaxel from clinical practice data from IDIBGI and PanCAN.
In IDIBGI we find $n=33$ patients treated with FOLFIRINOX and $n=192$ with
gemcitabine + nab-paclitaxel.
In PanCAN we find $n=91$ patients treated with FOLFIRINOX and $n=86$ with gemcitabine +
nab-paclitaxel patients that meet the criteria, totalling $n=177$ patients out of
$181$ originally available excluding ECOG 3 and 4. Among the $177$ patients,
$2$ are censored at the time the study starts. Therefore their data is not
informative for the Cox model fitting but might still be useful for the estimation
of the propensity model.
In addition, we identify in the PanCAN cohort the presence of an immortal time bias due to biopsy collection,
i.e., a patient will enter the PanCAN database only if they have had at least one biopsy before their last known follow-up.
Consequently, patients who died before having any biopsy will not be included in the database,
and patients in the database will be alive at least until their first biopsy.
The time interval between the start of treatment and the first biopsy is therefore an immortal time
for all patients in the PanCAN cohort.
Such immortal time bias will lead to inflated survival rates \cite{suissa2007} and is crucial to the present study.
To correct for this bias, we thus retrieve for all PanCAN patients the date of their first biopsy
and adjust their entry date into the study by taking the latest date between the first biopsy
and the start of first-line treatment for metastatic pancreatic cancer.

For each patient we access the following covariates: age at diagnosis,
ECOG performance status, biological gender determined by self-report and whether or not patients have liver metastasis
following the literature~\cite{klein2022comparison} and restrictions due
to data availability for covariates in each center.
We present baseline characteristics for each of the centers in ~\ref{tab:ffcd},\ref{tab:idibigi} and \ref{tab:pancan}
respectively for FFCD, IDIBGI and PanCAN.
We then filter these patients to remove non-informative patients, i.e., patients with
missing treatment or survival information.
The final full distributed cohort consists of $n=555$ patients ($n=153$ for FFCD,
$n=225$ for IDIBGI and $n=177$ for PanCAN).
We infer the missing covariates on a per-center basis using MissForest~\cite{stekhoven2012missforest}
considering ECOG as a numerical variable because it is ordered and apply
minimum-maximum normalization to numerical variables using $[0, 100]$ for age
values and $[0.0, 2.0]$ for ECOG loosely following~\cite{klein2022comparison}.

\paragraph*{Practical considerations associated with setting-up real-world federated learning collaborations}

While clinical trials data is well-standardized, real-world data from 
centers from multiple continents are not and need to be harmonized for the federation to be considered.
Clinical practice data has to be extracted by partners from different local sources
stored in different databases and accessed by different internal toolings leading to
a variety of extracted formats.
We ask the centers to align on a common data dictionary created from FFCD data,
which acts as the reference center as RCT data is already well-curated.
We share this dictionary as a Google Sheet to all partners specifying expected variables, units
formats and possible values.
Resulting data extracts have missing values and some data entries contain errors.
Thus, while some parts can be automated, the whole process from data extraction
to data harmonization involves some back and forth between data engineers from partner
centers, medical doctors, data stewarts and data scientists in order to perform thorough quality
checks of the input data.
It is interesting to note that, while we did not use large language models (LLMs)~\cite{vaswani2017attention}
in this work, they could certainly be useful to streamline parts of this process~\cite{kather2024large},
However, end-to-end automation seems out of reach with current technology~\cite{singhal2023large}.

\subsection{Real-world experiments setup details\label{subsec:real_world}}

All experiments in this article are simulated in-RAM with the exception of two experiments:
the first one which uses synthetic data and splits it into 10 cloud nodes and the
pancreatic adenocarcinoma use-case.

We refer to those two experiments as real-world in order to distinguish the
complexity of their deployment from in-RAM simulation cases.

In both cases, we use the Substra platform~\cite{galtier2019substra} to deploy the
federated learning network over secure cloud-based infrastructures.

Substra is distributed with Helm charts for each component. The charts package
all the files required for a deployment in a Kubernetes cluster.
Provisioning of the clusters and Substra deployment are performed using a private
Terraform module (known as infrastructure as-code).
We detail below the two different deployments.

\subsubsection{Synthetic data infrastructure setup \label{sec:gke_setup}}
For this experiment, the clusters are hosted on Google Kubernetes engine (GKE)
but Substra's deployment is cloud-agnostic. Provisioning of the GKE cluster and
Substra deployment are performed using a private Terraform module (known as
infrastructure as-code).
For this experiment, we used 11 Kubernetes clusters:
\begin{itemize}
\item 1 cluster is hosting the Substra orchestrator - single source of truth within the
federation - as well as a Substra Backend and Frontend, which makes it capable of
receiving and performing aggregation tasks. 
Substra's documentation refers to this cluster as ``AggregationNode".
\item 10 clusters are hosting a Substra Backend (and Frontend) only ; performing compute
tasks on local data. Substra's documentation refers to each of these clusters as ``TrainDataNode".
\end{itemize}
Clusters are physically in Belgium according to Google (``zone europe-west1"~\url{https://cloud.google.com/compute/docs/regions-zones?hl=en}).
GKE version used is \verb_1.27.2-gke.1200_ and the machines used are the ``n1-standard-16"~\url{https://cloud.google.com/compute/docs/general-purpose-machines?hl=en\#n1_machine_types}.
Regarding the communication protocol between centers, the organizations communicate
with the orchestrator via gRPC and over http(s) one to another. Since the experiment
is simulated in an internal environment using synthetic data we chose not to
enforce mutual transport layer security (mTLS).
More information can be found in Substra's documentation\url{https://docs.substra.org/en/latest/documentation/components.html}.

\subsubsection{Metastatic pancreatic adenocarcinoma data infrastructure setup\label{subsec:real_world_pancreatic}}

Similarly we deploy within Owkin Inc.'s own Federated Research Network cloud infrastructure
four nodes: a node for each of the three participating centers and an additional server
node the ``AggregationNode" with responsibilities described above.
Each node here is independent: the nodes are deployed in different regions with
different cloud providers. PanCan data are located in the US while Idibgi and FFCD data
are hosted in Europe.
The precise version information of the \href{https://github.com/Substra}{Substra}
versions used for this deployment are \verb_0.51.0+dev_ for \verb_substra-frontend_,
\verb_0.47.0+d5dfbdb6_ for \verb_substra-backend_ and
\verb_0.42.0+e6b1bddb_ for the \verb_orchestrator_ repository.

Partner centers uploaded their data to the corresponding nodes.

\subsection{Estimation of the treatment effect}

We compare FedECA to several competitors which, with the exception of pooled IPTW, are all adapted to the ECA setup,
where the IPD of a local cohort are accessible and only aggregated statistics are accessible for an external cohort.
Given the time-to-event nature of the outcome, we choose to estimate the hazard ratio
under the proportional hazards assumption as a measure of the treatment effect.
For all competitors, data is used to fit a Cox model as implemented in the \verb_lifelines_
library~\cite{davidson2019lifelines} to obtain the estimation.

\paragraph*{Unweighted Cox regression}
We implement a naïve Cox model regressing the observed outcome $Y$ on the treatment
allocation variable $A$, without using the weights of the samples.
This corresponds to an unadjusted comparison between the treated and untreated groups,
which would be valid in a randomized setting but not in an external control arm case.
In the ECA setup, this estimator corresponds to the WebDISCO method
and we use the implementation provided by the authors of this method.

\paragraph*{MAIC \label{sec:maic}}
Although not a method originally proposed for ECA, the MAIC method can be
adapted to perform ECA analysis under proper assumptions:
First, for the reweighting step, we make use of the implementation available in the
\verb_indcomp_ package (\url{https://github.com/AidanCooper/indcomp}).
Specifically, the IPD of the local cohort are reweighted so that a specified group
of covariates matches the external cohort in terms of means and variances,
creating, by design, reweighted data with zero SMD relative to the external cohort for each covariate.
Then, in the absence of IPD of the external cohort, in order to train the Cox model to estimate treatment effect,
we further assume that the (aggregated) risk set of the external cohort is also accessible.
In this case, methods based on the digitization of the Kaplan-Meier curve~\cite{guyot2012enhanced}
can be used to construct pseudo IPD as an approximation to the IPD of the external cohort.
The pseudo IPD are then assigned a uniform weight of one
and combined with the reweighted local cohort to estimate the treatment effect.
In our simulation experiments, for reasons of simplicity and without loss of validity,
we use the real IPD of the external cohort as an idealization of the pseudo IPD,
which sets the upper bound of MAIC's performance in the ECA setup.

\paragraph*{Pooled IPTW}
The general concept and strategy of IPTW has been described before (see Section~\ref{sec:method_overview}).
In the implementation, the core estimation process is divided into two key steps.
First, the propensity scores are estimated using unpenalized logistic regression
or, alternatively, they can be provided externally to the estimator. These scores are
then used to compute inverse probability weights tailored to the effect estimand.
For the average treatment effect (ATE), weights are based on the inverse of propensity
scores for both treated and control groups. For the average treatment effect on the
treated (ATT), the weights involve a combination of treatment indicators (for the
treated individuals) and inverse propensity scores (for the control individuals).
Second, the treatment effect estimation is performed by fitting a weighted Cox
proportional hazards model, where the inverse probability weights are incorporated in
the regression model of the observed outcome $Y$ on the treatment allocation $A$.

\paragraph*{Competing paradigms}
We already motivated the choice of IPTW as the best weighting method for small
sample sizes.
However other methods than weighting and matching could be considered for federation
as well such as G-computation~\cite{robins1986new,chatton2020g} or doubly debiased machine
learning~\cite{chernozhukov2018double,loiseau2022external} as their performance
should be comparable~\cite{loiseau2022external}. We leave their federation to future work.

\subsection{Experiments details}

\subsubsection{Early stopping within a static distributed framework \label{sec:static}}

As Substra is static and requires to fix the number of federated rounds a priori,
we implement early-stopping for the stopping criterion on the Hessian norm by
running up to $MAX_{iter}$ rounds ($20$ in practice) and backtrack to find the
first round where convergence was achieved.

\subsubsection{Federated hyper-parameters selection \label{sec:fedhp}}

We note that, due to distribution and privacy constraints, standard practices such as cross-validation
might become difficult to setup. In practice in the case there is one sample per patient, which is what we study here,
what we recommend is to treat the distributed datasets as a single distributed dataset, as in the
federated global bootstrap, and proceed to splitting accordingly. We further recommend per-center stratification
for fairness considerations and to avoid pathological cases. We note that, naïvely, any stratification on a private variable would require
exposing and sharing this variable at least to the server, which limits the kinds of cross-validation that can be applied.
We refer to the works of ~\cite{dai2020federated, khodak2021federated, pmlr-v202-wang23n} for going beyond those recommandations.

\subsubsection{Software and Reproducibility \label{sec:software}}

Following the recent trend of switching from R to Python for implementing statistical software~\cite{polsterl2020scikit, skglm, muzellec2022pydeseq2},
we chose Python as the base language for our implementation.
This choice is also motivated by the fact that most FL research implementation code
is written in Python.
We follow reference survival analysis packages implementation design choices such as
\verb_lifelines_~\cite{davidson2019lifelines} and \verb_scikit-survival_~\cite{polsterl2020scikit}.
We use the Substra software~\cite{galtier2019substra} which is an open-source software that has been audited and
validated by security teams of both hospitals and pharmaceutical companies across
different FL projects.
Substra has demonstrated its ability to be deployed in real-world conditions for
biomedical research purposes in the MELLODDY project~\cite{Oldenhof2023, Oldenhof2023},
as well as in the HealthChain project on breast cancer treatment response prediction~\cite{du2023federated}.

FedECA is available as a Python package on Github: \url{https://github.com/owkin/fedeca} for non-commercial use.

The code in this repository follows best practices such as continuous integration (CI), thorough code testing (coverage
of code at $82$\% on commit \verb_a6ec22c_), deployed documentation using Github pages as well as the use of pre-commit hooks
to help manage the repository's evolution.

The availability of the code not only ensures the reproducibility of the results presented in this article as well as the possibility
to audit its implementation, but also opens the possibility for other research teams to perform real-world
federated ECA.

Indeed, a user can launch FedECA running the exact same code either in-RAM for simulations,
or on a real deployed substra network in real conditions by modifying the backend type,
as shown in ~\ref{list:code_example}.

The \href{https://github.com/owkin/fedeca}{FedECA repository} contains a quickstart
as well as detailed documentation and comments, which should allow easy replication.

All quantitative figures in this article with synthetic data can be reproduced by
following instructions in \verb_experiments/README.md_. The associated yaml
configurations provide all hyper-parameters that were used.

For experiments on 10 centers replication involves deploying a substra network,
which require some development operations (DevOps) capabilities. However, details
in section~\ref{sec:gke_setup} should be sufficient to reproduce the results.
The associated experiment script is defined in \href{https://github.com/owkin/fedeca/blob/main/experiments/config/experiment/real_world_runtimes.yaml}{real\_world\_runtimes.yaml}.

For experiments on YODA data, we install the fedeca package within the YODA
platform, split the data in such a way that the control arm and the treatment arm are
in two separate groups, and run fedeca with bootstrap variance estimation.
Scripts used to preprocess the data and run the experiments are available in the
\href{https://github.com/owkin/fedeca/blob/main/experiments/yoda/}{yoda folder}
in the \verb_fedeca_ repository.

For experiments on metastatic pancreatic adenocarcinoma data, we use the fedeca
package unaltered on commit \verb_a6ec22c_ after having registered the data in the Substra
platform. Obfuscated versions of the scripts that ran on the deployed platform and that were
used to generate the related figures in the article are available in the
\href{https://github.com/owkin/fedeca/blob/main/experiments/pdac/}{pdac folder}
in the \verb_fedeca_ repository.
By obfuscated we mean that dataset hashes or urls in this script were
converted to random strings so they cannot be mapped to any of the original data
or servers.

The nature of this last experiment is such that replications require data access which
might be restricted, see Section~\ref{sec:data_availability}. However once access to
data is obtained and federated network is deployed all experiments should be easily
reproduced thanks to the above scripts.

Regarding the automatic tracing of all quantities communicated by FedECA's federated
algorithms such as the one reported in ~\ref{fig:fedeca_graph} the logging code originally developed
by~\cite{muzellec2024fedpydeseq2} and relying on remote methods decorators can be found in the \href{https://github.com/owkin/fedeca/pull/77}{jean/logging branch}.
Executing \verb+plot_graphs_and_tables.py+
automatically generates all corresponding grapĥs and tables in the article.

Further questions can be addressed to the corresponding author J.O.d.T. through
the creation of github issues or via direct e-mail.

\sectionwordcount
\section{Data availability \label{sec:data_availability}}
All data generated in this study can be re-generated using the scripts provided in https://github.com/owkin/fedeca.

Data are available from the Yale University Open Data Access (YODA) Project under restricted access to qualified researchers.
Access can be obtained by following instructions on the YODA Project website https://yoda.yale.edu/request/.
The entire data access process takes approximately 3 months.
Access is provided for 1 year but can be renewed for additional years.

The FFCD, IDIBGI data are available under restricted access to qualified researchers, access can be obtained by contacting
directly the main investigators in each center: J.-B. B. for FFCD and R. C. for IDIBGI.
Expected timeframe for access is a few months. Data can be available for any negotiated durations.

Data from PanCAN are available under restricted access to qualified researchers,
access can be obtained by submitting a proposal for review at www.pancan.org/spark.
Data will be provided to qualified and approved researchers within one month of request.
The duration of data access varies based on contractual agreement,
but generally is two or three years.

Source data are provided with this paper.

\sectionwordcount
\section{Code Availability}
The integrality of the code is publicly released and openly available for
research purposes under a research only license at the following URL: \url{https://github.com/owkin/fedeca}.

\renewcommand{\refname}{References}
\putbib[article]
\end{bibunit}

\section{Acknowledgements}
\sectionwordcount
This study, carried out under YODA Project 2023-5198, used data obtained from
the Yale University Open Data Access Project, which has an agreement with
JANSSEN RESEARCH \& DEVELOPMENT, L.L.C.. The interpretation and reporting of
research using this data are solely the responsibility of the authors and does
not necessarily represent the official views of the Yale University Open Data
Access Project or JANSSEN RESEARCH \& DEVELOPMENT, L.L.C..

The data coming from the Pancreatic Cancer Action Network was collected
through the \href{https://spark.sbgenomics.com/}{SPARK health data platform} and derives from the
"PanCAN Know Your Tumor Program". We especially thank the patients that
shared their data with this program.

Part of this work corresponding to work-package 4 of the RHU AI-TRIOMPH 
and carried out by Owkin France was supported by Agence Nationale de la Recherche
as part of the France 2030 plan with reference ANR-23-RHUS-0012 (H.L. and F.B).

We thank Cameron Davidson Pilon for developing the wonderful python package that is
\verb_lifelines_ and that was a source of inspiration for this work.
We thank Arthur Pignet for discussions on efficient bootstrap implementation
with substra as well as for his code allowing efficient distributed computation of
moments.
We thank Constance Beguier for her help in deriving the distributed computation of
the Kaplan-Meier estimator.
We thank Ulysse Marteau-Ferey for his help applying and modifying the Substra tracing code that
he developed for another concurrent research work.
We thank Andy Karabajakian for having helped answer questions on chemotherapies combinations for
metastatic pancreatic cancers.
We thank Maylis Largeteau and Parjeet Kaur for their help in writing the FedECA license.
Finally, we thank Jean-Philippe Vert and Nathan Noiry for their insightful comments
and suggestions.

\section{Author Contributions Statement}
\sectionwordcount
M.A. and F.B. conceived the idea of investigating federated methods for external
control arms and supervised the paper writing and both contributed equally.
J. O.d.T., Q.K. and H.L. wrote the paper and led the research.
J. O.d.T. designed and implemented all the federated algorithms.
Q.K. implemented the data generation process as well as the pooled baselines with the help of H.L.
and J. O.d.T implemented all federated learning related code. M.A. and J. O.d.T wrote
the differential privacy federated code together.
H.L. and I.M. helped writing the paper and performing experiments.
Notably H.L. performed the YODA experiments with the help of I.M. and J.O.d.T.
N. L. reviewed the statistical methodology and helped writing the paper.
M. H. provided support for creating the figures and feedbacks on the writing of the paper.
M. D. on one side and T. C. and T. F. on the other were in charge of respectively
administrating the federated learning network infrastructure and deploying Substra.
L. D., J. T., P. L.-P., J.-B. B., S. Z., R. N. and J. C. were responsible
for the data from the PRODIGE35 and PRODIGE37 trials held at FFCD. D. G. reviewed the paper.
In addition to providing data from listed clinical trials, J. T. discussed the study design
and reviewed the manuscript

R. C. T. and A. G. V. curated the data from IDIBGI, K. A. and S. D. curated the data from PanCAN
under the guidance of J. A. C. and Z. Y. in charge of data-harmonization and specifications.

\section{Competing Interests Statement}
The authors declare the existence of a financial competing interest.

Some authors are or were employed by Owkin, Inc. during their
time on the project (J. O.d.T, Q.K., M.A., H.L, I.M., N.L., M.H., M.D., T.C., T.F., F.B., J.A.C., Z.Y.).

P. L.-P. has received honoraria for consulting and/or advisory board for AMGEN,
Pierre Fabre, Biocartis, Servier and BMS.

J.B. Bachet has received personal fees from Amgen, Bayer, Bristol Myers Squibb,
GlaxoSmithKline, Merck Serono, Merck Sharp \& Dohme, Pierre Fabre, Sanofi, Servier,
and non-financial support from Amgen, Merck Serono, and Roche, outside the
submitted work.

J. T. has received honoraria as a speaker and/or in an advisory role from AMGEN,
Astelllas, Astra Zeneca, Boehringer, BMS, Merck KGaA, MSD, Novartis, ONO
pharmaceuticals, Pierre Fabre, Roche Genentech, Sanofi, Servier and Takeda.

A. G. V. has received honoraria as a speaker and/or in an advisory role from
Astra Zeneca, Merck Serono, MSD, Novartis, Roche Genentech, Sanofi, and Servier.

R. N. has received honoraria as a consultant from Cure51.

Part of this work corresponding to work-package 4 of the RHU AI-TRIOMPH and carried out
by Owkin France was supported by Agence Nationale de la Recherche as part of the France
2030 plan with reference ANR-23-RHUS-0012 (H.L and F.B.).

The remaining authors declare no competing interests.

\setcounter{figure}{0}

\section{Tables}
\FloatBarrier
\begin{table}
  \centering
  \resizebox{\columnwidth}{!}{
  \begin{tabular}{llllllll}
    \toprule
    \makecell{Experiment \\ (Treatment vs. Control)} & Method & \makecell{Treatment \\ data source} & \makecell{Control \\ data source} & \makecell{log(HR)} &  HR (95\% CI) & Z & p \\
    \midrule
    Apa-AA-P vs. AA-P & FedECA & Trial 1 & Trial 2 & \num{-0.416977} & \num{0.659036} (\num{0.554353}, \num{0.783488}) & \num{-4.724677} & \textless0.00001 \\
                      & Literature \cite{saad2021nct02257736} & Trial 1 & Trial 1 & \num{-0.356674} & 0.70 (\num{0.60}, \num{0.83}) & \num{-4.3086506} & \textless0.0001 \\
    \midrule
    AA-P vs. P        & FedECA & Trial 1 & Trial 2 & \num{-0.691208} & \num{0.500971} (\num{0.423294}, \num{0.592901}) & \num{-8.040978} & \textless0.000001 \\
                      & Literature \cite{ryan2013nct00887198} & Trial 2 & Trial 2 & \num{-0.634878} & 0.53 (\num{0.45}, \num{0.62}) & \num{-7.765664} & \textless0.001 \\
    \midrule
    AA-P vs. AA-P     & FedECA & Trial 1 & Trial 2 & \num{-0.106712} & \num{0.898784} (\num{0.767739}, \num{1.052198}) & \num{-1.327159} & \num{0.184456} \\
    \midrule
    Apa-AA-P vs. P    & FedECA & Trial 1 & Trial 2 & \num{-0.988639} & \num{0.372083} (\num{0.308969}, \num{0.448090}) & \num{-10.424608} & \textless0.000001 \\
    \bottomrule
    \end{tabular}}
\caption{
  Treatment effect estimation on radiographic progression-free survival by comparing different regimens across different trials.
  Trial 1 refers to NCT02257736, Trial 2 refers to NCT00887198.
  All $p$-values resulting from the Wald test performed on the entry of $\bm{\hat{\beta}}$ corresponding to the treatment allocation, assuming a $\chi^2$
  distribution with $1$ degree of freedom are obtained with bootstrap variance estimation.
  Exact $p$-values are from top to bottom (excluding results from literature) $2.3e^{-6}$, $8.9e^{-16}$, $0.184456$ and $1.9e^{-25}$.
  Source data are provided as a Source Data file.
}
\label{tab:real_world_stats_yoda}
\end{table}
\begin{table}[tbh]
  \centering
\begin{tabular}{llllll}
  \toprule
  Method & Data source & log(HR) & HR (95\% CI) & Z & p \\
  \midrule
  FedECA & {FFCD, IDIBGI} & \num{-0.170391} & \num{0.843335} (\num{0.680945}, \num{1.044451}) & \num{-1.561420} & \num[round-mode=figures,round-precision=3]{0.118425} \\

  IPTW   & {FFCD}         & \num{0.125947} & \num{1.134222} (\num{0.799698}, \num{1.608681}) & \num{0.706363} & \num[round-mode=figures,round-precision=3]{0.479963} \\

  IPTW   & {IDIBGI}       & \num{-0.117347} & \num{0.889276} (\num{0.646546}, \num{1.223134}) & \num{-0.721526} & \num[round-mode=figures,round-precision=3]{0.470586} \\

  \bottomrule
  \end{tabular}
\caption{
  Estimation of treatment effect on overall survival by comparing FOLFIRINOX against gemcitabine + nab placlitaxel.
  Results with single data source are obtained without FL.
  All $p$-values resulting from the Wald test performed on the entry of $\bm{\hat{\beta}}$ corresponding to the treatment allocation, assuming a $\chi^2$
  distribution with $1$ degree of freedom are obtained with bootstrap variance estimation.
  Exact $p$-values are from top to bottom $0.118425$, $0.479963$ and $0.470586$.
  Source data are provided as a Source Data file.
}
\label{tab:real_world_stats}
\end{table}
\FloatBarrier

\section{Figure Legends/Captions}

\begin{figure*}[!h]
  \centering
  \begin{subfigure}[b]{\textwidth}
    \centering
    \includegraphics[width=0.1\linewidth]{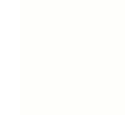}
    \includegraphics[width=0.1\linewidth]{placeholder.png}
    \includegraphics[width=0.1\linewidth]{placeholder.png}
    \caption{Illustration of randomized controlled trials (RCT) versus an external control arm (ECA) analysis.}
  \end{subfigure}
  \begin{subfigure}[b]{0.8\textwidth}
    \centering
    \includegraphics[width=0.3\linewidth]{placeholder.png}
    \caption{Federated ECA setup.}
  \end{subfigure}
  \caption{
      FedECA graphical abstract. (a) In an RCT, patients are randomly assigned to either the experimental (i.e. treatment) or the control arm. In an ECA, patients are assigned to the treatment arm, while the control arm is defined using historical data. Due to this absence of randomization and the resulting confounding, the two groups of patients cannot be compared directly.
      To overcome this issue, a model is used to capture the association between the treatment allocation and the confounding factors.
      From this model, weights are computed and are used to balance the two arms to ensure comparability.
      Then, the weights are incorporated into a Cox model to estimate the treatment effect.
      Finally a statistical test is performed to assess the significance of the measured treatment effect.
      (b) In the considered setting, patient data is stored in different geographically distinct centers
      and a similar analysis as in (a) is attempted thanks to our algorithm FedECA.
      A trusted third party is responsible for the orchestration of the training processes, which
      consists of exchanging model related quantities across the centers.
      No individual patient data is shared between the centers and only aggregated
      information is exchanged, which limits patient data exposure while producing equivalent
      results.
      Some of the symbols used in the figure have been bought to the Noun Project, Inc. by M.H. granting M.H. perpetual, 
      non-exclusive, worldwide rights to such symbols.
  }
  \label{fig:graphical_abstract}
\end{figure*}

\begin{figure*}[tb]
    \centering
    \includegraphics[width=0\linewidth]{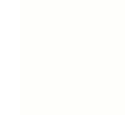}
    \caption{
        Pooled equivalence between IPTW and FedECA.
        Box- and swarm-plots of the relative errors between FedECA and the pooled IPTW on four different quantities:
        the hazard ratio of the treatment allocation covariate estimated from a Cox model,
        the partial likelihood of the Cox model,
        the p-value associated to the hazard ratio,
        and the propensity scores estimated from the logistic regression.
        For each quantity, relative error is defined as the absolute difference
        between the pooled IPTW value and the FedECA value, divided by the pooled IPTW value.
        Each quantity was computed from $n=100$ repetitions of the simulation that is computed by running
        FedECA and pooled IPTW on $n$ random draws of $1000$ samples with $10$ covariates.
        Red dotted line indicates a relative error of $0.2$\% between FedECA and the pooled IPTW.
        Boxplot and swarm-plot uses the seaborn python library's default settings that is: boxes are from the first to the third quartiles, the black line being the median,
        and whiskers extend to the lowest (resp. highest) data point still within $1.5$ inter quartile range of the lower (upper) quartile.
        No statistical test was used.
        Source data are provided as a Source Data file.
    }
    \label{fig:pooled_equivalent}
\end{figure*}

\begin{figure*}
  \centering
  \captionsetup{justification=centering}
  \medskip

    \centering
    \begin{subfigure}{0.3\linewidth}
    \includegraphics[width=\linewidth]{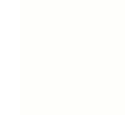}
    \subcaption{Mean absolute SMD as a function of the covariate shirt.}
    \label{fig:smd_a}
    \end{subfigure}
    \begin{subfigure}{0.3\linewidth}
    \includegraphics[width=\linewidth]{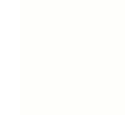}
    \subcaption{Absolute SMD for each covariate.}
    \label{fig:smd_b}
    \end{subfigure}
    \begin{subfigure}{0.3\linewidth}
      \includegraphics[width=\linewidth]{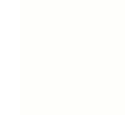}
      \subcaption{Comparison of different methods on statistical power and type I error of treatment
      effect estimation.}
      \label{fig:power_typeI}
    \end{subfigure}
    \caption{
          Comparison of different methods on statistical power, type I error of treatment effect estimates
          as well as standardized mean difference (SMD) of covariates between the two treatment arms.
        (a) Curves representing the mean absolute SMD computed on $10$ covariates as a function of the covariate shift for three different methods: FedECA, MAIC and the non-adjusted treatment effect estimation (unweighted) over $n=100$ repetitions.
        Shaded area is the two-sided $95$\% interval around the mean assuming standard normal distributions.
        (b) Boxplots representing the distribution of the absolute SMD over the $n=100$ repetitions for the first five covariates.
        Each estimation of SMD is based on $n=100$ repetitions of propensity score
        estimation. For all simulations, we generate $10$ covariates and $1000$ samples.
        Boxplot and swarmplot uses the seaborn python library's default settings that is: boxes are from the first to the third quartiles, the black line being the median,
        and whiskers extend to the lowest (resp. highest) data point still within $1.5$ inter quartile range of the lower (upper) quartile.
        (c) Different variance estimation methods leading to different p-values
        are given in parentheses after each method giving point estimates of
        the hazard ratio. In particular, the naive variance estimation is based
        on the simple inversion of the observed Fisher information. For statistical power,
        only results of methods that consistently control the type I error around/under
        $0.05$ (marked by grey dashed lines in top panels) are shown.
        Each estimation of statistical power or type I error is based on $n=1000$ repetitions
        of treatment effect estimation. For bootstrap-based variance estimating methods, the number
        of bootstrap resampling is set to $200$. For all simulations, we assume $10$ covariates.
        The hazard ratio of the simulated treatment effect is set to $0.4$ for the
        estimation of statistical power, and to $1.0$ for the estimation of type I error.
        For simulations with varying covariate shifts (the two panels on the left),
        the number of samples is fixed at $700$. For simulations with varying sample
        size (the two panels on the right), the covariate shift is fixed at $2.0$.
        The asterisk on FedECA indicates that, due to the
        time-consuming nature of the power analysis, their more lightweight pooled-equivalent 
        counterparts were used instead (pooled IPTW)
        For confidence intervals we use the central
        limit theorem applied to Bernoulli variables to compute parameters of the associated normal
        and plot the two-sided $95$\% intervals as error bars.
        No statistical test was used.
        Source data are provided as a Source Data file.
        \label{fig:smd}
    }
  \end{figure*}
 
  \begin{figure}
  \centering
  \includegraphics[width=0\linewidth]{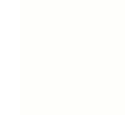}
  \caption{
        Real-world FOLFIRINOX effect estimation using FedECA versus local analyses.
      (a) SMD of covariates between the two arms of the combined FFCD+IDIBGI cohort, before and after weighting by FedECA's propensity model. Therefore each
      dot represents the SMD over $n=153+225=378$ samples.
      (b) Weighted Kaplan-Meier curves of the combined FFCD+IDIBGI cohort using FedECA's propensity model. Sample size is $n=378$. The $95$\% confidence intervals displayed are obtained using the exponential Greenwood formula.
      (c) Weighted Kaplan-Meier curves of the FFCD cohort using a local propensity model. Sample size is $n=225$. The $95$\% confidence intervals displayed are obtained using the exponential Greenwood formula.
      (d) Weighted Kaplan-Meier curves of the IDIBGI cohort using a local propensity model. Sample size is $n=153$. The $95$\% confidence intervals displayed are obtained using the exponential Greenwood formula.
      Associated $p$-values can be found in the associated table.
      Source data are provided as a Source Data file.
  }
  \label{fig:real_world}
\end{figure}
  \clearpage
\FloatBarrier

\newpage
\appendix
\clearpage
\newpage
\title{FedECA: Federated External Control Arms for Causal Inference with Time-To-Event Data in Distributed Settings}
\section*{Supplementary Material}
\FloatBarrier

\renewcommand\thefigure{Supplementary Figure \arabic{figure}} %
\renewcommand\thetable{Supplementary Table \arabic{table}}
\renewcommand\thelstlisting{Supplementary Figure \arabic{lstlisting}}

\renewcommand{\figurename}{}
\renewcommand{\tablename}{}
\renewcommand{\lstlistingname}{}

\addcontentsline{toc}{section}{Supplementary Material}

\setcounter{table}{0}
\setcounter{figure}{1}
\setcounter{lstlisting}{0}
\sectionwordcount
\setlength{\LTpre}{1pt} %
\setlength{\LTpost}{1pt}

\begin{table*}
  \centering
  \begin{tabular}{llrl}
    \toprule
    Method & Environment & \#centers & Runtime (s) \\
    \midrule
    FedECA (robust) & real-world setup & 2 & $4.48\cdot 10^{3} \pm 1.08\cdot 10^{2}$ \\
    FedECA (robust) & real-world setup & 3 & $4.57\cdot 10^{3} \pm 5.88\cdot 10^{1}$ \\
    FedECA (robust) & real-world setup & 5 & $4.58\cdot 10^{3} \pm 9.42\cdot 10^{1}$ \\
    FedECA (robust) & real-world setup & 8 & $4.56\cdot 10^{3} \pm 9.53\cdot 10^{1}$ \\
    FedECA (robust) & real-world setup & 10 & $5.00\cdot 10^{3} \pm 7.80\cdot 10^{2}$ \\
    FedECA (robust) & in-RAM & 2 & $4.95 \pm 8.21\cdot 10^{-1}$ \\
    FedECA (robust) & in-RAM & 3 & $6.72 \pm 4.92\cdot 10^{-1}$ \\
    FedECA (robust) & in-RAM & 5 & $1.27\cdot 10^{1} \pm 1.72$ \\
    FedECA (robust) & in-RAM & 8 & $1.43\cdot 10^{1} \pm 1.38$ \\
    FedECA (robust) & in-RAM & 10 & $1.93\cdot 10^{1} \pm 2.02$ \\
    IPTW & -- & -- & $2.34\cdot 10^{-1} \pm 2.41\cdot 10^{-2}$ \\
    \bottomrule
    \end{tabular}
\caption{Runtimes of different federated and pooled experiments in different conditions: in-RAM simulations or running in a deployed Substra network in the cloud (real-world setup).
$n=5$ repetitions per experiment are used to measure the mean and standard deviations.
Source data are provided as a Source Data file.
\label{tab:real_world_synth}}
\end{table*}

\begin{table*}[tbh]
  \small
  \centering
  \caption{Comparison of distributed ECA methods for time-to-event outcomes and generic data pooling alternative. Green color highlights
  methods compatible with distributed ECA (ATE:
  average treatment effect; ATT: average treatment effect on the treated; ATC:
  average treatment effect on the control; KM-type information: Kaplan-Meier-type information consisting of observed time, censorship status and potentially group assignment). The differential privacy check marks that authors do study the application of differential privacy to their base method not necessarily that their method uses DP by default. 
  Green shade is used to highlight the $3$ leading methods in terms of number of supported settings. \label{tab:relatedworks}} Source data are provided as a Source Data file.
  \renewcommand{\arraystretch}{1.2}
  \setlength{\tabcolsep}{1pt}
  \begin{adjustbox}{angle=90}
  \begin{tabular}{lgccccccgg}
    \multicolumn{1}{l}{} & \multicolumn{8}{c}{Methods} \\
    \rowcolor{white}                         \\
    \multicolumn{1}{l|}{} & \makecell{\cellcolor{white} FedECA\\\cellcolor{white}(Ours)} & \makecell{WebDISCO\\\cite{lu2015webdisco}} & \makecell{\, IPW Cox\\\cite{shu2020inverse}} & \makecell{\, FCI \,\\\cite{xiong2021federated}} & \makecell{DC-COX\\\cite{imakura2023dc}} &  \makecell{ODACH\\\cite{luo2022odach}} & \makecell{WICOX\\\cite{park2022wicox}} & \makecell{\cellcolor{white} Huang et al.\\\cellcolor{white} \cite{huang2023covariate}} & \makecell{\cellcolor{white} MAIC\\ \cellcolor{white} \cite{signorovitch2012matching}} \\ \midrule \midrule
    \rowcolor{white}
    \multicolumn{1}{l|}{Supported training settings} &           &           &           &         &         &         &         &      \\ \hline \hline
    \multicolumn{1}{l|}{\makecell[l]{\;\;- stratified Cox model training \\\;\;\;\;  w/o weights}} &    \cmark      &     \cmark      &     \cmark     &   \cmark   &    \cmark      &    \cmark    &     \cmark & \cmark  &     \cmark      \\
    \multicolumn{1}{l|}{\makecell[l]{\;\;- weighted stratified Cox model training}} &    \cmark      &     \xmark      &     \cmark     &   \cmark   &    \cmark      &    \xmark    &     \xmark &     \cmark   &     \cmark      \\
    \multicolumn{1}{l|}{\makecell[l]{\;\;- stratified IPTW analysis}} &     \cmark      &     \xmark      &     \cmark   &     \cmark   &    \cmark   &    \xmark   &    \xmark &     \cmark  &     \cmark     \\
    \multicolumn{1}{l|}{\makecell[l]{\;\;- distribution-independent Cox \\\;\;\;\;  model training w/o weights}} &    \cmark      &     \cmark      &   \xmark   &   \xmark   &    \cmark      &    \xmark    &     \xmark  &     \cmark  &     \cmark      \\
    \multicolumn{1}{l|}{\makecell[l]{\;\;- weighted distribution-independent \\\;\;\;\; Cox model training}} &    \cmark      &     \xmark      &   \xmark   &   \xmark   &    \cmark      &    \xmark    &     \xmark &     \cmark   &     \cmark     \\
    \multicolumn{1}{l|}{\makecell[l]{\;\;- distribution-independent IPTW analysis}} &     \cmark      &     \xmark      &     \cmark   &     \cmark   &    \cmark   &    \xmark   &    \xmark &     \cmark &     \cmark     \\
    \multicolumn{1}{l|}{\makecell[l]{\;\;- Pooled equivalence with classical IPTW}} &     \cmark      &     \cmark      &     \cmark      &    \cmark     &    \xmark    &   \cmark     &   \cmark  &    \xmark  &    \xmark     \\\midrule \midrule
    \rowcolor{white}
  \multicolumn{1}{l|}{Treatment groups} &           &           &           &         &         &         &         \\ \midrule \midrule
    \multicolumn{1}{l|}{\makecell[l]{\;\;- Treatment groups in distinct centers}} &     \cmark      &     \cmark   &    \xmark     &    \xmark     &     \cmark   &   \xmark    &     \xmark    &     \cmark   &     \cmark      \\
    \multicolumn{1}{l|}{\makecell[l]{\;\;- Correction for confounding}} &     \cmark      &    \xmark       &     \cmark  &     \cmark  &     \cmark     &   \xmark   &   \xmark    &     \cmark    &     \cmark   \\
    \multicolumn{1}{l|}{\makecell[l]{\;\;- Multiple external control centers}} &    \cmark      &     \cmark      &     \xmark   &     \xmark   &    \cmark      &    \xmark    &     \xmark  &     \cmark   &     \xmark     \\ \midrule \midrule
    \rowcolor{white}
    \multicolumn{1}{l|}{Causal estimands} &           &           &           &         &         &         &          \\ \midrule \midrule
    \multicolumn{1}{l|}{\;\;- ATE} &     \cmark      &     \xmark      &     \cmark   &     \cmark   &     \cmark   &   \xmark   &   \xmark  &     \cmark  &     \xmark       \\
    \multicolumn{1}{l|}{\;\;- ATT} &     \cmark      &     \xmark      &     \cmark   &    \cmark   &     \cmark   &   \xmark    &   \xmark   &     \cmark &    \xmark       \\
    \multicolumn{1}{l|}{\;\;- ATC} &     \cmark      &     \xmark     &     \cmark   &     \cmark   &     \cmark   &   \xmark    &   \xmark  &     \cmark  &     \cmark      \\ \midrule \midrule
    \rowcolor{white}
    \multicolumn{1}{l|}{Privacy} &           &           &           &         &         &         &            \\ \midrule \midrule
    \multicolumn{1}{l|}{\makecell[l]{\;\;- Does not require pooling data}} &     \cmark      &     \cmark      &     \cmark      &    \cmark     &    \xmark    &   \cmark     &   \cmark   &     \cmark  &    \cmark     \\
    \multicolumn{1}{l|}{\makecell[l]{\;\;- Only KM-type\\\;\;\;\; information is shared}} &     \cmark      &     \cmark      &     \cmark   &     \cmark      &     \xmark     &     \cmark      &     \cmark    &     \cmark  &     \cmark     \\
    \multicolumn{1}{l|}{\;\;- Differential privacy} &     \cmark      &     \xmark      &     \xmark  &     \xmark  &     \xmark    &    \xmark  &    \xmark  &     \xmark  &     \xmark     \\ \midrule \midrule
    \multicolumn{1}{l|}{\makecell[l]{Available implementation}} &     \cmark      &     \cmark      &     \cmark      &    \cmark     &    \xmark    &   \cmark     &   \xmark   &     \xmark  &    \cmark      \\
  \end{tabular}
\end{adjustbox}

\end{table*}

\begin{figure}[tb]
  \centering
  \includegraphics[width=1.\linewidth]{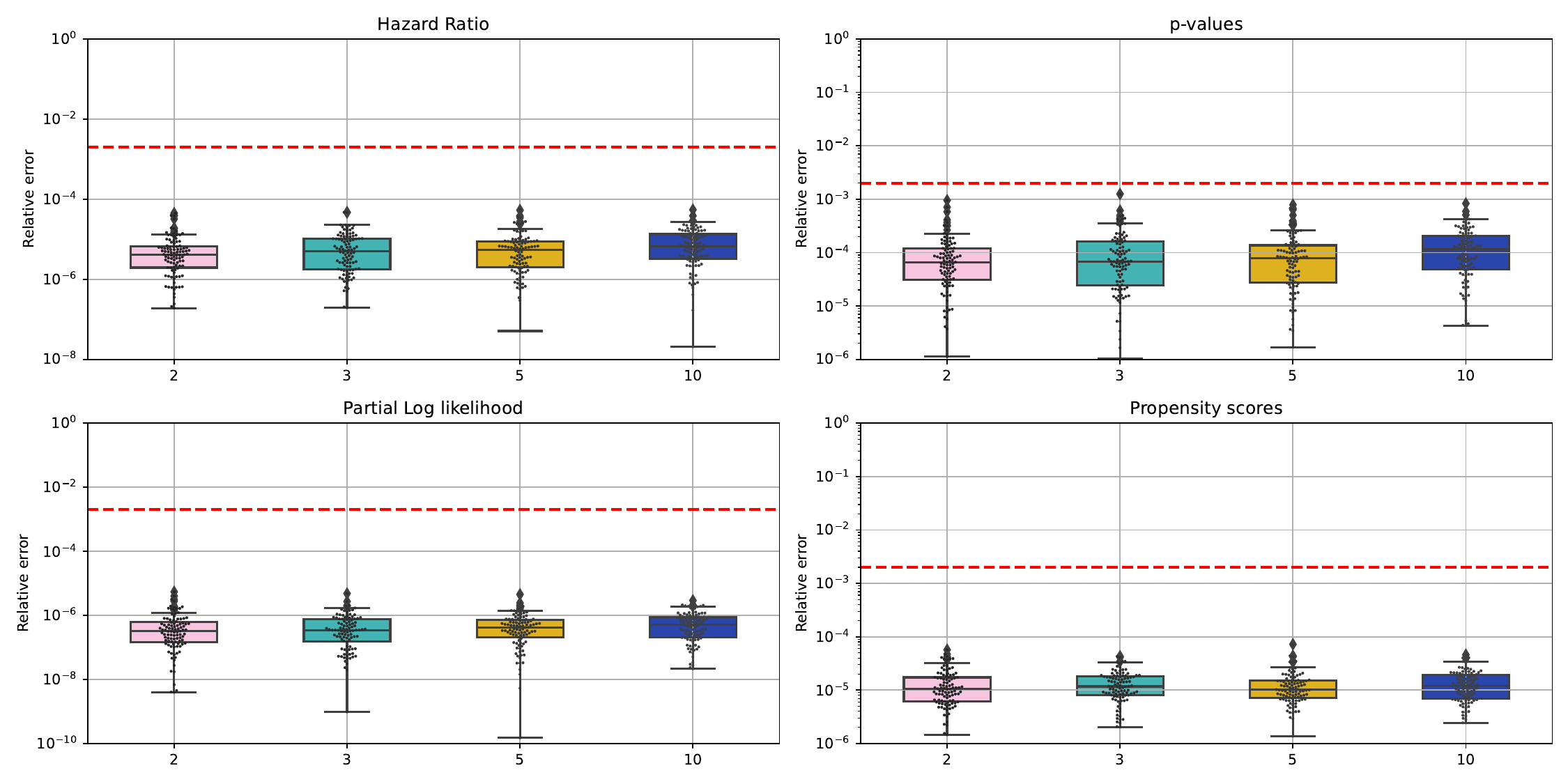}
  \caption{
      Pooled equivalent with varying number centers.
      Boxplots of the relative error between the pooled IPTW and the FedECA algorithm on four different
      quantities. The propensity scores estimated from the logistic regression, the hazard ratio (the treatment effect)
      the p-values associated to the treatment allocation variable (Wald test) and the partial likelihood resulting from the Cox model.
      Each of these quantities was monitored as we increased the number of centers across which the data is split from 2 to 10 centers.
      The errors were computed on simulated data with 100 repetitions.
      The red dotted line represents a relative error of 1\% between pooled IPTW and FedECA.
      $n=100$ repetitions are used similarly.
      Boxplot and swarm-plot uses the seaborn python library's default settings that is: boxes are from the first to the third quartiles, the black line being the median,
      and whiskers extend to the lowest (resp. highest) data point still within $1.5$ inter quartile range of the lower (upper) quartile.
      No statistical test was used.
      Source data are provided as a Source Data file.
  }
  \label{fig:pooled_equivalent_nb_clients}
\end{figure}

\begin{figure}[t]
  \centering
  \begin{subfigure}[b]{0.49\textwidth}
    \centering
    \includegraphics[width=1.\linewidth]{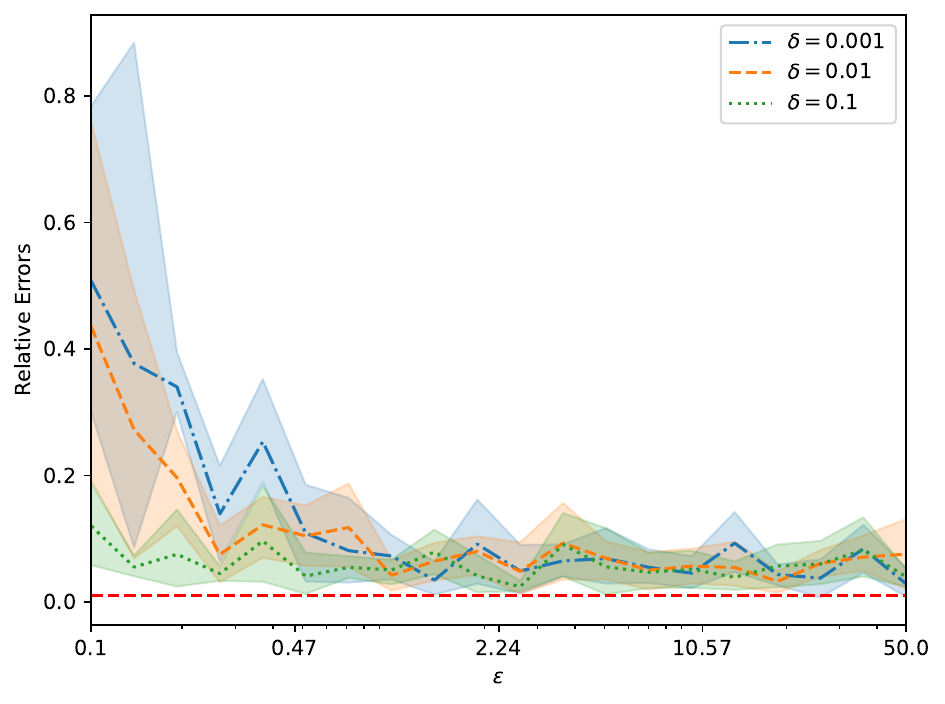}
    \caption{Hazard-Ratios}
  \end{subfigure}
  \begin{subfigure}[b]{0.49\textwidth}
    \centering
    \includegraphics[width=1.\linewidth]{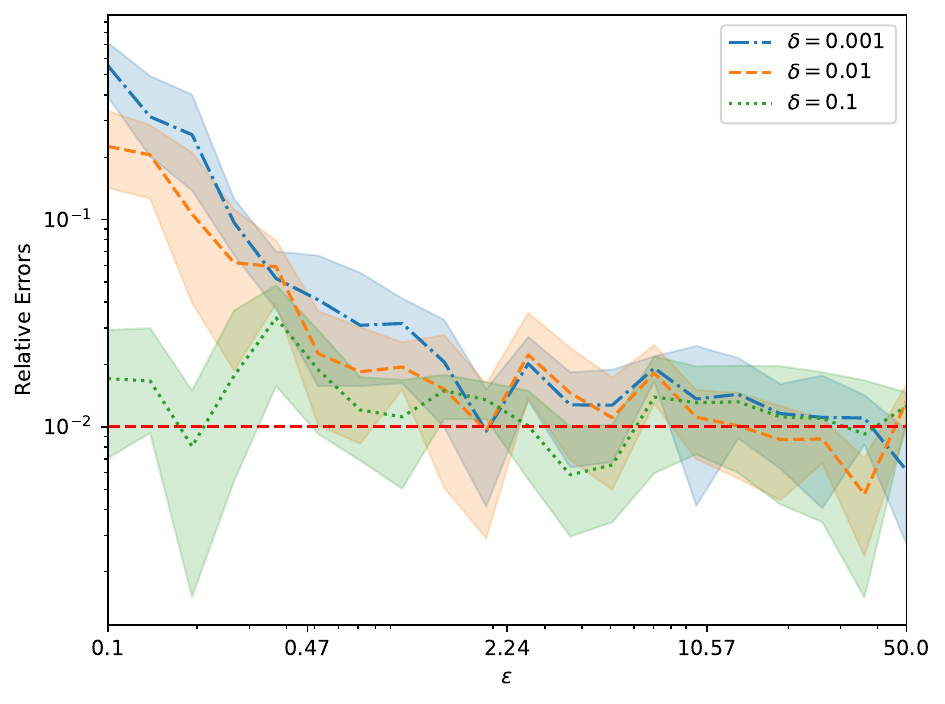}
    \caption{Partial Log-likelihood}
  \end{subfigure}
  \begin{subfigure}[b]{0.49\textwidth}
    \centering
    \includegraphics[width=1.\linewidth]{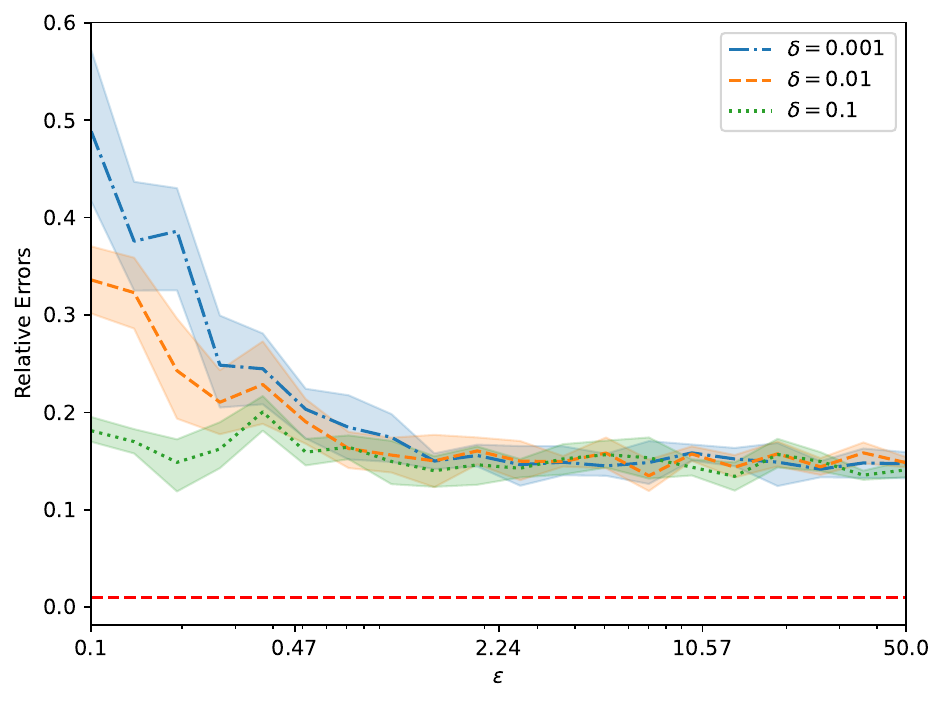}
    \caption{Propensity scores}
  \end{subfigure}
  \begin{subfigure}[b]{0.49\textwidth}
    \centering
    \includegraphics[width=1.\linewidth]{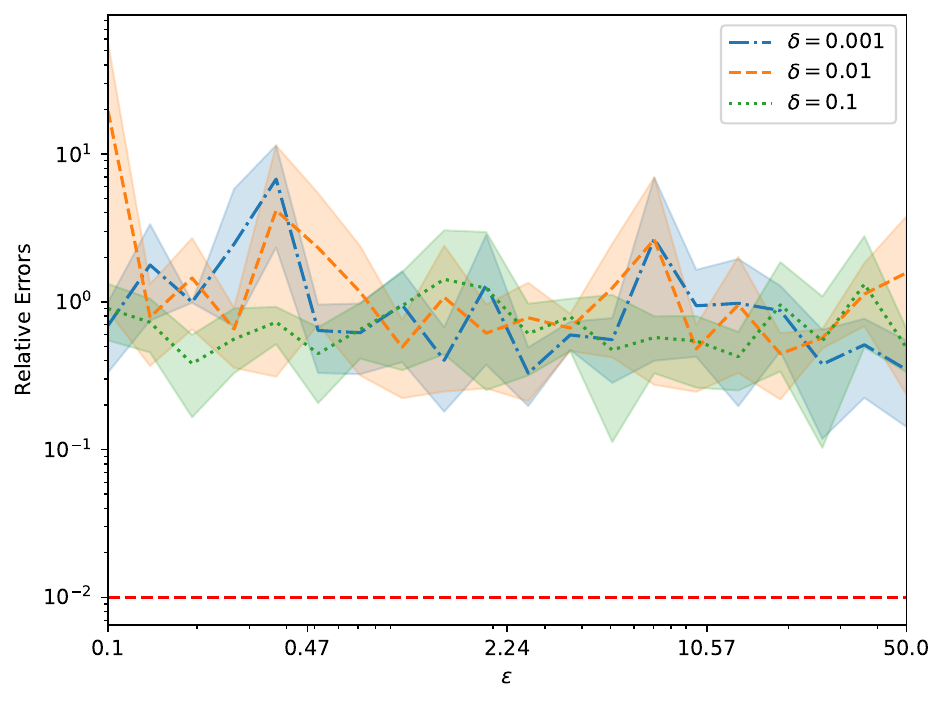}
    \caption{P-values}
  \end{subfigure}
  \caption{
      DP-FedECA. Adding differential privacy into FedECA.
      Comparison of the results of running DP-FedECA with respect to the pooled baseline
      with no privacy. We see that even for large $\epsilon$ that correspond to lower
      amount of noise, the relative difference between the $p$-values produced by DP-FedECA
      and the true $p$-value is high even if the propensity weights are relatively close.
      The final operation to build the $p$-value involves a second-order term which is highly sensitive
      to the precise value of the propensity scores.
      $n=5$ repetitions were used for each experiment.
      Error bars are $95$\% confidence intervals obtained via $1000$ bootstraps of those $5$ samples as per lineplot's defaults.
      Source data are provided as a Source Data file.
  }
  \label{fig:dp_results}
\end{figure}

\begin{figure*}[t]
  \centering
  \begin{subfigure}[b]{0.49\textwidth}
    \centering
    \includegraphics[width=1.\linewidth]{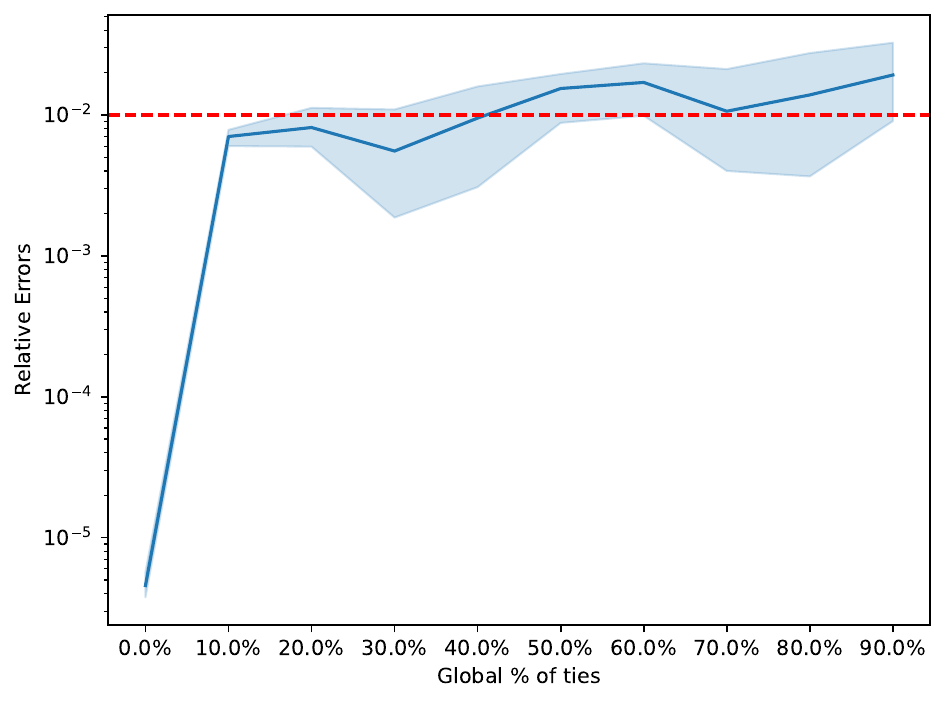}
    \caption{Hazard-Ratios}
  \end{subfigure}
  \begin{subfigure}[b]{0.49\textwidth}
    \centering
    \includegraphics[width=1.\linewidth]{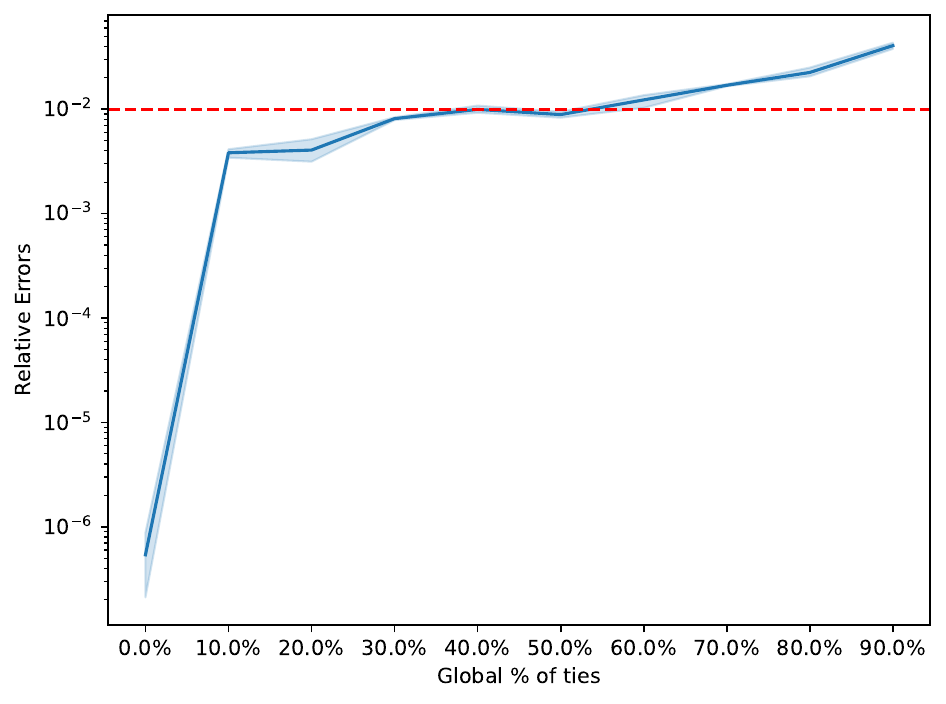}
    \caption{Partial Log-likelihood}
  \end{subfigure}
  \begin{subfigure}[b]{0.49\textwidth}
    \centering
    \includegraphics[width=1.\linewidth]{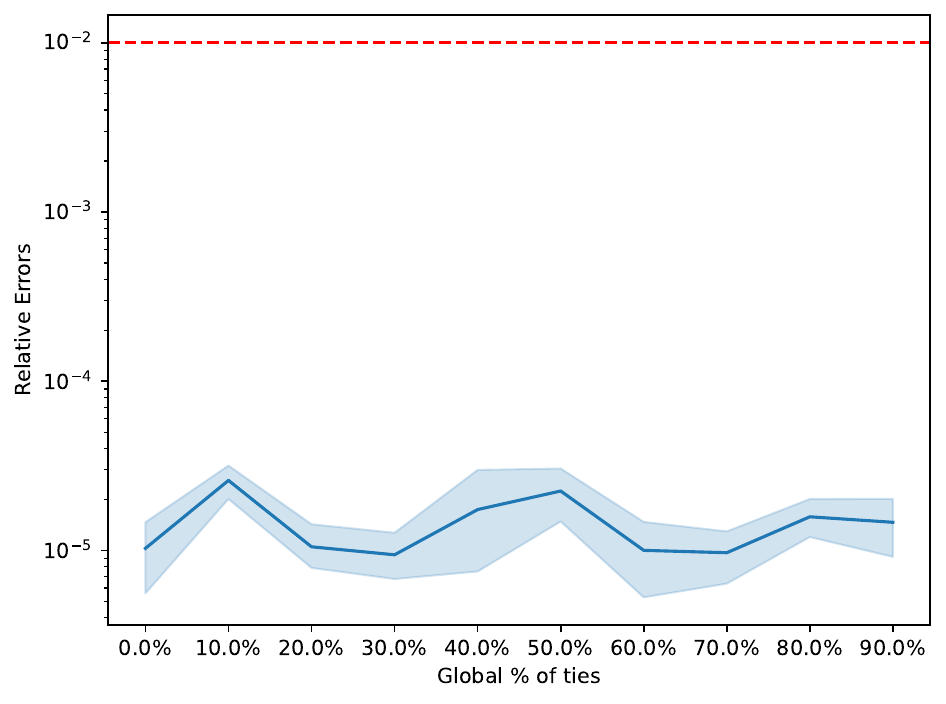}
    \caption{Propensity scores}
  \end{subfigure}
  \begin{subfigure}[b]{0.49\textwidth}
    \centering
    \includegraphics[width=1.\linewidth]{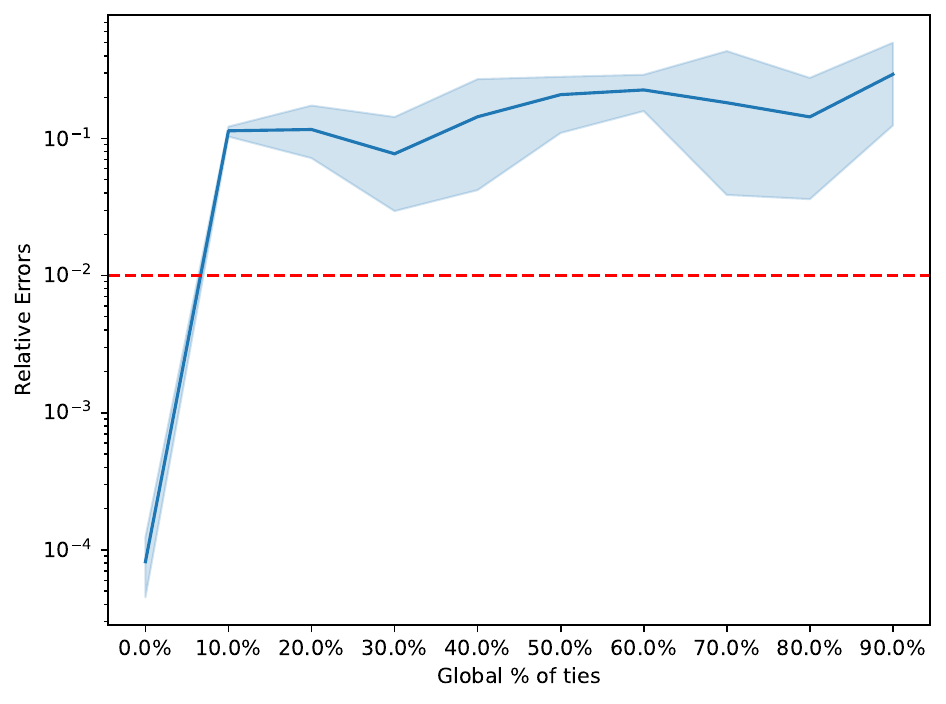}
    \caption{P-values}
  \end{subfigure}
  \caption{
      Influence of ties on FedECA accuracy.
      Comparison of the results of running FedECA with respect to the pooled baseline
      using Efron's approximation. Performance degrades with the number of ties.
      For realistic number of ties errors are below $1$\%.
      $n=5$ repetitions were used in this plot.
      Error bars are $95$\% confidence intervals of $n=1000$ bootstraps as per lineplot's defaults.
      Source data are provided as a Source Data file.
  }
  \label{fig:ties}
\end{figure*}

\begin{lstlisting}[language=Python, caption=Python code to launch FedECA on simulated data using any type of deployment., label=list:code_example, captionpos=b]
  from fedeca import FedECA
  from fedeca.survival_utils import CoxData

  SEED = 42  # Seed for the generation of synthetic data
  NSAMPLES = 1000  # Number of samples in total
  N_CLIENTS = 5  # Number of simulated centers
  N_COV = 10  # Number of covariates
  # Types of backend used for the FL, simu is the most lightweight,
  # real-world FL is "remote"
  BACKEND_TYPE = "simu"
  # Simulates FL by splitting a dataframe across centers and register
  # each dataset into Substra. In case of a real deployment, private
  # datasets are registered by each organization's data engineers.
  data = CoxData(seed=SEED, n_samples=NSAMPLES, ndim=N_COV)
  df = data.generate_dataframe()
  df.drop(columns=["propensity_scores"], axis=1, inplace=True)
  # As in sklearn we first instantiate an object
  fedeca = FedECA(N_COV, treated_col="treated", duration_col="T", event_col="E",
                  num_rounds_list=[50, 50], variance_method="robust")
  # We then call the fit method of the object to launch the FL
  fedeca.fit(df, n_clients=N_CLIENTS, backend_type=BACKEND_TYPE)
  \end{lstlisting}

\begin{figure}[bt!]
    \centering
    \includegraphics[width=1.\linewidth]{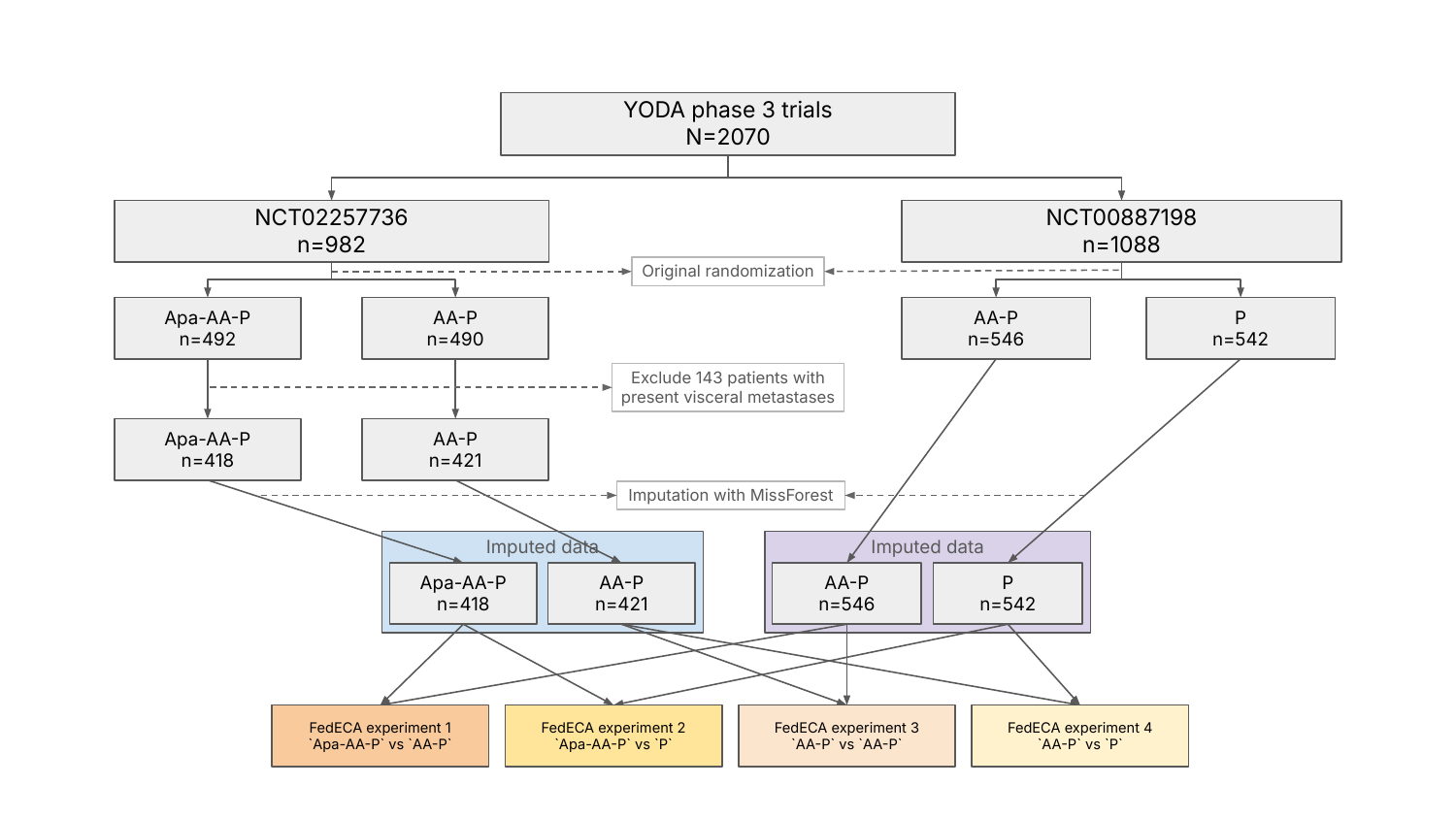}
    \caption{
        Flow diagram of YODA clinical trial cohort construction. Source data are provided as a Source Data file.
    }
    \label{fig:yoda_flow}
\end{figure}

\begin{table}[h]
  \footnotesize
  \centering
\begin{tabular}{rlll}
  \toprule
  & Apa-AA-P (n=418) & AA-P (n=421) & All (n=839) \\
  \midrule
  Age \\
  Median (IQR) & 71.0 (66.0-78.0) & 71.0 (65.0-77.0) & 71.0 (66.0-77.0) \\
  Percent missing & 0.0\% & 0.0\% & 0.0\% \\
  \midrule
  Body Mass Index (BMI) \\
  Median (IQR) & 27.5 (24.9-30.9) & 27.9 (25.1-31.3) & 27.72 (24.9-31.14) \\
  Percent missing & 2.1\% & 2.9\% & 2.5\% \\
  \midrule
  Performance Status (ECOG) \\
  0 & 290 (69.4\%) & 299 (71.0\%) & 589 (70.2\%) \\
  1 & 128 (30.6\%) & 122 (29.0\%) & 250 (29.8\%) \\
  2 & 0 (0\%) & 0 (0\%) & 0 (0\%) \\
  Percent missing & 0 (0\%) & 0 (0\%) & 0 (0\%) \\
  \midrule
  Brief Pain Inventory (BPI) score \\
  $\leq$ 1 & 317 (75.8\%) & 295 (70.1\%) & 612 (72.9\%) \\
  $>$ 1 & 95 (22.7\%) & 118 (28.0\%) & 213 (25.4\%) \\
  Percent missing & 6 (1.4\%) & 8 (1.9\%) & 14 (1.7\%) \\
  \midrule
  Bone metastasis only \\
  True  & 207 (49.5\%) & 205 (48.7\%) & 412 (49.1\%) \\
  False & 211 (50.5\%) & 216 (51.3\%) & 427 (50.9\%) \\
  Percent missing & 0 (0.0\%) & 0 (0.0\%) & 0 (0.0\%) \\
  \bottomrule
\end{tabular}
\caption{Baseline characteristics of NCT02257736. Source data are provided as a Source Data file.}
\label{tab:nct02257736}
\end{table}

\begin{table}[th]
  \footnotesize
  \centering
\begin{tabular}{rlll}
  \toprule
  & AA-P (n=546) & P (n=542) & All (n=1088) \\
  \midrule
  Age \\
  Median (IQR) & 68.0 (64.0-76.0) & 68.0 (60.0-76.0) & 68.0 (64.0-76.0) \\
  Percent missing & 0.0\% & 0.0\% & 0.0\% \\
  \midrule
  Body Mass Index (BMI) \\
  Median (IQR) & 28.4 (26.0-31.5) & 28.4 (25.8-31.8) & 28.4 (25.8-31.6) \\
  Percent missing & 2.2\% & 2.2\% & 2.2\% \\
  \midrule
  Performance Status (ECOG) \\
  0 & 411 (75.3\%) & 409 (75.5\%) & 820 (75.4\%) \\
  1 & 134 (24.5\%) & 133 (24.5\%) & 267 (24.5\%) \\
  2 & 1 (0.2\%) & 0 (0\%) & 1 (0.1\%) \\
  Percent missing & 0 (0\%) & 0 (0\%) & 0 (0\%) \\
  \midrule
  Brief Pain Inventory (BPI) score \\
  $\leq$ 1 & 370 (67.8\%) & 346 (63.8\%) & 716 (65.8\%) \\
  $>$ 1 & 129 (23.6\%) & 147 (27.1\%) & 276 (25.4\%) \\
  Percent missing & 47 (8.6\%) & 49 (9\%) & 96 (8.8\%) \\
  \midrule
  Bone metastasis only \\
  True  & 238 (43.6\%) & 241 (44.5\%) & 479 (44.0\%) \\
  False & 286 (52.4\%) & 281 (51.8\%) & 567 (52.1\%) \\
  Percent missing & 22 (4.0\%) & 20 (3.7\%) & 42 (3.9\%) \\
  \bottomrule
\end{tabular}
\caption{Baseline characteristics of NCT00887198. Source data are provided as a Source Data file.}
\label{tab:nct00887198}
\end{table}

\begin{table}[h]
  \footnotesize
  \centering
\begin{tabular}{rlll}
  \toprule
  & FOLFIRINOX (n=92) & Gemcitabine + Nab-Paclitaxel (n=61) & All (n=153) \\
  \midrule
  Age \\
  Median (IQR) & 64.82 (56.08-69.61) & 65.99 (59.59-69.46) & 65.21 (58.08-69.53) \\
  Percent missing & 0.0\% & 0.0\% & 0.0\% \\
  \midrule
  Performance Status (ECOG) \\
  0 & 36 (39.1\%) & 23 (37.7\%) & 59 (38.6\%) \\
  1 & 56 (60.9\%) & 31 (50.8\%) & 87 (56.9\%) \\
  2 & 0 (0.0\%) & 7 (11.5\%) & 7 (4.6\%) \\
  Percent missing & 0 (0.0\%) & 0 (0.0\%) & 0 (0.0\%) \\
  \midrule
  Biological gender \\
  F & 36 (39.1\%) & 33 (54.1\%) & 69 (45.1\%) \\
  M & 56 (60.9\%) & 28 (45.9\%) & 84 (54.9\%) \\
  Percent missing & 0 (0.0\%) & 0 (0.0\%) & 0 (0.0\%) \\
  \midrule
  Liver metastasis \\
  True & 76 (82.6\%) & 51 (83.6\%) & 127 (83.0\%) \\
  False & 16 (17.4\%) & 10 (16.4\%) & 26 (17.0\%) \\
  Percent missing & 0 (0.0\%) & 0 (0.0\%) & 0 (0.0\%) \\
  \bottomrule
\end{tabular}
\caption{Baseline characteristics of FFCD. Source data are provided as a Source Data file.}
\label{tab:ffcd}
\end{table}

\begin{table}[h]
  \footnotesize
  \centering
\begin{tabular}{rlll}
  \toprule
  & FOLFIRINOX (n=33) & Gemcitabine + Nab-Paclitaxel (n=192) & All (n=225) \\
  \midrule
  Age \\
  Median (IQR) & 59.00 (52.00-64.00) & 66.00 (61.00-73.00) & 65.00 (60.00-72.00) \\
  Percent missing & 0.0\% & 0.0\% & 0.0\% \\
  \midrule
  Performance Status (ECOG) \\
  0 & 10 (30.3\%) & 39 (20.3\%) & 49 (21.8\%) \\
  1 & 18 (54.5\%) & 119 (62.0\%) & 137 (60.9\%) \\
  2 & 1 (3.0\%) & 19 (9.9\%) & 20 (8.9\%) \\
  Percent missing & 4 (12.1\%) & 15 (7.8\%) & 19 (8.4\%) \\
  \midrule
  Biological gender \\
  F & 17 (51.5\%) & 88 (45.8\%) & 105 (46.7\%) \\
  M & 16 (48.5\%) & 104 (54.2\%) & 120 (53.3\%) \\
  Percent missing & 0 (0.0\%) & 0 (0.0\%) & 0 (0.0\%) \\
  \midrule
  Liver metastasis \\
  True & 27 (81.8\%) & 141 (73.4\%) & 168 (74.7\%) \\
  False & 6 (18.2\%) & 51 (26.6\%) & 57 (25.3\%) \\
  Percent missing & 0 (0.0\%) & 0 (0.0\%) & 0 (0.0\%) \\
  \bottomrule
\end{tabular}
\caption{Baseline characteristics of IDIBGI. Source data are provided as a Source Data file.}
\label{tab:idibigi}
\end{table}

\begin{table}[h]
  \footnotesize
  \centering
\begin{tabular}{rlll}
  \toprule
  & FOLFIRINOX (n=91) & Gemcitabine + Nab-Paclitaxel (n=86) & All (n=177) \\
  \midrule
  Age \\
  Median (IQR) & 61.51 (57.20-68.11) & 66.25 (57.83-74.19) & 63.61 (57.47-70.40) \\
  Percent missing & 0.0\% & 0.0\% & 0.0\% \\
  \midrule
  Performance Status (ECOG) \\
  0 & 29 (31.9\%) & 19 (22.1\%) & 48 (27.1\%) \\
  1 & 37 (40.7\%) & 41 (47.7\%) & 78 (44.1\%) \\
  2 & 2 (2.2\%) & 10 (11.6\%) & 12 (6.8\%) \\
  Percent missing & 23 (25.3\%) & 16 (18.6\%) & 39 (22.0\%) \\
  \midrule
  Biological gender \\
  F & 30 (33.0\%) & 40 (46.5\%) & 70 (39.5\%) \\
  M & 61 (67.0\%) & 46 (53.5\%) & 107 (60.5\%) \\
  Percent missing & 0 (0.0\%) & 0 (0.0\%) & 0 (0.0\%) \\
  \midrule
  Liver metastasis \\
  True & 66 (72.5\%) & 53 (61.6\%) & 119 (67.2\%) \\
  False & 25 (27.5\%) & 33 (38.4\%) & 58 (32.8\%) \\
  Percent missing & 0 (0.0\%) & 0 (0.0\%) & 0 (0.0\%) \\
  \bottomrule
\end{tabular}
\caption{Baseline characteristics of PanCAN. Source data are provided as a Source Data file.}
\label{tab:pancan}
\end{table}

\begin{table}[tbh]
  \centering
\begin{tabular}{llllll}
  \toprule
  Method & Data source & log(HR) & HR (95\% CI) & Z & p \\
  \midrule
  FedECA & {FFCD, IDIBGI, PanCAN} & \num{-0.209824} & \num{0.810727} (\num{0.661086}, \num{0.994240}) & \num{-2.015453} & \num[round-mode=figures,round-precision=3]{0.043857} \\
  IPTW   & {PanCAN}         & \num{-0.136442} & \num{0.872457} (\num{0.608231}, \num{1.251468}) & \num{-0.741272} & \num[round-mode=figures,round-precision=3]{0.458528} \\
  \bottomrule
  \end{tabular}
\caption{
  Estimation of treatment effect on overall survival by comparing FOLFIRINOX against gemcitabine + nab placlitaxel,
  using data from all three centers and from PanCAN alone (no FL involved).
    All $p$-values resulting from the Wald test performed on the entry of $\bm{\hat{\beta}}$ corresponding to the treatment allocation, assuming a $\chi^2$
  distribution with $1$ degree of freedom are obtained with bootstrap variance estimation.
  Exact $p$-values are from top to bottom $0.043857$ and $0.458528$.
  Source data are provided as a Source Data file.
}
\label{tab:real_world_all_centers}
\end{table}

\begin{figure}
  \centering
  \includegraphics[width=\linewidth]{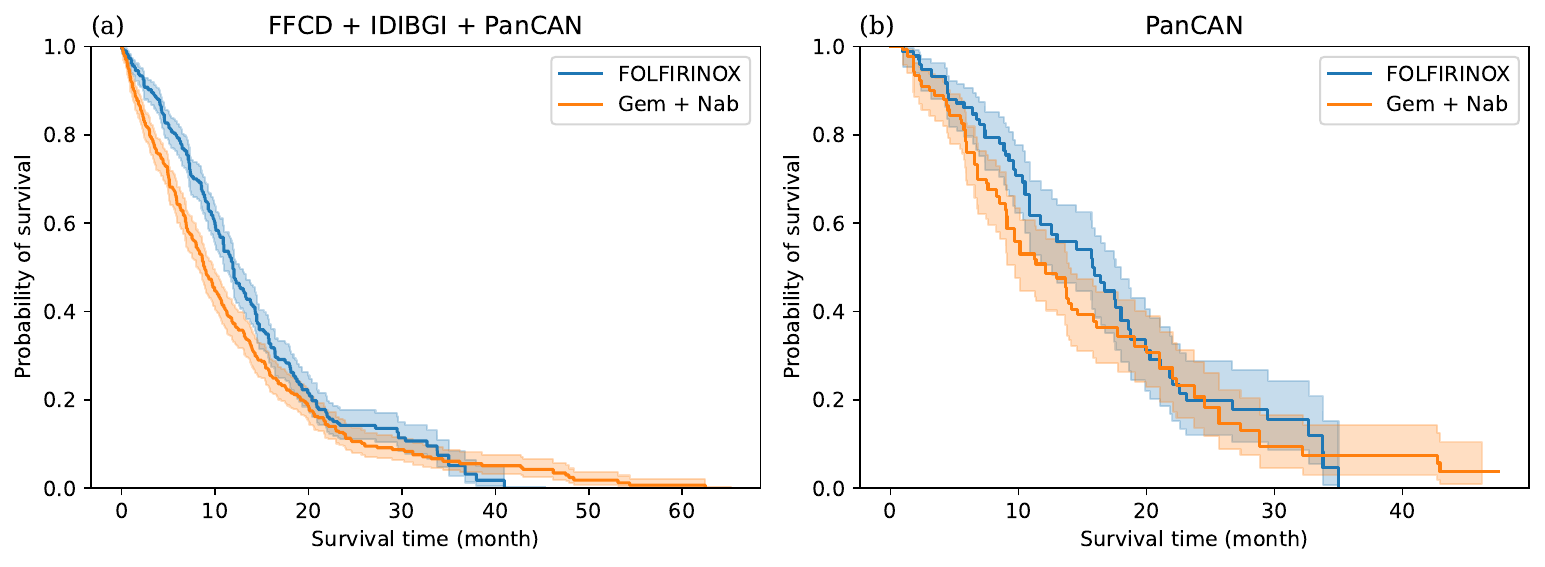}
  \caption{
        Real-world FOLFIRINOX effect estimation using FedECA versus local analyses.
      (a) Weighted Kaplan-Meier curves of the combined FFCD+IDIBGI+PanCAN cohort using FedECA's propensity model. Sample size is $n=153+225+177=555$. The $95$\% confidence intervals displayed are obtained using the exponential Greenwood formula.
      (b) Weighted Kaplan-Meier curves of the PanCAN cohort using a local propensity model. Sample size is $n=177$. The $95$\% confidence intervals displayed are obtained using the exponential Greenwood formula.
      \label{fig:real_world_all_and_pancan}
      Associated $p$-values can be found in the associated table.
      Source data are provided as a Source Data file.
  }
  
\end{figure}

\begin{table}[tbh]
  \centering
  \begin{tabular}{lllllll}
    \toprule
    \makecell{Pair of centers \\ A vs. B} & Treatment & \makecell{Number of patients \\ A / B (overall)} & log(HR) & HR (95\% CI) & Z & p \\
    \midrule
    FFCD vs. IDIBGI   & FOLFIRINOX & 92 / 33 (125)  & \num{-0.072794} & \num{0.929792} (\num{0.647945}, \num{1.334240}) & \num{-0.395048} & \num[round-mode=figures,round-precision=3]{0.692807} \\
                      & Gem + Nab  & 61 / 192 (253) & \num{-0.320409} & \num{0.725852} (\num{0.544807}, \num{0.967061}) & \num{-2.188766} & \num[round-mode=figures,round-precision=3]{0.057228} \\  %
    \midrule
    PanCAN vs. FFCD   & FOLFIRINOX & 91 / 92 (183)  & \num{-0.417989} & \num{0.658370} (\num{0.476833}, \num{0.909021}) & \num{-2.539491} & \num[round-mode=figures,round-precision=3]{0.022202} \\  %
                      & Gem + Nab  & 86 / 61 (147)  & \num{-0.190725} & \num{0.826360} (\num{0.558495}, \num{1.222698}) & \num{-0.954131} & \num[round-mode=figures,round-precision=3]{0.340017} \\
    \midrule
    PanCAN vs. IDIBGI & FOLFIRINOX & 91 / 33 (124)  & \num{-0.548156} & \num{0.578015} (\num{0.371733}, \num{0.898766}) & \num{-2.433868} & \num[round-mode=figures,round-precision=3]{0.014938} \\
                      & Gem + Nab  & 86 / 192 (278) & \num{-0.453865} & \num{0.635169} (\num{0.472882}, \num{0.853149}) & \num{-3.015006} & \num[round-mode=figures,round-precision=3]{0.005140} \\  %
    \bottomrule
  \end{tabular}
  \caption{
    Exchangeability tests for each pair of centers.
    Note that p-values are adjusted using the Holm-Bonferroni precedure
    to account for multiple testing of exchangeability at pair level.
    In other words, the exchangeability between a pair of centers only holds
    if tests in both treatment groups are not rejected.
    The test results indicate that exchangeability holds only between FFCD and IDIBGI.
    All $p$-values resulting from the Wald test performed on the entry of $\bm{\hat{\beta}}$ corresponding to the treatment allocation, assuming a $\chi^2$
    distribution with $1$ degree of freedom are obtained with bootstrap variance estimation.
    Exact $p$-values are from top to bottom $0.692807$, $0.057228$, $0.022202$, $0.340017$, $0.014938$ and $0.005140$.
    Source data are provided as a Source Data file.
  }
  \label{tab:exchang}
\end{table}

\begin{figure}
  \centering
  \includegraphics[width=\linewidth]{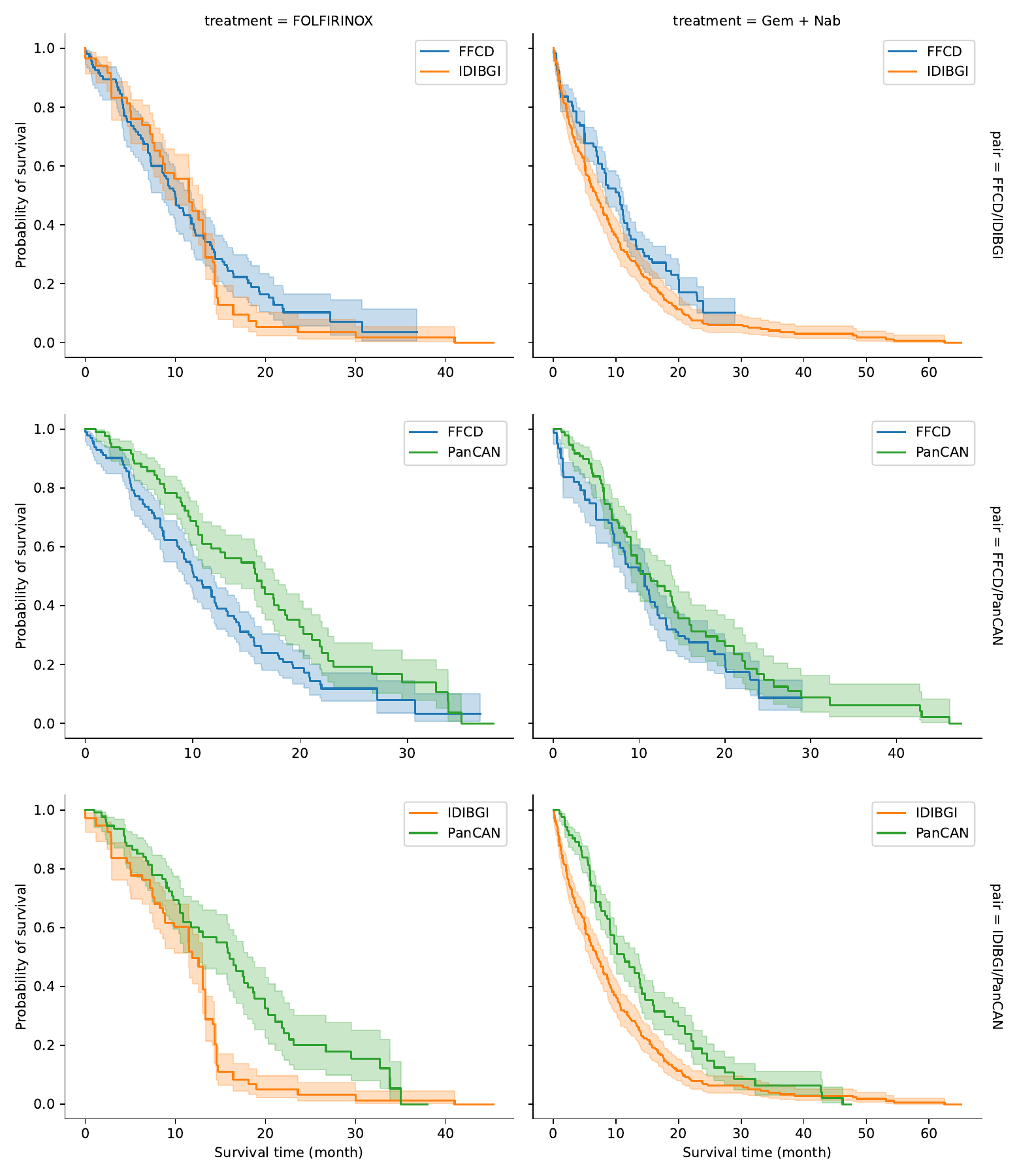}
  \caption{
          Comparison of federated weighted Kaplan-Meier curves between pair of centers
          in the same treatment group.
          Sample sizes are for top to bottom first and then from left to right:
          $n=125$, $n=253$, $n=183$, $n=147$, $n=124$ and $n=278$.
          The $95$\% confidence intervals displayed are obtained using the exponential Greenwood formula.
          Source data are provided as a Source Data file.
  }
  \label{fig:exchang_km}
\end{figure}

\begin{figure}
  \centering
\begin{subfigure}{\linewidth}
    \includegraphics[width=\linewidth]{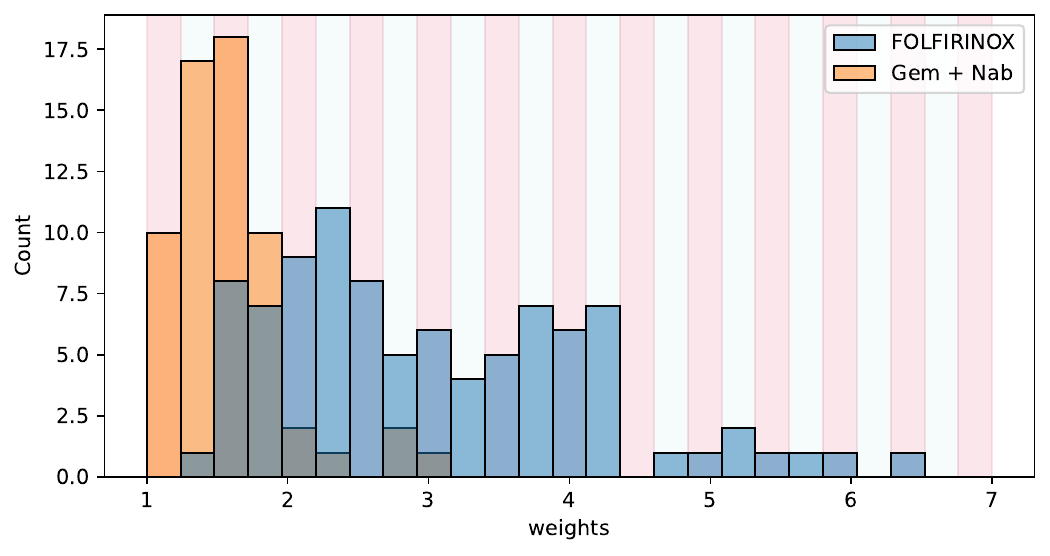}
    \subcaption{Histogram of weights in FFCD. Sample size $n=153$.}
    \label{fig:weights_hist_ffcdb}
\end{subfigure}
\begin{subfigure}{\linewidth}
    \includegraphics[width=\linewidth]{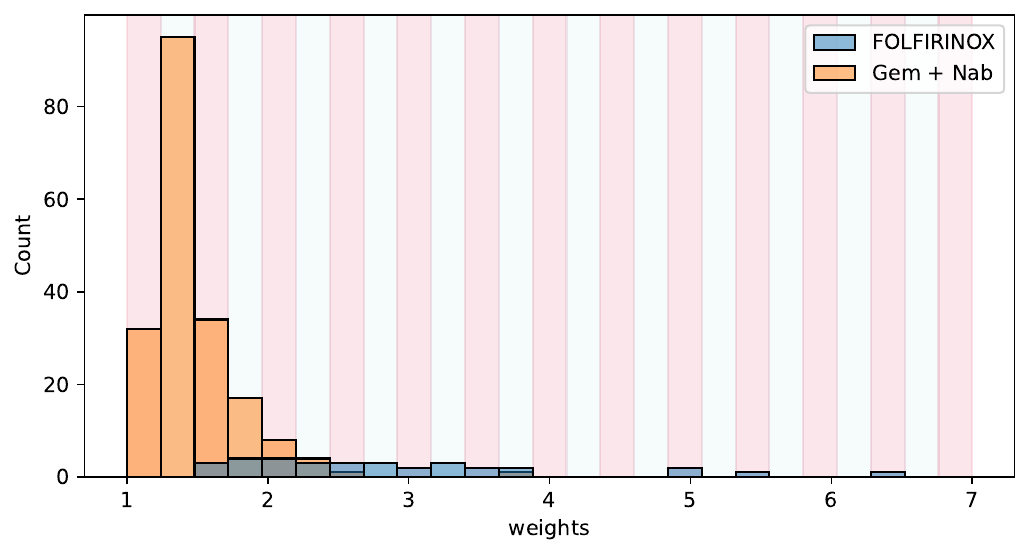}
    \subcaption{Histogram of weights in IDIBGI. Sample size $n=225$.}
    \label{fig:weights_hist_idibigi}
\end{subfigure}
  \caption{
        Weights quantized distributions in the different centers.
        We use a fixed number of $25$ histogran buckets based on global minimum and maximum
        in order to limit patient leakage. We plot all patients
        $n = 153$ for FFCD, $n = 225$ for IDIBGI.
        Source data are provided as a Source Data file.
  }
  \label{fig:hist_weights}
\end{figure}

\newpage
\begin{figure*}
  \centering
  \begin{subfigure}[b]{0.325\textwidth}
    \centering
    \includegraphics[page=1, width=1.\linewidth]{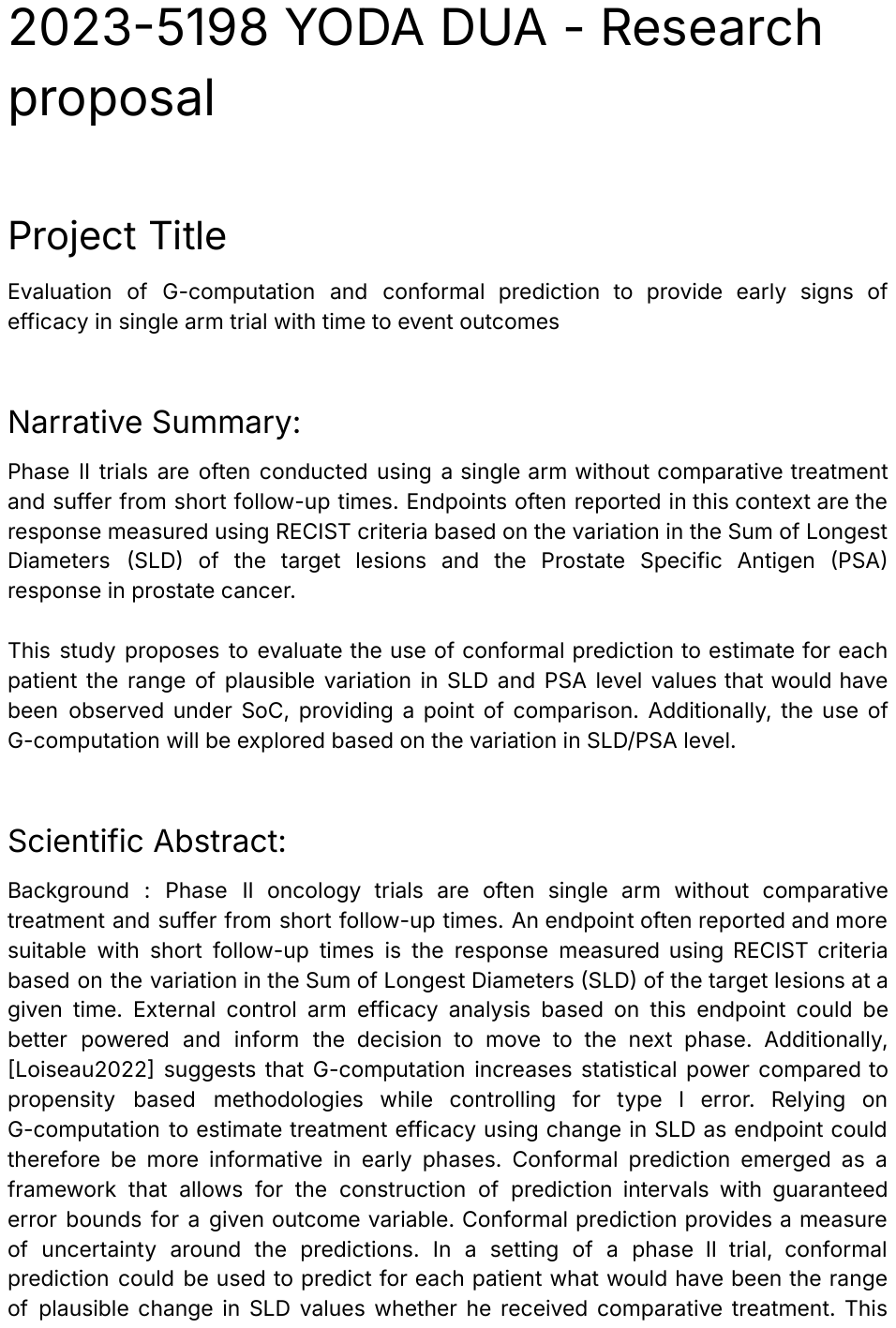}
  \end{subfigure}
  \begin{subfigure}[b]{0.325\textwidth}
    \centering
    \includegraphics[page=2, width=1.\linewidth]{YODA_UA_Research_proposal.pdf}
  \end{subfigure}
  \begin{subfigure}[b]{0.325\textwidth}
    \centering
    \includegraphics[page=3, width=1.\linewidth]{YODA_UA_Research_proposal.pdf}
  \end{subfigure}
  \begin{subfigure}[b]{0.325\textwidth}
    \centering
    \includegraphics[page=4, width=1.\linewidth]{YODA_UA_Research_proposal.pdf}
  \end{subfigure}
  \begin{subfigure}[b]{0.325\textwidth}
    \centering
    \includegraphics[page=5, width=1.\linewidth]{YODA_UA_Research_proposal.pdf}
  \end{subfigure}
  \begin{subfigure}[b]{0.325\textwidth}
    \centering
    \includegraphics[page=6, width=1.\linewidth]{YODA_UA_Research_proposal.pdf}
  \end{subfigure}
  \begin{subfigure}[b]{0.325\textwidth}
    \centering
    \includegraphics[page=7, width=1.\linewidth]{YODA_UA_Research_proposal.pdf}
  \end{subfigure}

  \caption{
      YODA DUA Research Proposal.
      The research proposal submitted to the YODA DUA program. Source data are provided as a Source Data file.
  }
  \label{fig:yoda_propal}
\end{figure*}

\begin{figure}[tb]
  \centering
  \includegraphics[width=.25\linewidth]{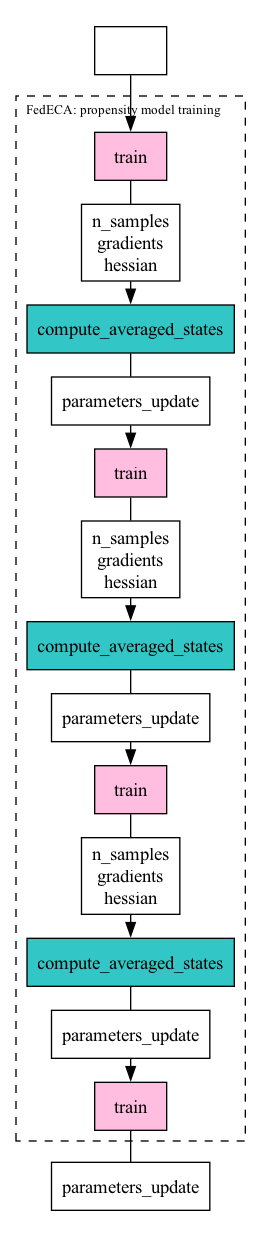}
  \includegraphics[width=.15\linewidth]{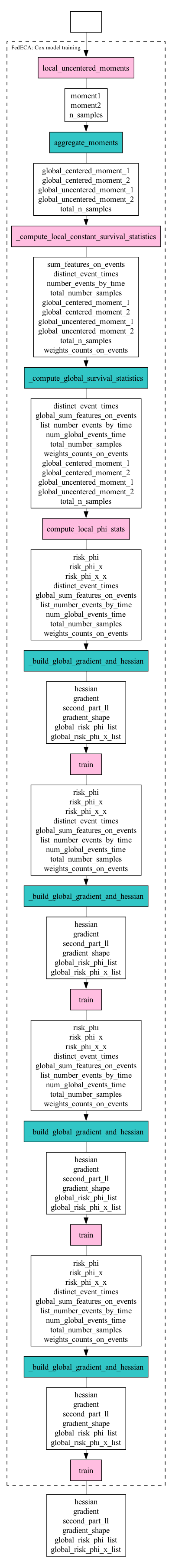}
  \caption{
      FedECA: communicated quantities during 3 rounds: non robust variance estimation. Source data are provided as a Source Data file.
  }
  \label{fig:fedeca_graph}
\end{figure}
\begin{figure}[tb]
  \centering
  \includegraphics[width=.25\linewidth]{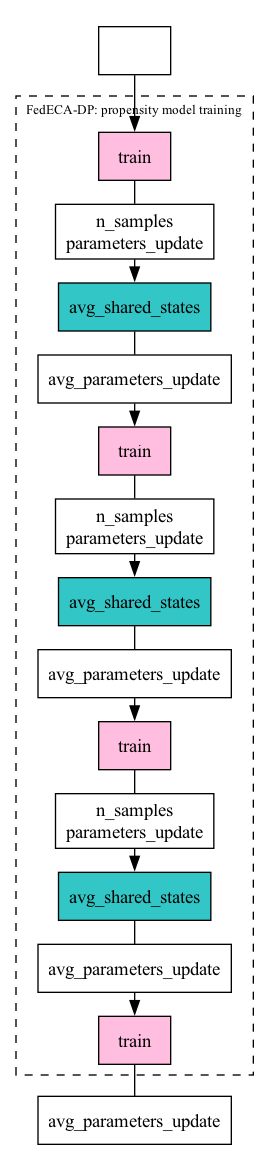}
  \includegraphics[width=.15\linewidth]{fedeca_cox_model_graph.png}
  \caption{
      DP-FedECA: communicated quantities during 3 rounds: non robust variance estimation. Source data are provided as a Source Data file.
  }
  \label{fig:fedeca_dp_graph}
\end{figure}
\begin{figure}[tb]
  \centering
  \includegraphics[width=.3\linewidth]{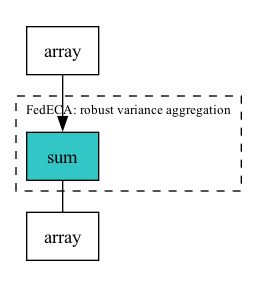}
  \caption{
      Robust Variance Estimation. Source data are provided as a Source Data file.
  }
  \label{fig:robust_cox_var}
\end{figure}
\begin{figure}[tb]
  \centering
  \includegraphics[width=.3\linewidth]{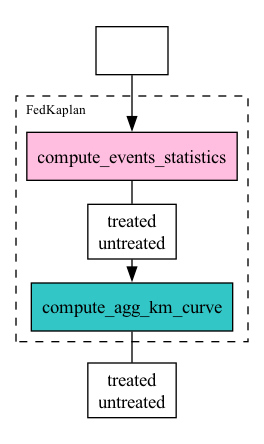}
  \caption{
      Fed-Kaplan Meier. Source data are provided as a Source Data file.
  }
  \label{fig:fed_kaplan_graph}
\end{figure}
\begin{figure}[tb]
  \centering
  \includegraphics[width=.3\linewidth]{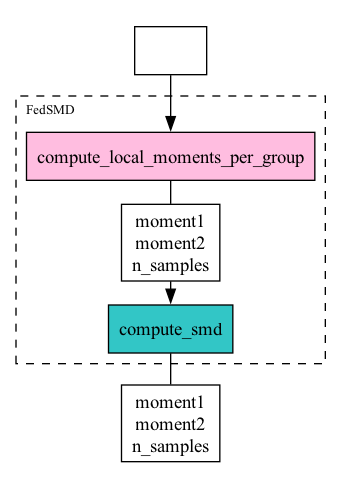}
  \caption{
      Fed-SMD. Source data are provided as a Source Data file.
  }
  \label{fig:fed_smd_graph}
\end{figure}

\newpage
\FloatBarrier
\begin{tiny}
  \begin{longtable}{l @{\hspace{0.75\tabcolsep}}l@{\hspace{0.75\tabcolsep}}l@{\hspace{0.75\tabcolsep}}l@{\hspace{0.75\tabcolsep}}p{6cm}@{\hspace{0.75\tabcolsep}}l}
  \toprule
   &  & Type & Shape & Description & Shared with \\
  ID & Name &  &  &  &  \\
  \midrule
  \multirow[t]{3}{*}{1} & n\_samples & int &  & The number of samples (scalar) of each center & Server \\
   & gradients & list &  & The local gradient of the propensity model of dimension the number of covariates. & Server \\
   & hessian & nparray & $(11, 11)$ & The hessian of the model of dimension the number of covariates squared. & Server \\
  \cline{1-6}
  2 & parameters\_update & list &  & The Newton-Raphson step computed with the global hessian. & Center \\
  \cline{1-6}
  \multirow[t]{3}{*}{3} & n\_samples & int &  & The number of samples (scalar) of each center & Server \\
   & gradients & list &  & The local gradient of the propensity model of dimension the number of covariates. & Server \\
   & hessian & nparray & $(11, 11)$ & The hessian of the model of dimension the number of covariates squared. & Server \\
  \cline{1-6}
  4 & parameters\_update & list &  & The Newton-Raphson step computed with the global hessian. & Center \\
  \cline{1-6}
  \multirow[t]{3}{*}{5} & n\_samples & int &  & The number of samples (scalar) of each center & Server \\
   & gradients & list &  & The local gradient of the propensity model of dimension the number of covariates. & Server \\
   & hessian & nparray & $(11, 11)$ & The hessian of the model of dimension the number of covariates squared. & Server \\
  \cline{1-6}
  6 & parameters\_update & list &  & The Newton-Raphson step computed with the global hessian. & Center \\
  \cline{1-6}
  7 & parameters\_update & list &  & The Newton-Raphson step computed with the global hessian. & All \\
  \cline{1-6}
  \bottomrule
  \caption{FedECA propensity model training: variables definitions. Source data are provided as a Source Data file.}
  \label{tab:prop_vars}
  \end{longtable}
  
  \end{tiny}

\newpage

\begin{tiny}
  \begin{longtable}{l @{\hspace{0.75\tabcolsep}}l@{\hspace{0.75\tabcolsep}}l@{\hspace{0.75\tabcolsep}}l@{\hspace{0.75\tabcolsep}}p{6cm}@{\hspace{0.75\tabcolsep}}l}
  \toprule
   &  & Type & Shape & Description & Shared with \\
  ID & Name &  &  &  &  \\
  \midrule
  \multirow[t]{3}{*}{1} & moment1 & Series & $(1,)$ & Uncentered (scalar) moments of order 1 for each covariate used by the Cox model. Therefore it is a list of scalars of size the number of covariates. In the case of IPTW the only covariate is the treatment. & Server \\
   & moment2 & Series & $(1,)$ & Uncentered (scalar) moments of order 2 for each covariate used by the Cox model. Therefore it is a list of scalars of size the number of covariates. In the case of IPTW the only covariate is the treatment. & Server \\
   & n\_samples & Series & $(1,)$ & The number of samples (scalar) of each center & Server \\
  \cline{1-6}
  \multirow[t]{5}{*}{2} & global\_centered\_moment\_1 & Series & $(1,)$ & Global centered moment of order 1 for each covariate used by the Cox model. Therefore it is a list of scalars of size the number of covariates. In the case of IPTW the only covariate is the treatment. & Center \\
   & global\_centered\_moment\_2 & Series & $(1,)$ & Global centered moment of order 2 for each covariate used by the Cox model. Therefore it is a list of scalars of size the number of covariates. In the case of IPTW the only covariate is the treatment. & Center \\
   & global\_uncentered\_moment\_1 & Series & $(1,)$ & Global uncentered moment of order 1 for each covariate used by the Cox model. Therefore it is a list of scalars of size the number of covariates. In the case of IPTW the only covariate is the treatment. Those values are not used in fact and are computed for testing purposes. & Center \\
   & global\_uncentered\_moment\_2 & Series & $(1,)$ & Global uncentered moment of order 2 for each covariate used by the Cox model. Therefore it is a list of scalars of size the number of covariates. In the case of IPTW the only covariate is the treatment. Those values are not used in fact and are computed for testing purposes. & Center \\
   & total\_n\_samples & int & $()$ & The total number of samples (scalar) across centers. & Center \\
  \cline{1-6}
  \multirow[t]{10}{*}{3} & weights\_counts\_on\_events & list &  & The weighted sum of samples on each distinct event times (list of scalars). & Server \\
   & total\_n\_samples & int & $()$ & The total number of samples (scalar) across centers. & Server \\
   & global\_uncentered\_moment\_2 & Series & $(1,)$ & Global uncentered moment of order 2 for each covariate used by the Cox model. Therefore it is a list of scalars of size the number of covariates. In the case of IPTW the only covariate is the treatment. Those values are not used in fact and are computed for testing purposes. & Server \\
   & global\_uncentered\_moment\_1 & Series & $(1,)$ & Global uncentered moment of order 1 for each covariate used by the Cox model. Therefore it is a list of scalars of size the number of covariates. In the case of IPTW the only covariate is the treatment. Those values are not used in fact and are computed for testing purposes. & Server \\
   & global\_centered\_moment\_2 & Series & $(1,)$ & Global centered moment of order 2 for each covariate used by the Cox model. Therefore it is a list of scalars of size the number of covariates. In the case of IPTW the only covariate is the treatment. & Server \\
   & total\_number\_samples & int &  & The total number of samples (scalar). & Server \\
   & number\_events\_by\_time & list &  & The number of events per distinct times (scalar). & Server \\
   & distinct\_event\_times & list &  & All distinct event times, which is list of scalars. & Server \\
   & sum\_features\_on\_events & nparray & $(1,)$ & the global sum of features of all samples on each distinct event times & Server \\
   & global\_centered\_moment\_1 & Series & $(1,)$ & Global centered moment of order 1 for each covariate used by the Cox model. Therefore it is a list of scalars of size the number of covariates. In the case of IPTW the only covariate is the treatment. & Server \\
  \cline{1-6}
  \multirow[t]{11}{*}{4} & global\_centered\_moment\_2 & Series & $(1,)$ & Global centered moment of order 2 for each covariate used by the Cox model. Therefore it is a list of scalars of size the number of covariates. In the case of IPTW the only covariate is the treatment. & Center \\
   & total\_n\_samples & int & $()$ & The total number of samples (scalar) across centers. & Center \\
   & global\_uncentered\_moment\_2 & Series & $(1,)$ & Global uncentered moment of order 2 for each covariate used by the Cox model. Therefore it is a list of scalars of size the number of covariates. In the case of IPTW the only covariate is the treatment. Those values are not used in fact and are computed for testing purposes. & Center \\
   & global\_uncentered\_moment\_1 & Series & $(1,)$ & Global uncentered moment of order 1 for each covariate used by the Cox model. Therefore it is a list of scalars of size the number of covariates. In the case of IPTW the only covariate is the treatment. Those values are not used in fact and are computed for testing purposes. & Center \\
   & global\_centered\_moment\_1 & Series & $(1,)$ & Global centered moment of order 1 for each covariate used by the Cox model. Therefore it is a list of scalars of size the number of covariates. In the case of IPTW the only covariate is the treatment. & Center \\
   & weights\_counts\_on\_events & list &  & The weighted sum of samples on each distinct event times (list of scalars). & Center \\
   & total\_number\_samples & int &  & The total number of samples (scalar). & Center \\
   & num\_global\_events\_time & int &  & The number of distinct event times globally, which is also the size of distinct\_event\_times list. & Center \\
   & list\_number\_events\_by\_time & list &  & The number of events at each distinct event times that is a list of scalars. & Center \\
   & global\_sum\_features\_on\_events & nparray & $(1,)$ & The global sum of covariates on each event times, a list of inputs of dimensions the number of covariates. Quantity which doesn't change through times. & Center \\
   & distinct\_event\_times & list &  & All distinct event times, which is list of scalars. & Center \\
  \cline{1-6}
  \multirow[t]{9}{*}{5} & weights\_counts\_on\_events & list &  & The weighted sum of samples on each distinct event times (list of scalars). & Server \\
   & total\_number\_samples & int &  & The total number of samples (scalar). & Server \\
   & list\_number\_events\_by\_time & list &  & The number of events at each distinct event times that is a list of scalars. & Server \\
   & global\_sum\_features\_on\_events & nparray & $(1,)$ & The global sum of covariates on each event times, a list of inputs of dimensions the number of covariates. Quantity which doesn't change through times. & Server \\
   & num\_global\_events\_time & int &  & The number of distinct event times globally, which is also the size of distinct\_event\_times list. & Server \\
   & risk\_phi\_x\_x & list &  & Local sums on each risk sets of risk\_phi\_x\_x globally that is a list of vector of dimensions number of features squared that is involved in the hessian computation. & Server \\
   & risk\_phi\_x & list &  & Local sums on each risk sets of risk\_phi\_x globally that is a list of vector of dimensions number of features and that is involved in the gradient computation. & Server \\
   & risk\_phi & list &  & Local sums on each risk sets of risk\_phi globally that is a list of scalars. & Server \\
   & distinct\_event\_times & list &  & All distinct event times, which is list of scalars. & Server \\
  \cline{1-6}
  \multirow[t]{6}{*}{6} & hessian & nparray & $(1, 1)$ & The hessian of the model of dimension the number of covariates squared. & Center \\
   & gradient & nparray & $(1,)$ & The global gradient of the Cox model of dimension the number of covariates. & Center \\
   & second\_part\_ll & nparray & $(1,)$ & Quantity necessary to compute the log-likelihood. & Center \\
   & gradient\_shape & int &  & The shape of the gradient (scalar). & Center \\
   & global\_risk\_phi\_list & list &  & Global sums on each risk sets of risk\_phi globally that is a list of scalars. & Center \\
   & global\_risk\_phi\_x\_list & list &  & Global sums on each risk sets of risk\_phi\_x globally that is a list of vector of dimensions number of features. & Center \\
  \cline{1-6}
  \multirow[t]{9}{*}{7} & num\_global\_events\_time & int &  & The number of distinct event times globally, which is also the size of distinct\_event\_times list. & Server \\
   & weights\_counts\_on\_events & list &  & The weighted sum of samples on each distinct event times (list of scalars). & Server \\
   & total\_number\_samples & int &  & The total number of samples (scalar). & Server \\
   & list\_number\_events\_by\_time & list &  & The number of events at each distinct event times that is a list of scalars. & Server \\
   & distinct\_event\_times & list &  & All distinct event times, which is list of scalars. & Server \\
   & global\_sum\_features\_on\_events & nparray & $(1,)$ & The global sum of covariates on each event times, a list of inputs of dimensions the number of covariates. Quantity which doesn't change through times. & Server \\
   & risk\_phi\_x\_x & list &  & Local sums on each risk sets of risk\_phi\_x\_x globally that is a list of vector of dimensions number of features squared that is involved in the hessian computation. & Server \\
   & risk\_phi\_x & list &  & Local sums on each risk sets of risk\_phi\_x globally that is a list of vector of dimensions number of features and that is involved in the gradient computation. & Server \\
   & risk\_phi & list &  & Local sums on each risk sets of risk\_phi globally that is a list of scalars. & Server \\
  \cline{1-6}
  \multirow[t]{6}{*}{8} & hessian & nparray & $(1, 1)$ & The hessian of the model of dimension the number of covariates squared. & Center \\
   & gradient & nparray & $(1,)$ & The global gradient of the Cox model of dimension the number of covariates. & Center \\
   & second\_part\_ll & nparray & $(1,)$ & Quantity necessary to compute the log-likelihood. & Center \\
   & gradient\_shape & int &  & The shape of the gradient (scalar). & Center \\
   & global\_risk\_phi\_list & list &  & Global sums on each risk sets of risk\_phi globally that is a list of scalars. & Center \\
   & global\_risk\_phi\_x\_list & list &  & Global sums on each risk sets of risk\_phi\_x globally that is a list of vector of dimensions number of features. & Center \\
  \cline{1-6}
  \multirow[t]{9}{*}{9} & weights\_counts\_on\_events & list &  & The weighted sum of samples on each distinct event times (list of scalars). & Server \\
   & total\_number\_samples & int &  & The total number of samples (scalar). & Server \\
   & num\_global\_events\_time & int &  & The number of distinct event times globally, which is also the size of distinct\_event\_times list. & Server \\
   & list\_number\_events\_by\_time & list &  & The number of events at each distinct event times that is a list of scalars. & Server \\
   & risk\_phi & list &  & Local sums on each risk sets of risk\_phi globally that is a list of scalars. & Server \\
   & distinct\_event\_times & list &  & All distinct event times, which is list of scalars. & Server \\
   & risk\_phi\_x\_x & list &  & Local sums on each risk sets of risk\_phi\_x\_x globally that is a list of vector of dimensions number of features squared that is involved in the hessian computation. & Server \\
   & risk\_phi\_x & list &  & Local sums on each risk sets of risk\_phi\_x globally that is a list of vector of dimensions number of features and that is involved in the gradient computation. & Server \\
   & global\_sum\_features\_on\_events & nparray & $(1,)$ & The global sum of covariates on each event times, a list of inputs of dimensions the number of covariates. Quantity which doesn't change through times. & Server \\
  \cline{1-6}
  \multirow[t]{6}{*}{10} & global\_risk\_phi\_list & list &  & Global sums on each risk sets of risk\_phi globally that is a list of scalars. & Center \\
   & global\_risk\_phi\_x\_list & list &  & Global sums on each risk sets of risk\_phi\_x globally that is a list of vector of dimensions number of features. & Center \\
   & gradient\_shape & int &  & The shape of the gradient (scalar). & Center \\
   & gradient & nparray & $(1,)$ & The global gradient of the Cox model of dimension the number of covariates. & Center \\
   & hessian & nparray & $(1, 1)$ & The hessian of the model of dimension the number of covariates squared. & Center \\
   & second\_part\_ll & nparray & $(1,)$ & Quantity necessary to compute the log-likelihood. & Center \\
  \cline{1-6}
  \multirow[t]{9}{*}{11} & total\_number\_samples & int &  & The total number of samples (scalar). & Server \\
   & weights\_counts\_on\_events & list &  & The weighted sum of samples on each distinct event times (list of scalars). & Server \\
   & list\_number\_events\_by\_time & list &  & The number of events at each distinct event times that is a list of scalars. & Server \\
   & global\_sum\_features\_on\_events & nparray & $(1,)$ & The global sum of covariates on each event times, a list of inputs of dimensions the number of covariates. Quantity which doesn't change through times. & Server \\
   & num\_global\_events\_time & int &  & The number of distinct event times globally, which is also the size of distinct\_event\_times list. & Server \\
   & risk\_phi\_x\_x & list &  & Local sums on each risk sets of risk\_phi\_x\_x globally that is a list of vector of dimensions number of features squared that is involved in the hessian computation. & Server \\
   & risk\_phi\_x & list &  & Local sums on each risk sets of risk\_phi\_x globally that is a list of vector of dimensions number of features and that is involved in the gradient computation. & Server \\
   & risk\_phi & list &  & Local sums on each risk sets of risk\_phi globally that is a list of scalars. & Server \\
   & distinct\_event\_times & list &  & All distinct event times, which is list of scalars. & Server \\
  \cline{1-6}
  \multirow[t]{6}{*}{12} & global\_risk\_phi\_list & list &  & Global sums on each risk sets of risk\_phi globally that is a list of scalars. & Center \\
   & hessian & nparray & $(1, 1)$ & The hessian of the model of dimension the number of covariates squared. & Center \\
   & gradient & nparray & $(1,)$ & The global gradient of the Cox model of dimension the number of covariates. & Center \\
   & second\_part\_ll & nparray & $(1,)$ & Quantity necessary to compute the log-likelihood. & Center \\
   & gradient\_shape & int &  & The shape of the gradient (scalar). & Center \\
   & global\_risk\_phi\_x\_list & list &  & Global sums on each risk sets of risk\_phi\_x globally that is a list of vector of dimensions number of features. & Center \\
  \cline{1-6}
  \multirow[t]{6}{*}{13} & gradient\_shape & int &  & The shape of the gradient (scalar). & All \\
   & second\_part\_ll & nparray & $(1,)$ & Quantity necessary to compute the log-likelihood. & All \\
   & global\_risk\_phi\_list & list &  & Global sums on each risk sets of risk\_phi globally that is a list of scalars. & All \\
   & hessian & nparray & $(1, 1)$ & The hessian of the model of dimension the number of covariates squared. & All \\
   & gradient & nparray & $(1,)$ & The global gradient of the Cox model of dimension the number of covariates. & All \\
   & global\_risk\_phi\_x\_list & list &  & Global sums on each risk sets of risk\_phi\_x globally that is a list of vector of dimensions number of features. & All \\
  \cline{1-6}
  \bottomrule
  \caption{FedECA Cox model training: variables definitions. Source data are provided as a Source Data file.}
  \label{tab:webdisco_vars}
  \end{longtable}
  \end{tiny}

  \newpage

  \begin{tiny}
    \begin{longtable}{l @{\hspace{0.75\tabcolsep}}l@{\hspace{0.75\tabcolsep}}l@{\hspace{0.75\tabcolsep}}l@{\hspace{0.75\tabcolsep}}p{6cm}@{\hspace{0.75\tabcolsep}}l}
    \toprule
     &  & Type & Shape & Description & Shared with \\
    ID & Name &  &  &  &  \\
    \midrule
    0 & array & nparray & $(1, 1)$ & The Qk matrices on each client. & Server \\
    \cline{1-6}
    1 & array & nparray & $(1, 1)$ & Sum of the Qk matrices across clients. & All \\
    \cline{1-6}
    \bottomrule
    \caption{FedECA Robust Cox variance estimation. Source data are provided as a Source Data file.}
    \label{tab:cox_robust_var}
    \end{longtable}
    
    \end{tiny}

\newpage

  \begin{tiny}
    \begin{longtable}{l @{\hspace{0.75\tabcolsep}}l@{\hspace{0.75\tabcolsep}}l@{\hspace{0.75\tabcolsep}}l@{\hspace{0.75\tabcolsep}}p{6cm}@{\hspace{0.75\tabcolsep}}l}
    \toprule
     &  & Type & Shape & Description & Shared with \\
    ID & Name &  &  &  &  \\
    \midrule
    \multirow[t]{2}{*}{1} & n\_samples & int &  & The number of samples (scalar) of each center & Server \\
     & parameters\_update & list &  & The Newton-Raphson step computed with the global hessian. & Server \\
    \cline{1-6}
    2 & avg\_parameters\_update & list &  & The average of the gradient from each center. & Center \\
    \cline{1-6}
    \multirow[t]{2}{*}{3} & n\_samples & int &  & The number of samples (scalar) of each center & Server \\
     & parameters\_update & list &  & The Newton-Raphson step computed with the global hessian. & Server \\
    \cline{1-6}
    4 & avg\_parameters\_update & list &  & The average of the gradient from each center. & Center \\
    \cline{1-6}
    \multirow[t]{2}{*}{5} & n\_samples & int &  & The number of samples (scalar) of each center & Server \\
     & parameters\_update & list &  & The Newton-Raphson step computed with the global hessian. & Server \\
    \cline{1-6}
    6 & avg\_parameters\_update & list &  & The average of the gradient from each center. & Center \\
    \cline{1-6}
    7 & avg\_parameters\_update & list &  & The average of the gradient from each center. & All \\
    \cline{1-6}
    \bottomrule
    \caption{FedECA-DP propensity model training: variables definitions}
    \label{tab:dp_prop_vars}
    \end{longtable}
    \end{tiny}

  \newpage

    \begin{tiny}
      \begin{longtable}{l @{\hspace{0.75\tabcolsep}}l@{\hspace{0.75\tabcolsep}}l@{\hspace{0.75\tabcolsep}}l@{\hspace{0.75\tabcolsep}}p{6cm}@{\hspace{0.75\tabcolsep}}l}
      \toprule
       &  & Type & Shape & Description & Shared with \\
      ID & Name &  &  &  &  \\
      \midrule
      \multirow[t]{2}{*}{1} & treated & tuple &  & Statistics dict computed on the treated population and containing 1. unique times of events in ascending order (list of scalars), the (weighted) number of individual at risks at each corresponding unique times (list of scalars), the (weighted) number of individuals with an event (death) at each corresponding unique times (list of scalars). & Server \\
       & untreated & tuple &  & Statistics dict computed on the treated population and containing 1. unique times of events in ascending order (list of scalars), the (weighted) number of individual at risks at each corresponding unique times (list of scalars), the (weighted) number of individuals with an event (death) at each corresponding unique times (list of scalars). & Server \\
      \cline{1-6}
      \multirow[t]{2}{*}{2} & treated & tuple &  & Statistics dict computed on the treated population and containing 1. unique times of events in ascending order (list of scalars), the (weighted) number of individual at risks at each corresponding unique times (list of scalars), the (weighted) number of individuals with an event (death) at each corresponding unique times (list of scalars). & All \\
       & untreated & tuple &  & Statistics dict computed on the treated population and containing 1. unique times of events in ascending order (list of scalars), the (weighted) number of individual at risks at each corresponding unique times (list of scalars), the (weighted) number of individuals with an event (death) at each corresponding unique times (list of scalars). & All \\
      \cline{1-6}
      \bottomrule
      \caption{Federated Kaplan Meier analytics: variables definition }
      \label{tab:fed_kaplan_vars}
      \end{longtable}
      \end{tiny}
\newpage
\begin{tiny}
  \begin{longtable}{l @{\hspace{0.75\tabcolsep}}l@{\hspace{0.75\tabcolsep}}l@{\hspace{0.75\tabcolsep}}l@{\hspace{0.75\tabcolsep}}p{6cm}@{\hspace{0.75\tabcolsep}}l}
  \toprule
    &  & Type & Shape & Description & Shared with \\
  ID & Name &  &  &  &  \\
  \midrule
  \multirow[t]{3}{*}{1} & moment1 & Series &  & Local uncentered weighted (scalar) moments of order 1 for each covariate used by the Cox model for both populations. Therefore it is a list of scalars of size the number of covariates. In the case of IPTW the only covariate is the treatment. & Server \\
    & moment2 & Series &  & Local uncentered weighted (scalar) moments of order 2 for each covariate used by the Cox model for both populations.. Therefore it is a list of scalars of size the number of covariates. In the case of IPTW the only covariate is the treatment. & Server \\
    & n\_samples & Series &  & The number of samples (scalar) of each center locally & Server \\
  \cline{1-6}
  \multirow[t]{3}{*}{2} & moment1 & Series &  & Uncentered weighted (scalar) moments of order 1 for each covariate used by the Cox model for both populations. Therefore it is a list of scalars of size the number of covariates. In the case of IPTW the only covariate is the treatment. & All \\
    & moment2 & Series &  & Global uncentered weighted (scalar) moments of order 2 for each covariate used by the Cox model for both populations. Therefore it is a list of scalars of size the number of covariates. In the case of IPTW the only covariate is the treatment. & All \\
    & n\_samples & Series &  & The global number of samples (scalar) of each center & All \\
  \cline{1-6}
  \bottomrule
  \caption{Federated SMD analytics: variables definition }
  \label{tab:fed_smd_vars}
  \end{longtable}
  
  \end{tiny}

\end{document}